\documentclass[iop]{emulateapj} 

\usepackage{graphicx}
\usepackage{epstopdf}
\newcommand{\hour}{\hbox{$^{\rm h}$}}
\newcommand{\minute}{\hbox{$^{\rm m}$}}

\usepackage{lscape}

\begin{document}

\title{$350\,\micron$ Polarimetry from the Caltech
  Submillimeter Observatory}

\author{
Jessie~L.~Dotson,\altaffilmark{1}
John~E.~Vaillancourt,\altaffilmark{2}
Larry~Kirby,\altaffilmark{3}
C.~Darren~Dowell,\altaffilmark{4}
Roger~H.~Hildebrand,\altaffilmark{3,5}
\\and
Jacqueline~A.~Davidson\altaffilmark{6}
}
\altaffiltext{1}{NASA Ames Research Center, Astrophysics Branch, MS
  245-6, Moffett Field, CA 94035 \email{jessie.dotson@nasa.gov}}
\altaffiltext{2}{Division of Physics, Mathematics, \& Astronomy,
  California Institute of Technology \email{johnv@submm.caltech.edu}
  Current address: Universities Space Research Association, NASA Ames
  Research Center, MS 211-3, Moffet Field, CA 94035}
\altaffiltext{3}{Enrico Fermi Institute and Department of Astronomy \&
  Astrophysics, University of Chicago}
\altaffiltext{4}{Jet Propulsion Laboratory}
\altaffiltext{5}{also Deptartment of Physics}
\altaffiltext{6}{University of Western Australia}

\begin{abstract}
We present a summary of data obtained with the $350\,\micron$
polarimeter, Hertz, at the Caltech Submillimeter Observatory. We give
tabulated results and maps showing polarization vectors and flux
contours. The summary includes over 4300 individual measurements in 56
Galactic sources and 2 galaxies.  Of these measurements, 2153 have $P
\geq 3\sigma_p$ statistical significance.  The median polarization of
the entire data set is 1.46\%.
\end{abstract}


\section{Introduction}

The University of Chicago $350\,\micron$ polarimeter, Hertz\footnote{%
Now replaced by a new polarimeter, SHARP \citep{gn04,li07}.}, was
operated at the Caltech Submillimeter Observatory (CSO) from 1994
to 2003. Many of the results have been published in some form but
usually without giving complete information. In this paper we
present a catalog of all the measurements giving coordinates,
position angles, degrees of polarization, and uncertainties as well as maps
showing polarization vectors and flux contours derived from the same
measurements. An earlier paper by \citet{jld00} presented a similar
summary for results\ at $60\,\micron$ and $100\,\micron$ from the
far-infrared polarimeter, Stokes, operated on the Kuiper Airborne
Observatory.

For these bright sources we attribute the polarized flux to thermal
emission from magnetically aligned non-spherical dust grains (e.g.,
\citealt{rhh88}).  The degree and angle of the polarization are useful
for investigating the intrinsic properties, alignment mechanism, and
populations of dust grains (e.g., \citealt{rhh99}, \citealt{rhh95})
and the structure of magnetic fields (e.g., \citealt{rhh90},
\citealt{das00}, \citealt{houde02}).

We discuss the instrument and observations in Section~2, and the
instrumental and systematic effects in Section~3.  In Section~4 we
present the object list (Table~1) and table of results (Table~2).
In Table~1 
we also give references where earlier papers have been based on the
same data.  In all other cases, this is the first presentation of
complete data sets.

All results in Table~2 and the figures are for polarization vectors.
The inferred directions of the magnetic field vectors (not shown) are
orthogonal to the polarization vectors.  The table contains all the
data satisfying the criterion $F > 3\sigma_F$, where $F$ is the total
unpolarized flux density.

\section{Instrument and Observations}

\label{sec-instNobs} The instrument, calibrations, and data analysis
techniques have been described in detail by \citet{das97},
\citet{cdd98}, and \citet{lk05}. Here we give a brief overview of the
instrument and observing techniques.

An interference filter provided peak transmission at $353\,\micron$
and a bandwidth of $62\,\micron$ (see Figure 1 in \citealt{cdd98}).
After passing through a quartz half-wave plate turned by a stepper
motor, the incoming radiation was separated into orthogonal components
of polarization by a polarizing grid inclined at $45\arcdeg$ to the
optic axis.  The reflected and transmitted components were detected
simultaneously for each pixel on the sky by corresponding pairs of NTD
germanium bolometers in arrays of 32 pixels each, arranged in a
6$\times $6 matrix with the corners omitted. Winston cones
concentrated the radiation into cylindrical cavities containing
silicon bolometers cooled to 0.26\,K by a dual stage $^{3}$He
cryostat.

The half-wave plate was moved in $30\arcdeg$ steps through six
positions. At each position standard chopped and beam-switched
photometric observations were performed. The chop throw varied between
$2\arcmin$ and $8\arcmin$ depending on the size of the object. The chop
frequency was $\sim 3$\,Hz.  Each data file took approximately 4 -- 5
minutes including time spent integrating, beam switching, and stepping
the halfwave plate.  In excellent weather conditions the noise
equivalent flux density for measurement of polarized flux was 3 -- 4
Jy\,Hz$^{-1/2}$.

Relative photometric measurements were obtained simultaneously with
the polarimetric measurements.  Array flat fielding was obtained by
pixel dithering as described by \citet{cdd98}. The individual data
files were corrected for their gain by normalizing to the bright flux
peak. In most cases, we did not Nyquist sample the image plane; the
center-to-center spacing of the detector array was measured to be
$17\farcs8$.  As a result, the resolution in our flux maps is $\sim
\sqrt{2}\ \times$ our nominal beamsize of 20\arcsec.

Peak fluxes for all objects presented here are given in
Table~\ref{tbl-object} and in the captions of Figures \ref{fig5} --
\ref{fig59}.  Fluxes for eleven objects were found in the literature
(see Table~\ref{tbl-object}, ``Flux Reference'' column).  These eleven
objects, along with observations of W3OH, Mars, Uranus, and Jupiter,
were taken as flux standards for calibrating all remaining objects.
The calibrations were performed with respect to the standards within
each observing run.  Average fluxes are reported for objects observed
in multiple runs.

\subsection{Instrumental Effects}

All of the measurements made during an observing run were combined to
calculate the telescope and instrumental polarization as described by
\citet{srp91}.  The mean instrumental polarization across the array
varied from 0.23\% -- 0.38\% on different observing runs.  The
telescope polarization varied from 0.04\% -- 0.48\% on different runs,
with a mean value of 0.22\%.  All results were corrected for telescope
and instrument polarizations as well as a polarization efficiency of
95\%.

\subsection{Systematic Effects}
A major obstacle to the measurement of submillimeter polarization is
the inevitable fluctuations in atmospheric transmission and emission.
The removal of background sky emission is mostly achieved by observing
reference regions away from the source of interest and subtracting the
reference position measurement from the on-source position
measurement.  Rapid (3 -- 4 Hz) modulation (``chopping'') of the
telescope secondary mirror between off-source and on-source positions
allows removal of most fluctuations in sky emission.  Slow
($\sim0.05$\,Hz) beam switching (``nodding'') facilitates removal of
linear gradients in the sky emission.  The effects of fluctuations in
the transmission of the atmosphere are largely removed by observing
two orthogonal components of polarization simultaneously.  While the
instrumental design of Hertz (two-component polarimetry) and our
observing strategy (chop-nod) allow us to remove many of the effects
of the sky fluctuations, these strategies do not allow us to correct
for fluctuations on all the relevant physical and time scales, so
there is still a residual noise in our measurements due to sky
fluctuations.  A relatively recent data analysis approach, developed
by \citet{lk05}, capitalizes on making repeated observations of the
source and measurements of the atmospheric optical depth to further
reduce the residual effects of sky fluctuation.  While our traditional
two-array measurement of the polarization (e.g., \citealt{srp91}) is
quite effective at measuring polarized flux, the \citet{lk05} approach
significantly improves the estimation of the reduced Stokes
parameters (polarized flux divided by total flux).  This new approach
has allowed us to extend measurements to fainter sources.

If the distance between the source and the reference positions is
small compared to the size of the source, then it is possible to
erroneously subtract source flux along with the sky background.  Large
chop throws ($\sim 6\arcmin$) were chosen for most Hertz observations
in order to mitigate this problem.  The size and range of directions
of the chop used for each object are given in Table~\ref{tbl-object}.

Despite the large chop throws available at the Caltech Submillimeter
Observatory, it is not always possible to avoid the problem of source
flux in the reference beam.  This is not a significant problem for
submillimeter photometry as one can at least report the flux
difference between the source and reference positions.  However, this
is not the case for polarimetry because the polarizations do not
subtract simply.  Methods to estimate these systematic uncertainties
have been developed by \citet{das97} and \citet{gn97}.

Estimates for the limits to these uncertainties include: (1) $\Delta
P_\mathrm{sys}^+$, the maximum amount by which the actual value of
$P$\/ could be larger than the measured value, (2) $\Delta
P_\mathrm{sys}^-$, the maximum amount by which it could be smaller,
and (3) $\Delta\phi_\mathrm{sys}$, the maximum amount by which the
measured position angle could be in error.  The magnitudes of these
limits are dependent on the ratio of flux in the source beam to that
in the reference beam and the ratio of the polarization in the source
and reference beams (for more details see \citealt{das97} and
\citealt{gn97}).  For any given polarization measurement, these
maximum uncertainties can be estimated if the polarized flux is known
in both the source and reference positions.  However, few large scale
$350\,\micron$ photometric maps, and no large scale submillimeter
polarimetric maps, exist from which to make these estimates.
Therefore, we have made no effort to estimate these systematics here.

\subsection{Positive Bias}

Since polarization is an inherently positive quantity, a positive bias
is introduced into any measurement with noise.  For high signal-to-noise ratios
($P/\sigma_p \gtrsim 4$) measurements can be best corrected for this
bias using the relation
\begin{equation}
P_0 \approx \sqrt{P_m^2 - \sigma_p^2}
\label{eq-bias}
\end{equation}
where $P_0$ is the corrected polarization, $P_m$ the measured
polarization, and $\sigma_p$ the measured uncertainty in the
polarization (e.g., \citealt{simmons85}).  Extending this relation to
slightly lower signal-to-noise yields a bias less than $0.06\times
P_m$ for $P_m > 3\sigma_p$.  Given this small level of bias, and the
fact that equation (\ref{eq-bias}) is not an exact solution for
$P/\sigma_p \lesssim 4$, we have made no attempt to correct the
polarizations presented in Table~2.  However, we caution that readers
wishing to make precise comparisons between this archive and other
polarization measurements should consider that some bias-correcting
scheme may be appropriate for some studies.

The bias correction \emph{has}\/ been applied to all plotted vectors
in Figures \ref{fig5} -- \ref{fig59}, including those with $P/\sigma_p
< 4$.  With this correction, any measurement with $P_m < \sigma_p$
yields $P_0 = 0$.  However, upper limits on the polarization can still
be placed on these measurements \citep{vaillancourt06}.  These points
are shown by circles on the maps when $P+2\sigma_p < 1\%$. This
$2\sigma$ criterion represents confidence levels between 95 -- 99\%
depending on the exact level of the polarization signal-to-noise.

\section{Results}

We have made two cuts to the polarization data set in order to
eliminate excessively noisy points. First we require that each
position be measured with at least 3 polarization files.  Second,
every measured point must have a flux signal-to-noise ratio $\geq$ 3.
When the flux S/N $<$ 3, we have not successfully measured total flux,
making it impossible to have successfully measured the polarization.
Even with flux S/N just above 3, it is unlikely that we have
successfully measured the source polarization, but in many cases the
flux information may be of use.  We have made no cuts based on the
value of the polarization or of the statistical uncertainty of the
polarization measurements; such cuts could bias the statistics of the
polarization distribution.  The cuts we did apply should only
reinforce the already existing bias of this data set towards brighter
regions of submillimeter flux.

\begin{figure}
\plotone{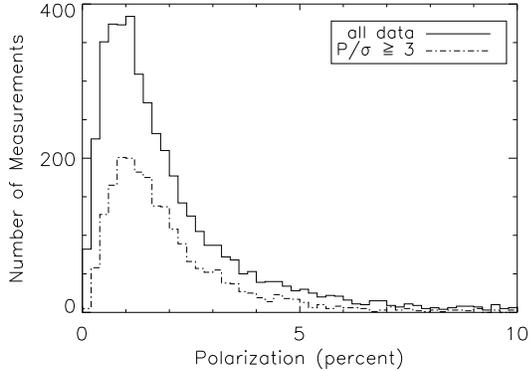} 
\caption{Distribution of all measured polarizations (solid line) and of
  measurements with a $3\sigma_p$ or better statisitical
  significance (dash-dotted line).  No correction has been made for
  positive bias (see Section~2.3).
\label{histogram}}
\end{figure}

The distribution in degrees of polarization is shown in
Figure~\ref{histogram}.  Among the 4372 measured points, 2153 have
3$\sigma$ or greater statistical significance, 990 $\geq$ 6$\sigma $,
and 350 $\geq$ 12$\sigma $.  The median $P$ and $P/\sigma_p$ for all
of the measurements are 1.46\% and 2.93, respectively. The
distribution of measured polarizations and how the distribution
changes with wavelength are discussed by Hildebrand et al.\ (1999).

\begin{figure*}
\plotone{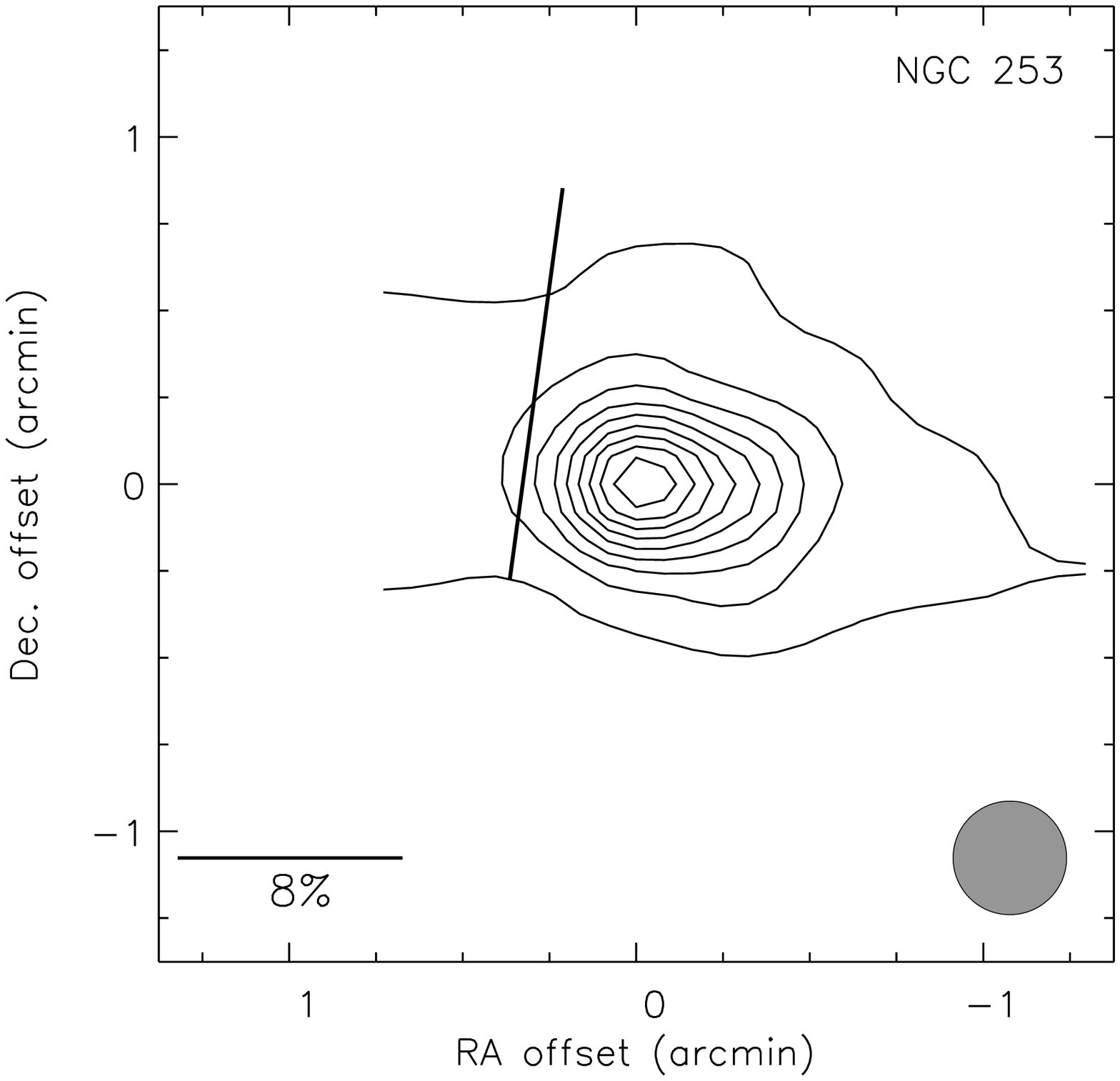}
\caption{NGC 253.
Offsets from $0\hour47\minute33\fs1$, $
-25\arcdeg17\arcmin15\arcsec$ (J2000).
Contours at 10, 20, ..., 90\% of the peak flux of 110\,Jy.
\label{fig5}}
\end{figure*}

\begin{figure*}
\plotone{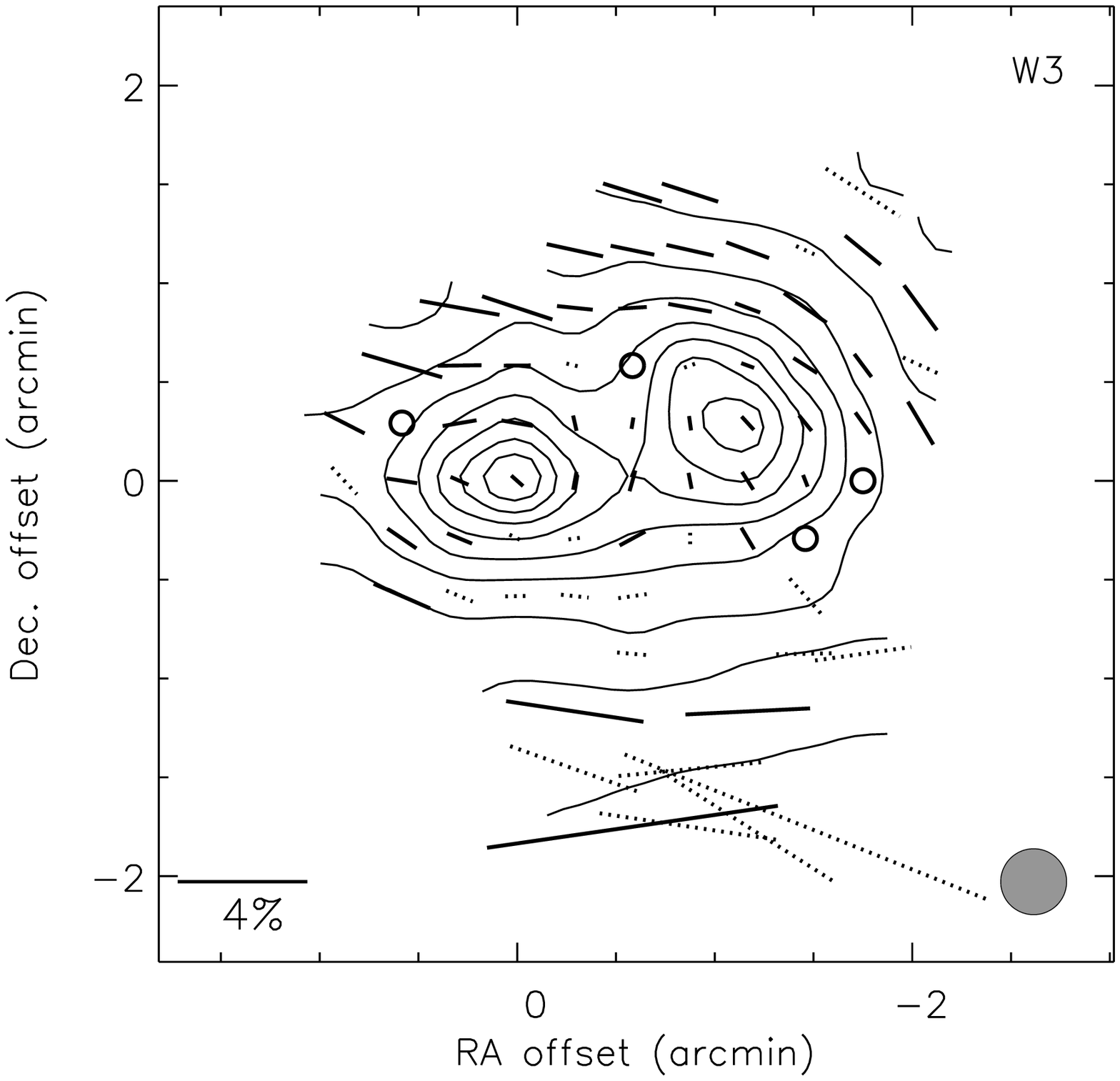}
\caption{W3.
Offsets from $2\hour25\minute40\fs7$, $
62\arcdeg5\arcmin52\arcsec$ (J2000).
Contours at 10, 20, ..., 90\% of the peak flux of 480\,Jy.
\label{fig6}}
\end{figure*}

\begin{figure*}
\plotone{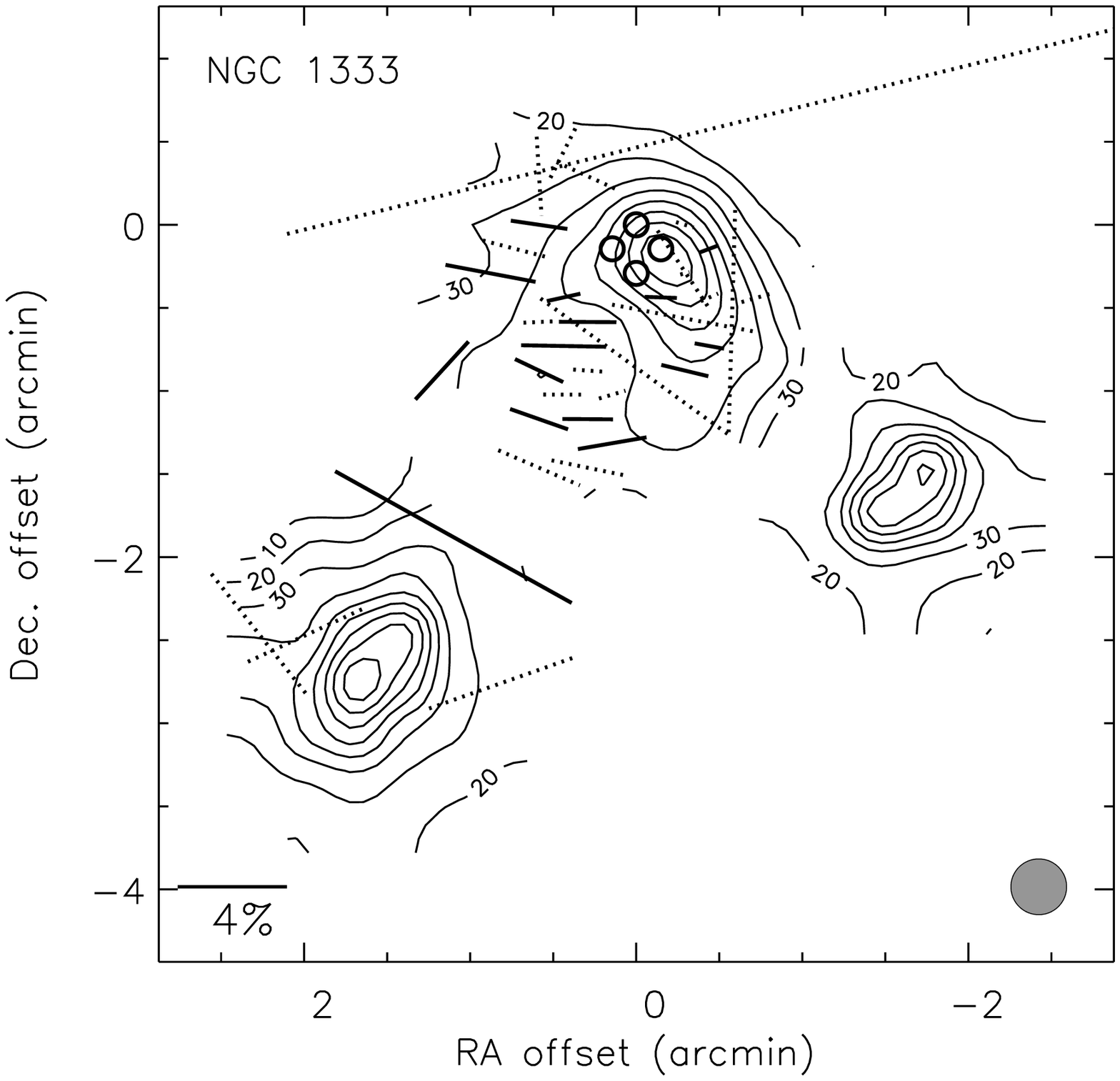}
\caption{NGC 1333.
Offsets from $3\hour29\minute3\fs8$, $
31\arcdeg16\arcmin3\arcsec$ (J2000).
Contours at 10, 20, ..., 90\% of the peak flux of 75\,Jy.
\label{fig7}}
\end{figure*}

\begin{figure*}
\plotone{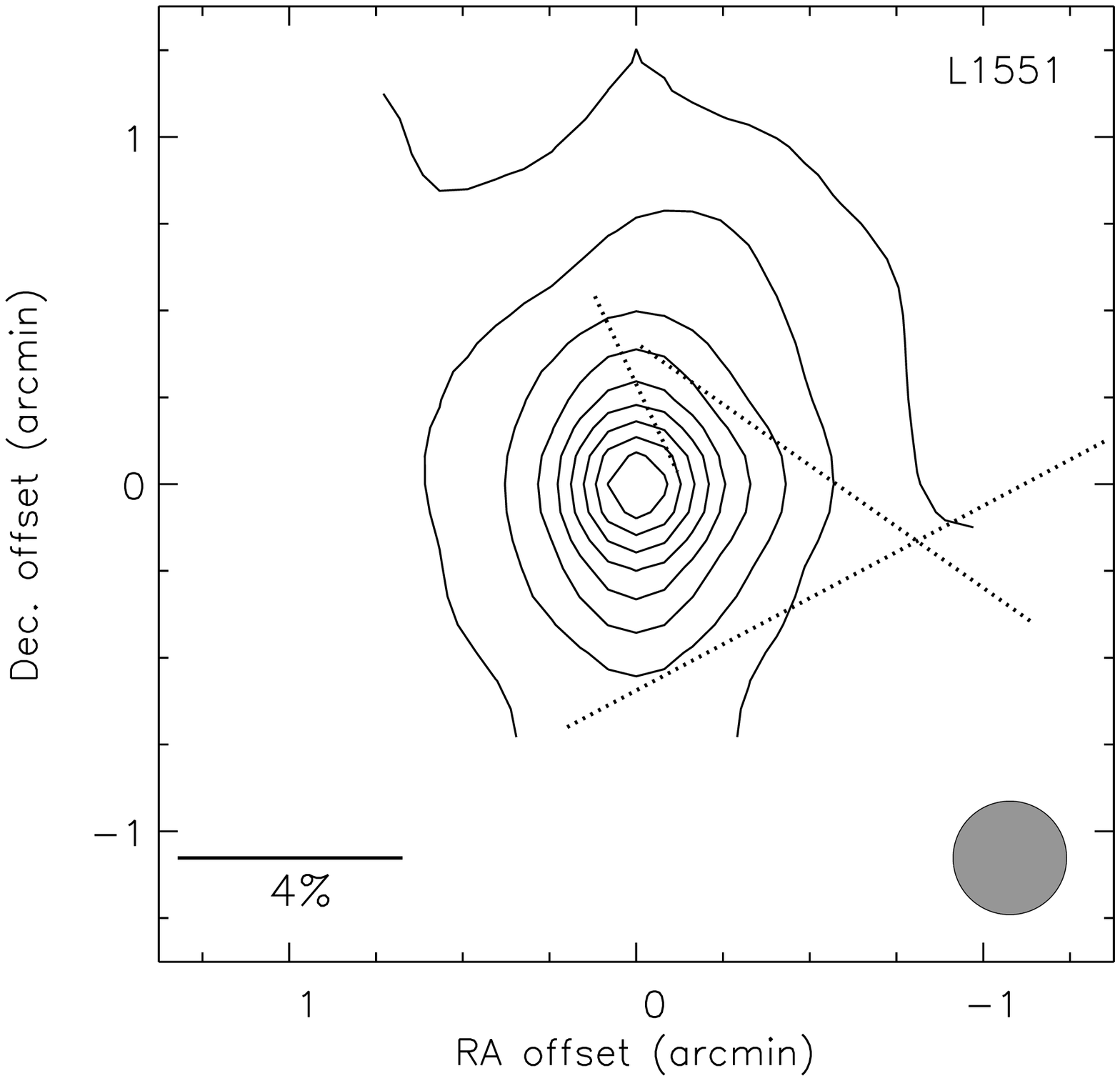}
\caption{L1551.
Offsets from $4\hour31\minute34\fs2$, $
18\arcdeg8\arcmin5\arcsec$ (J2000).
Contours at 10, 20, ..., 90\% of the peak flux of 95\,Jy.
\label{fig8}}
\end{figure*}

\begin{figure*}
\plotone{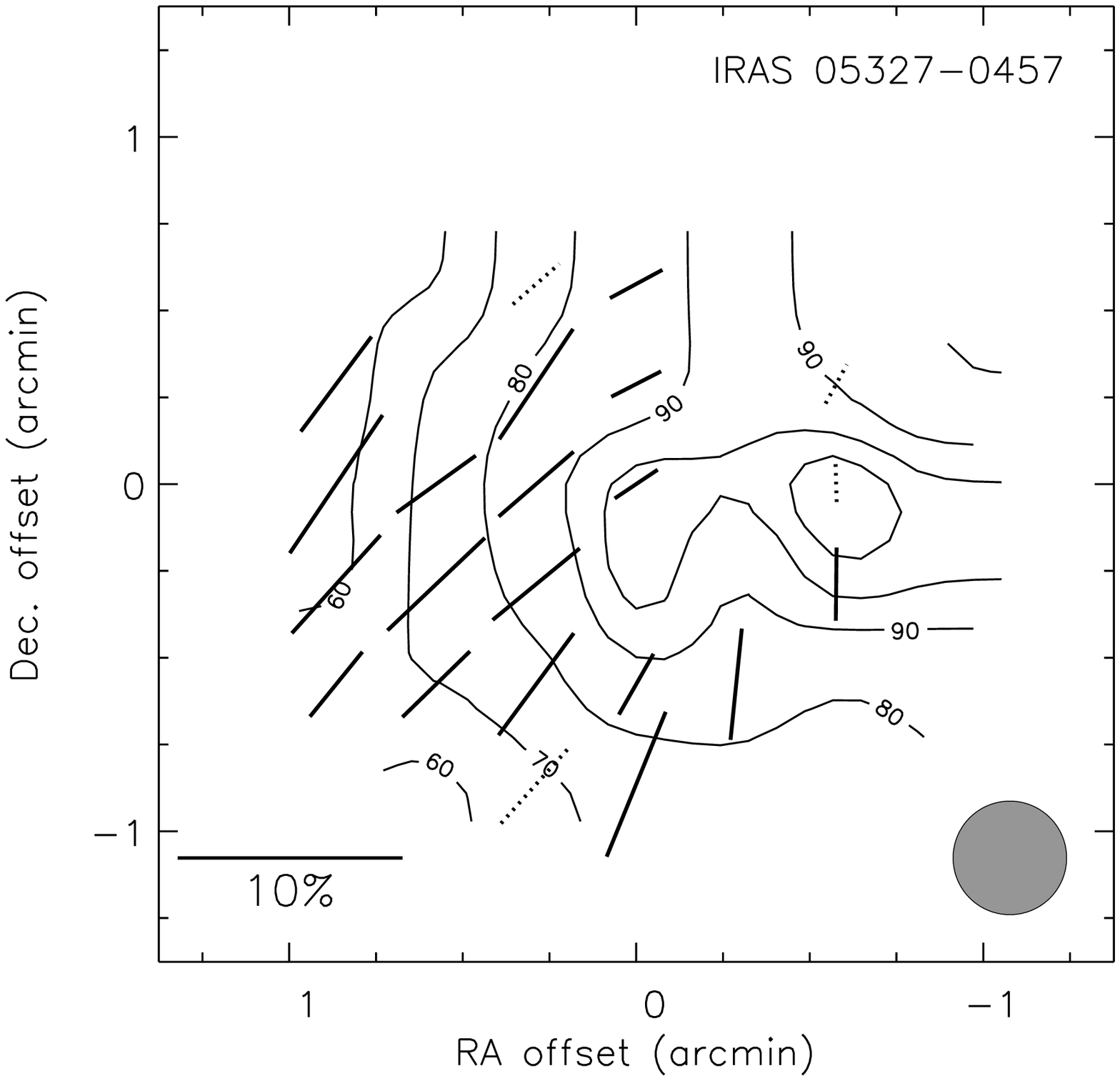}
\caption{IRAS 05327-0457.
Offsets from $5\hour35\minute14\fs4$, $
-4\arcdeg55\arcmin45\arcsec$ (J2000).
Contours at 60, 70, 80, 90, 95, and 98\% of the peak flux of 25\,Jy. The length of a 10\% vector is shown for scale
\label{fig9}}
\end{figure*}

\begin{figure*}
\plotone{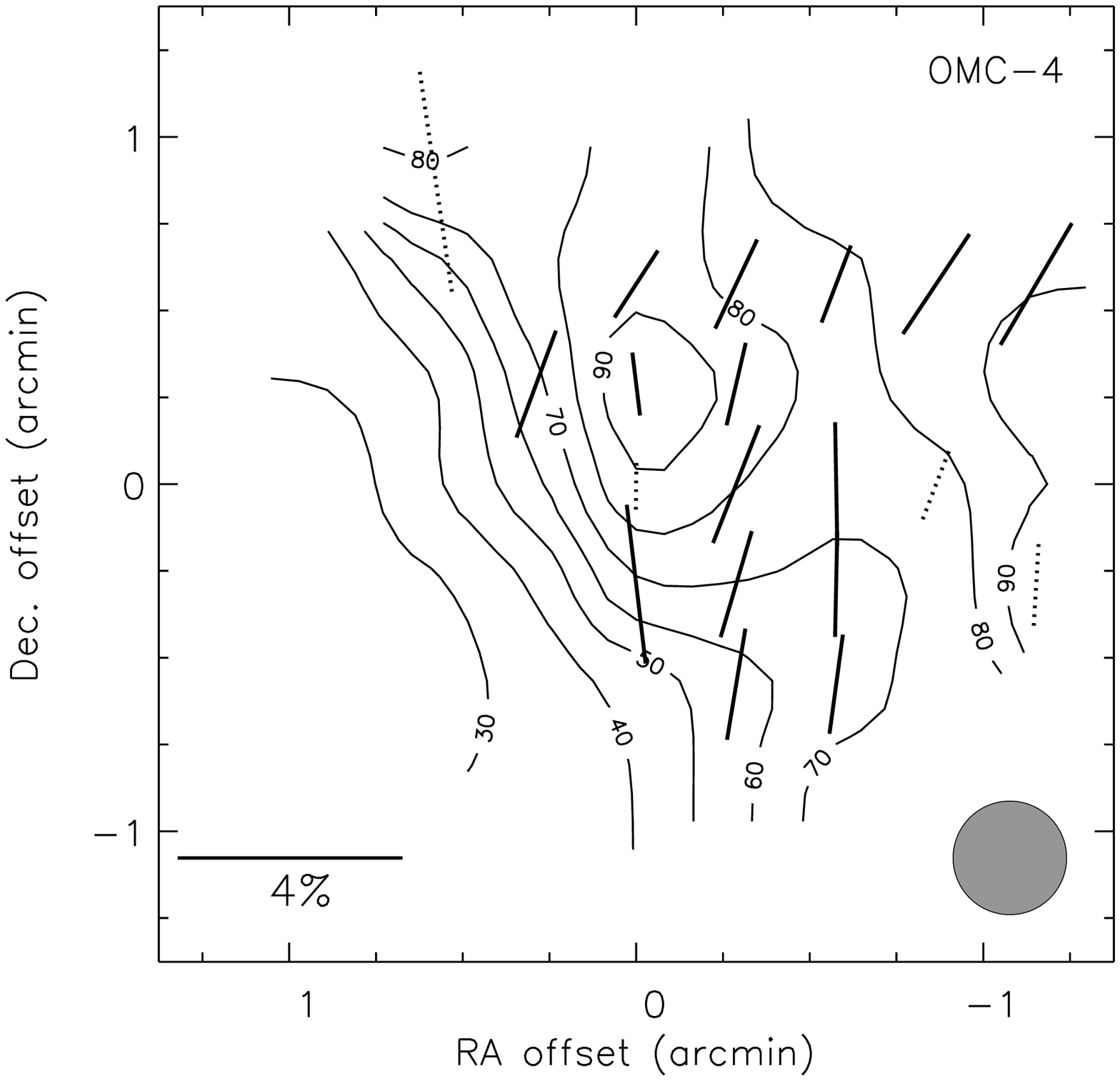}
\caption{Orion Molecular Cloud (OMC-4).
Offsets from $5\hour35\minute8\fs2$, $
-5\arcdeg35\arcmin56\arcsec$ (J2000).
Contours at 30, 40, ..., 90\% of the peak flux of 37\,Jy.
\label{fig10}}
\end{figure*}

\begin{figure*}
\plotone{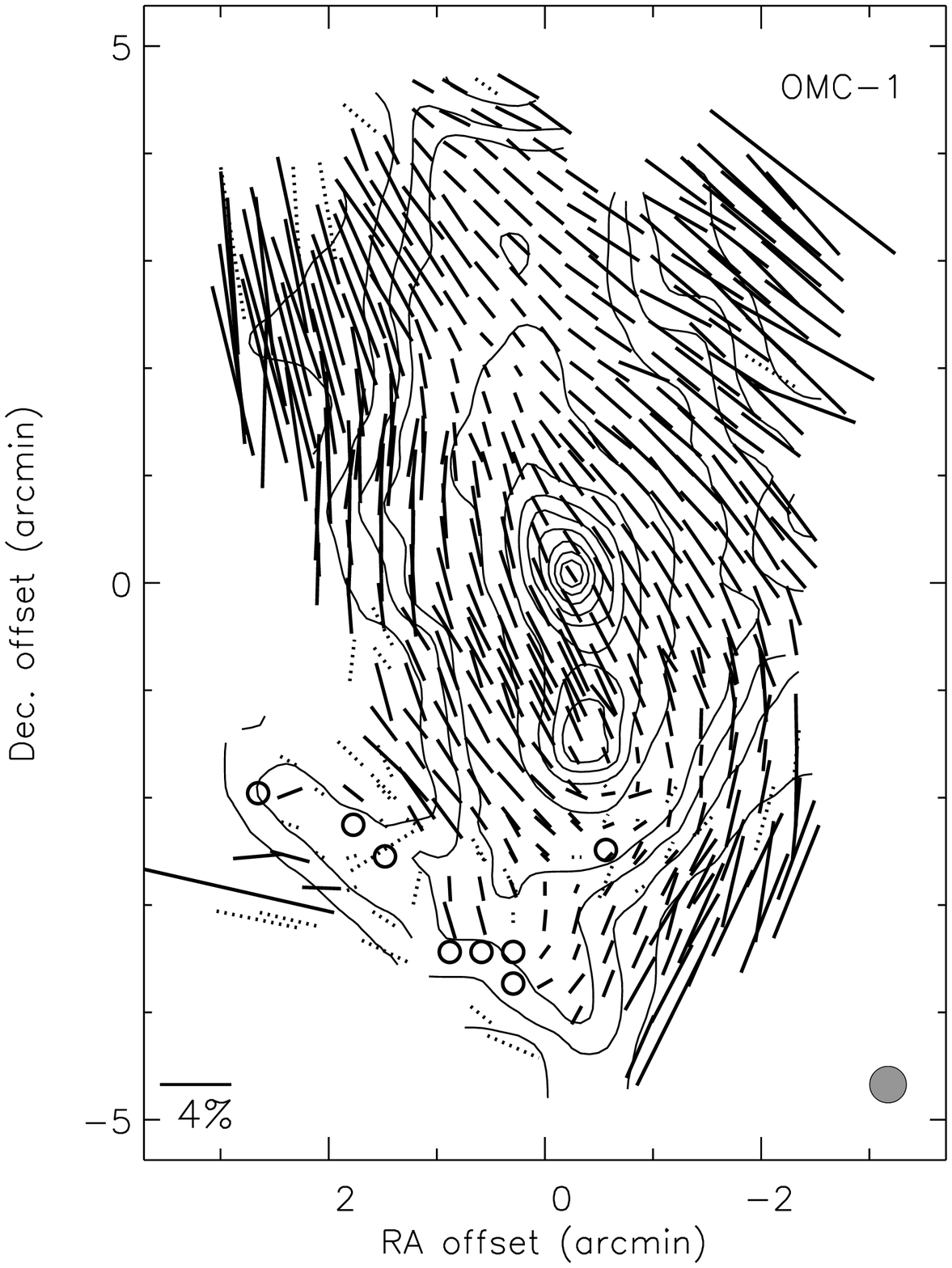}
\caption{Orion Molecular Cloud (OMC-1).
Offsets from $5\hour35\minute14\fs5$, $
-5\arcdeg22\arcmin32\arcsec$ (J2000).
Contours at 4, 6, 8, 10, 20, ..., 90\% of the peak flux of 2100\,Jy.
\label{fig11}}
\end{figure*}

\begin{figure*}
\plotone{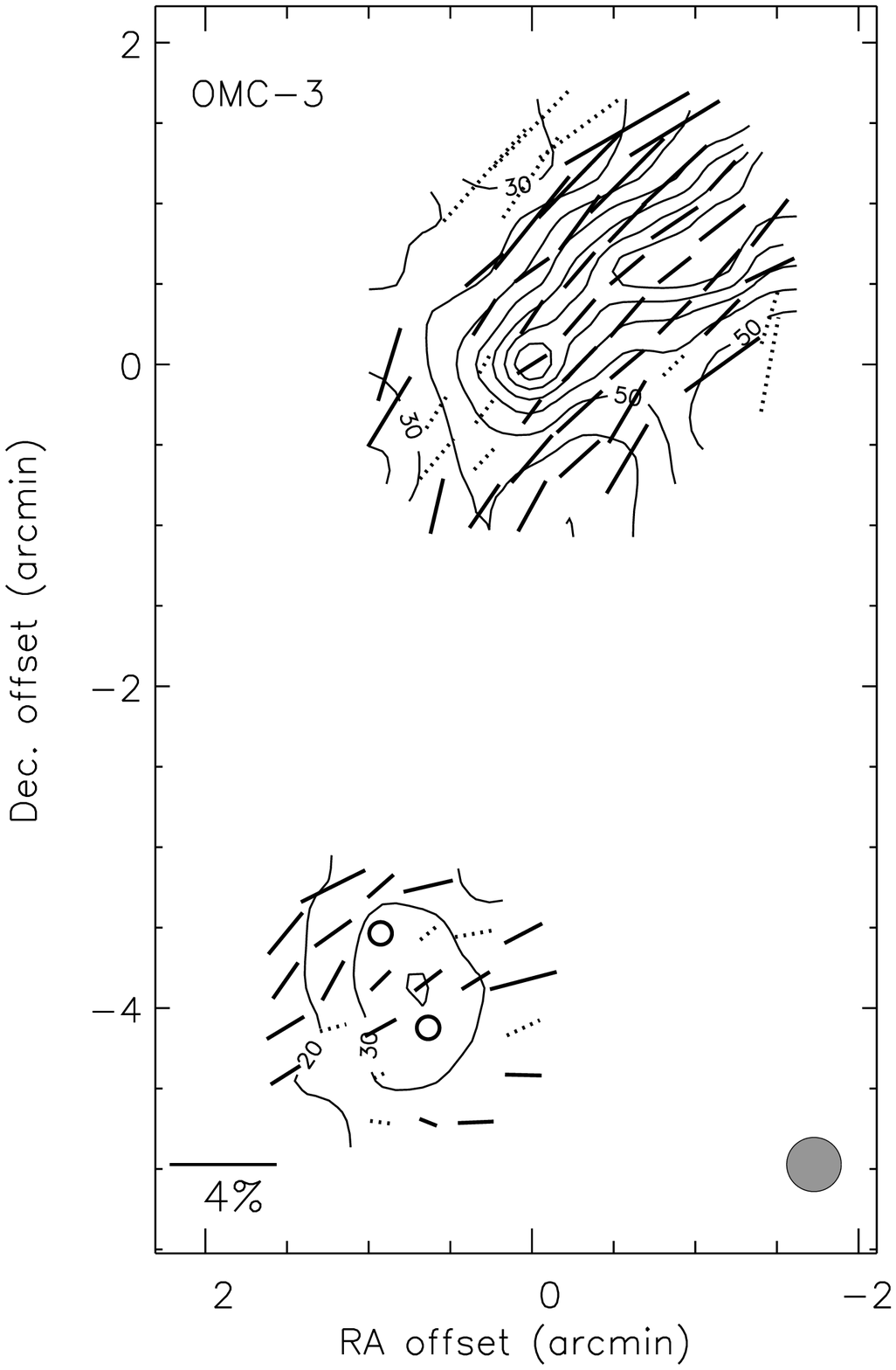}
\caption{Orion Molecular Cloud (OMC-3).
Offsets from $5\hour35\minute23\fs5$, $
-5\arcdeg1\arcmin32\arcsec$ (J2000).
Contours 20, ..., 90\% of the peak flux of 110\,Jy.
\label{fig12}}
\end{figure*}

\begin{figure*}
\plotone{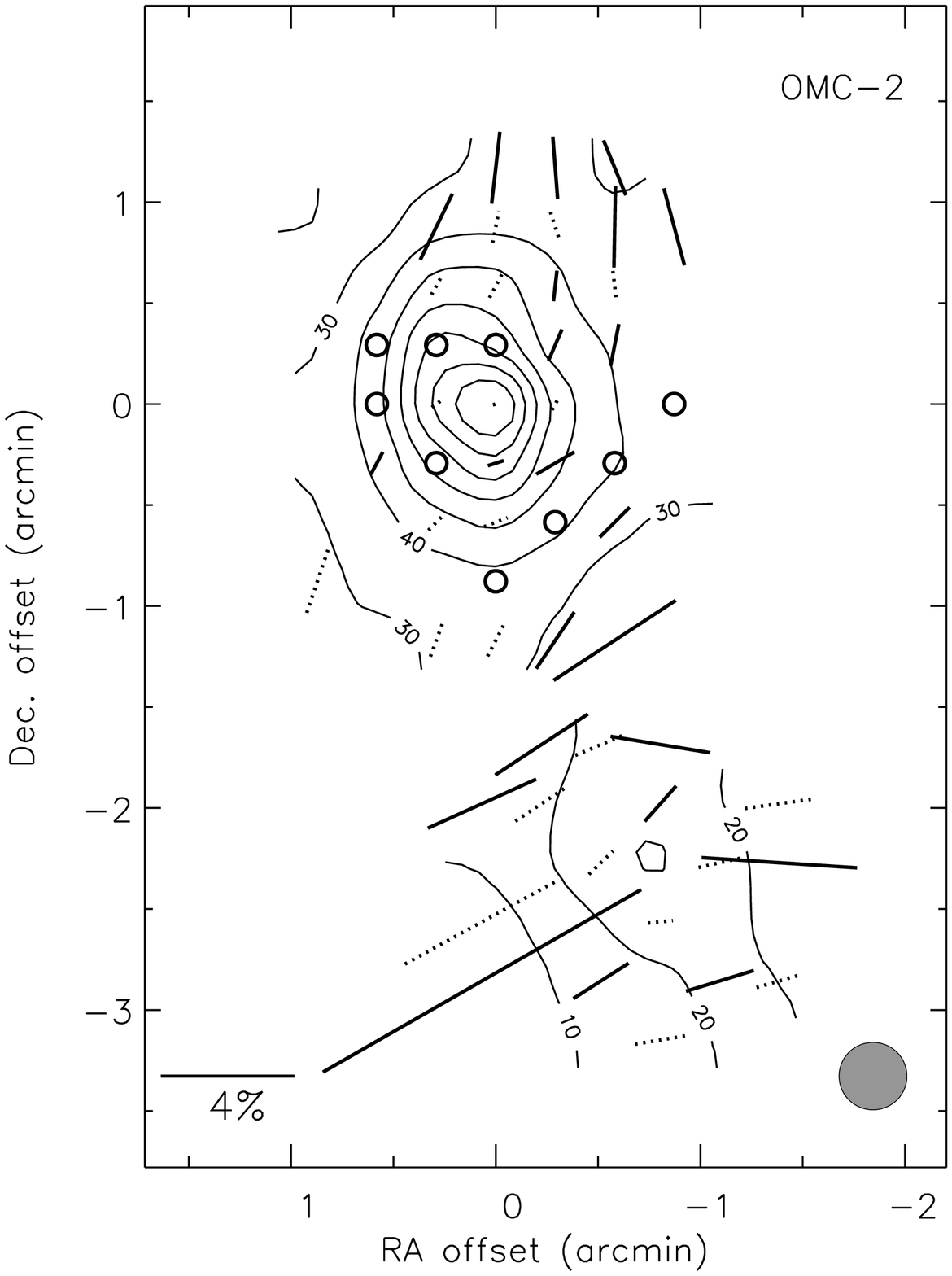}
\caption{Orion Molecular Cloud (OMC-2).
Offsets from $5\hour35\minute26\fs7$, $
-5\arcdeg10\arcmin0\arcsec$ (J2000).
Contours at 10, 20, ..., 90\% of the peak flux of 200\,Jy.
\label{fig13}}
\end{figure*}

\begin{figure*}
\plotone{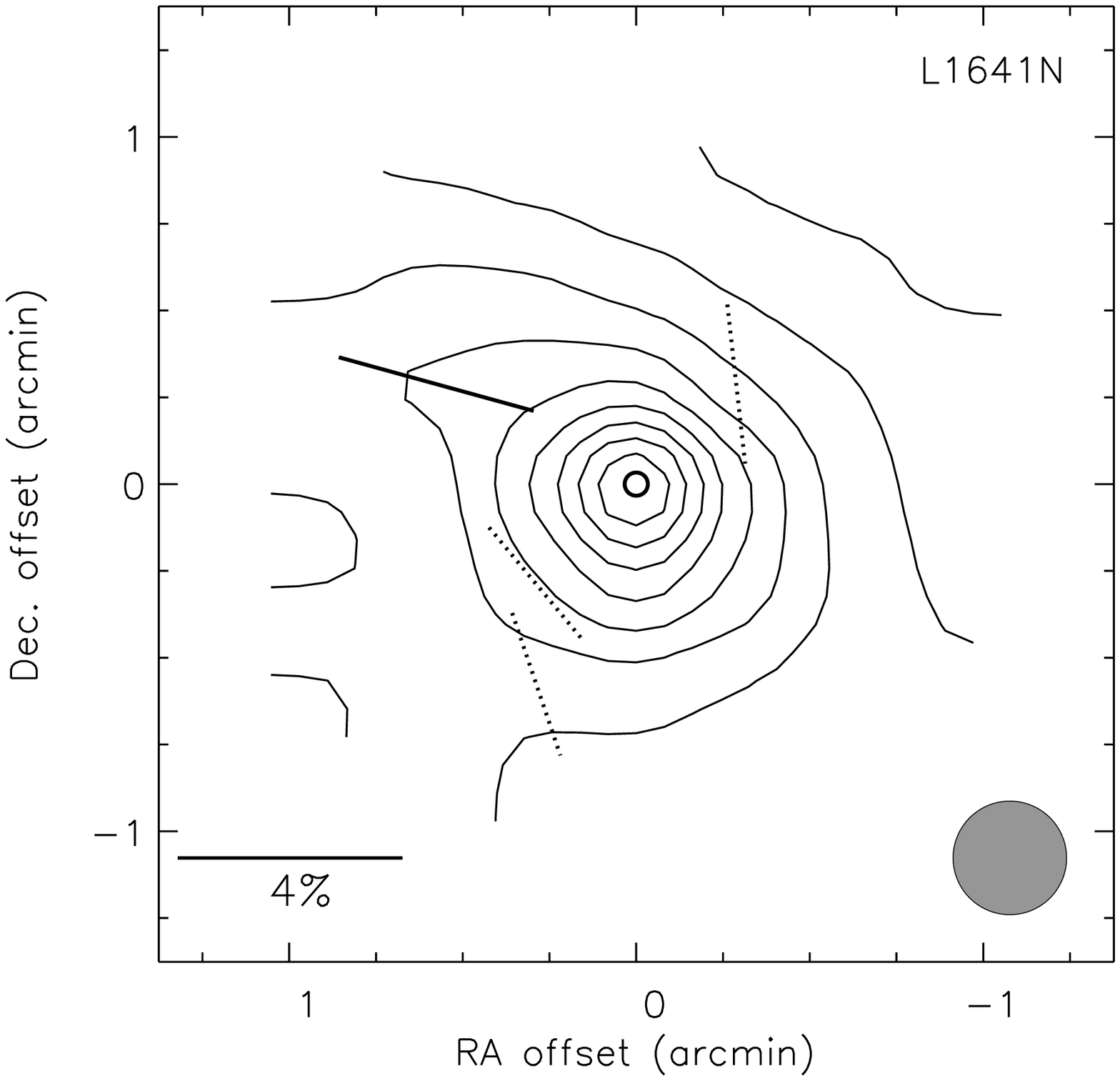}
\caption{L1641N.
Offsets from $5\hour36\minute18\fs8$, $
-6\arcdeg22\arcmin11\arcsec$ (J2000).
Contours at 10, 20, ..., 90\% of the peak flux of 89\,Jy.
\label{fig14}}
\end{figure*}

\begin{figure*}
\plotone{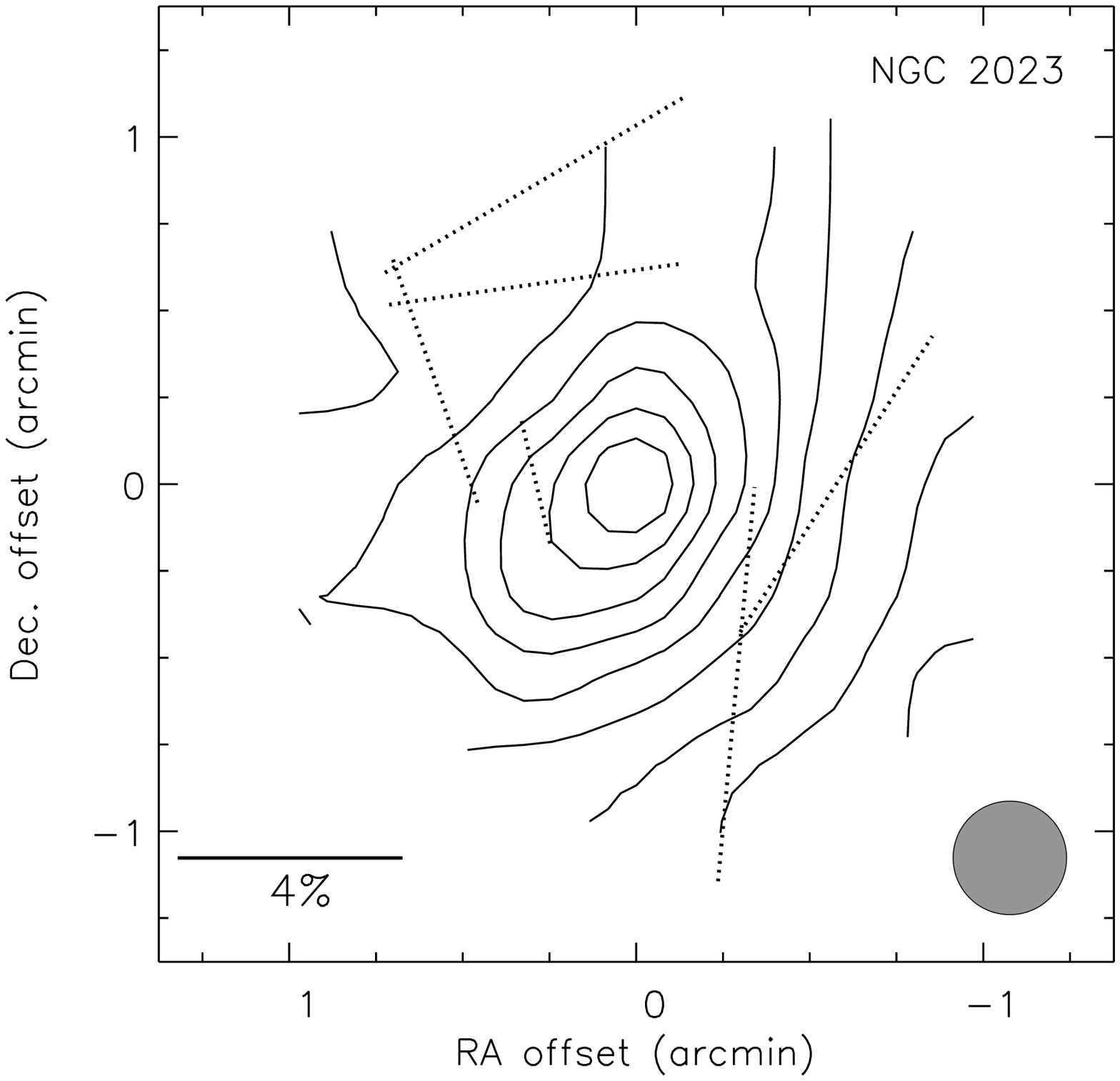}
\caption{NGC 2023.
Offsets from $5\hour41\minute25\fs4$, $
-2\arcdeg18\arcmin6\arcsec$ (J2000).
Contours at 20, 30, ..., 90\% of the peak flux of 43\,Jy.
\label{fig15}}
\end{figure*}

\begin{figure*}
\plotone{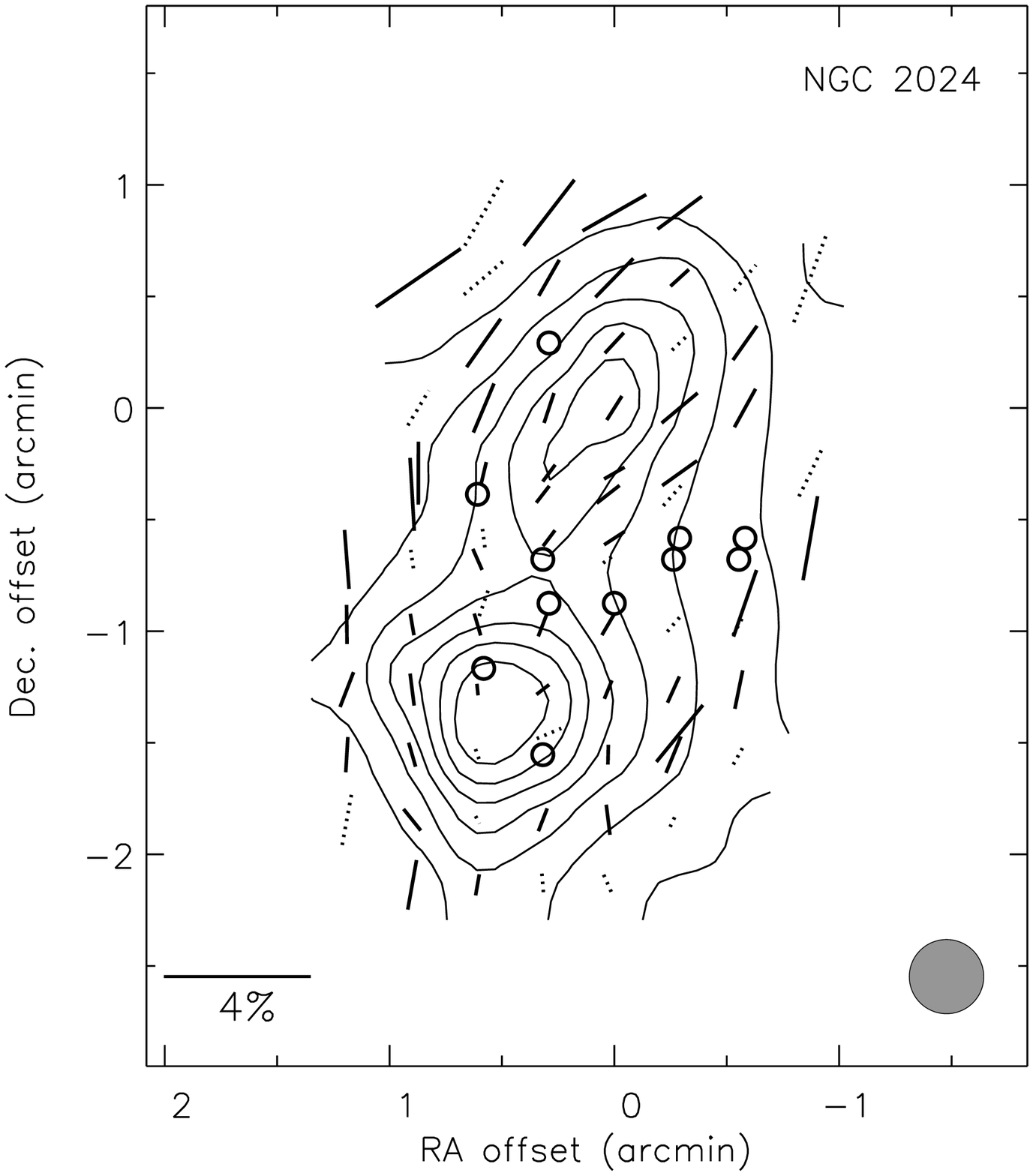}
\caption{NGC 2024.
Offsets from $5\hour41\minute43\fs0$, $
-1\arcdeg54\arcmin22\arcsec$ (J2000).
Contours at 20, 30, ..., 90\% of the peak flux of 470\,Jy.
\label{fig16}}
\end{figure*}

\begin{figure*}
\plotone{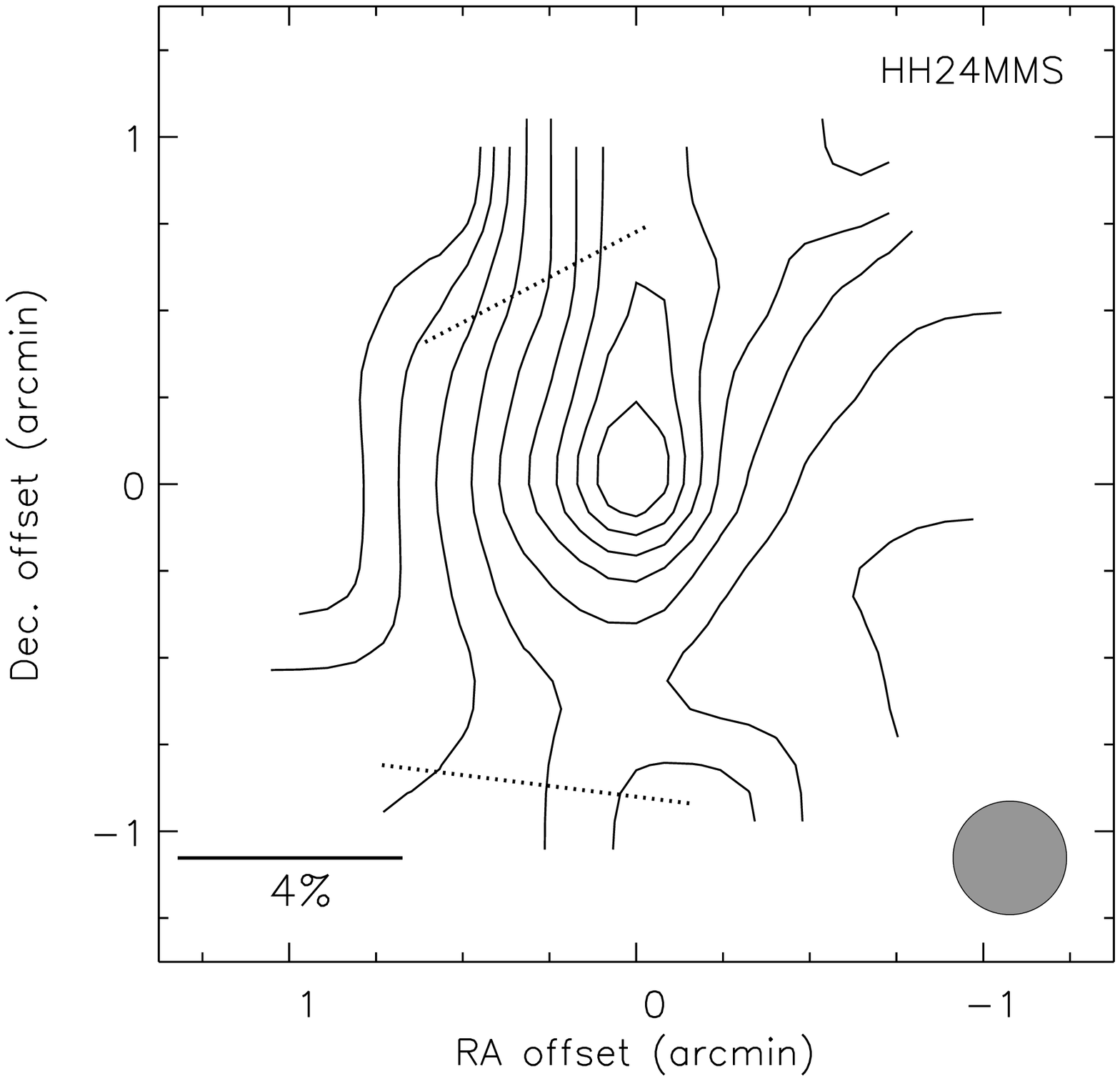}
\caption{HH24MMS.
Offsets from $5\hour46\minute8\fs4$, $
-0\arcdeg10\arcmin43\arcsec$ (J2000).
Contours at 10, 20, ..., 90\% of the peak flux of 26\,Jy.
\label{fig17}}
\end{figure*}

\begin{figure*}
\plotone{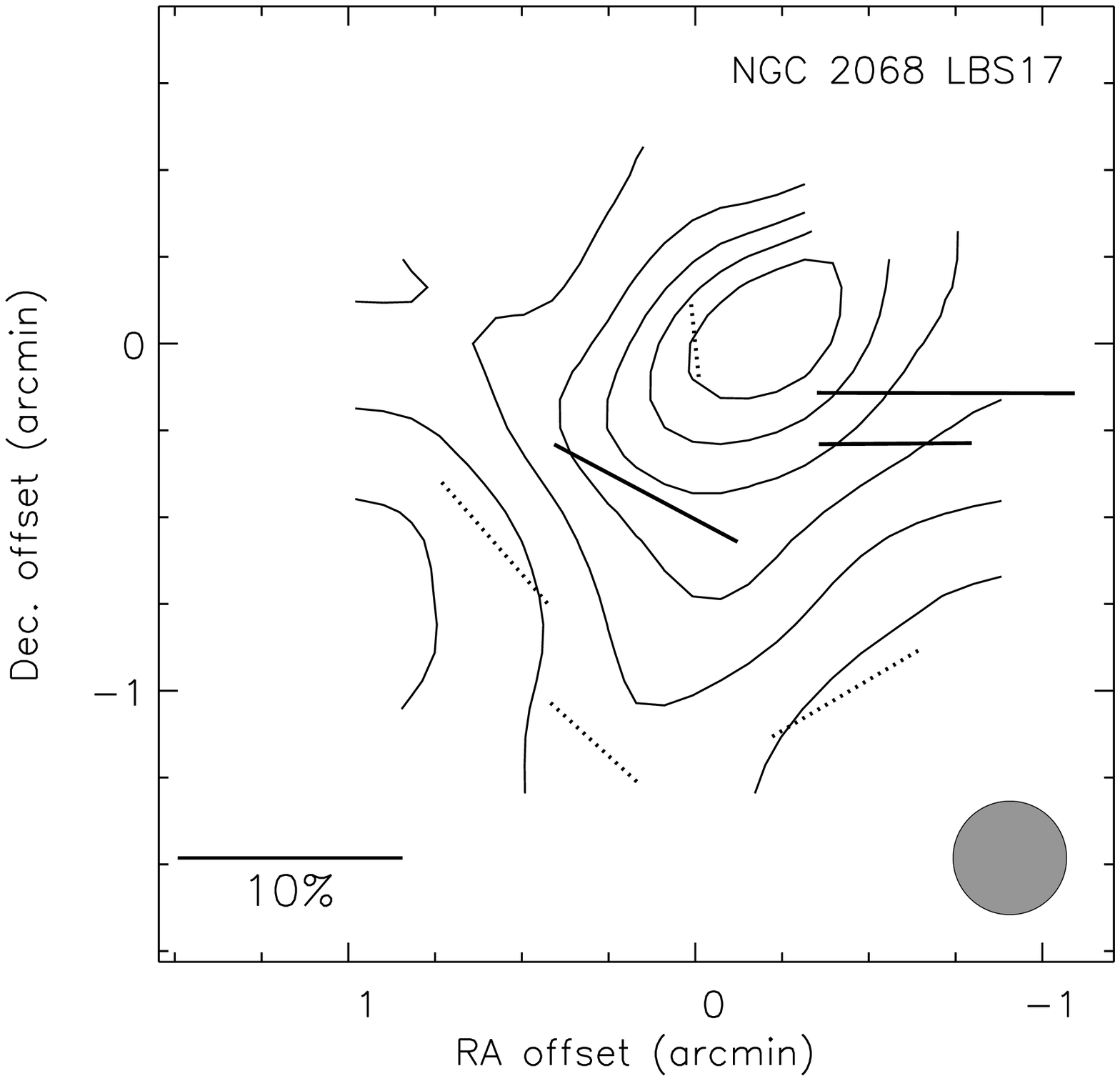}
\caption{NGC 2068 LBS17.
Offsets from $5\hour46\minute28\fs0$, $
-0\arcdeg0\arcmin54\arcsec$ (J2000).
Contours at 30, 40, ..., 90\% of the peak flux of 17\,Jy.  \label{fig18}}
\end{figure*}

\begin{figure*}
\plotone{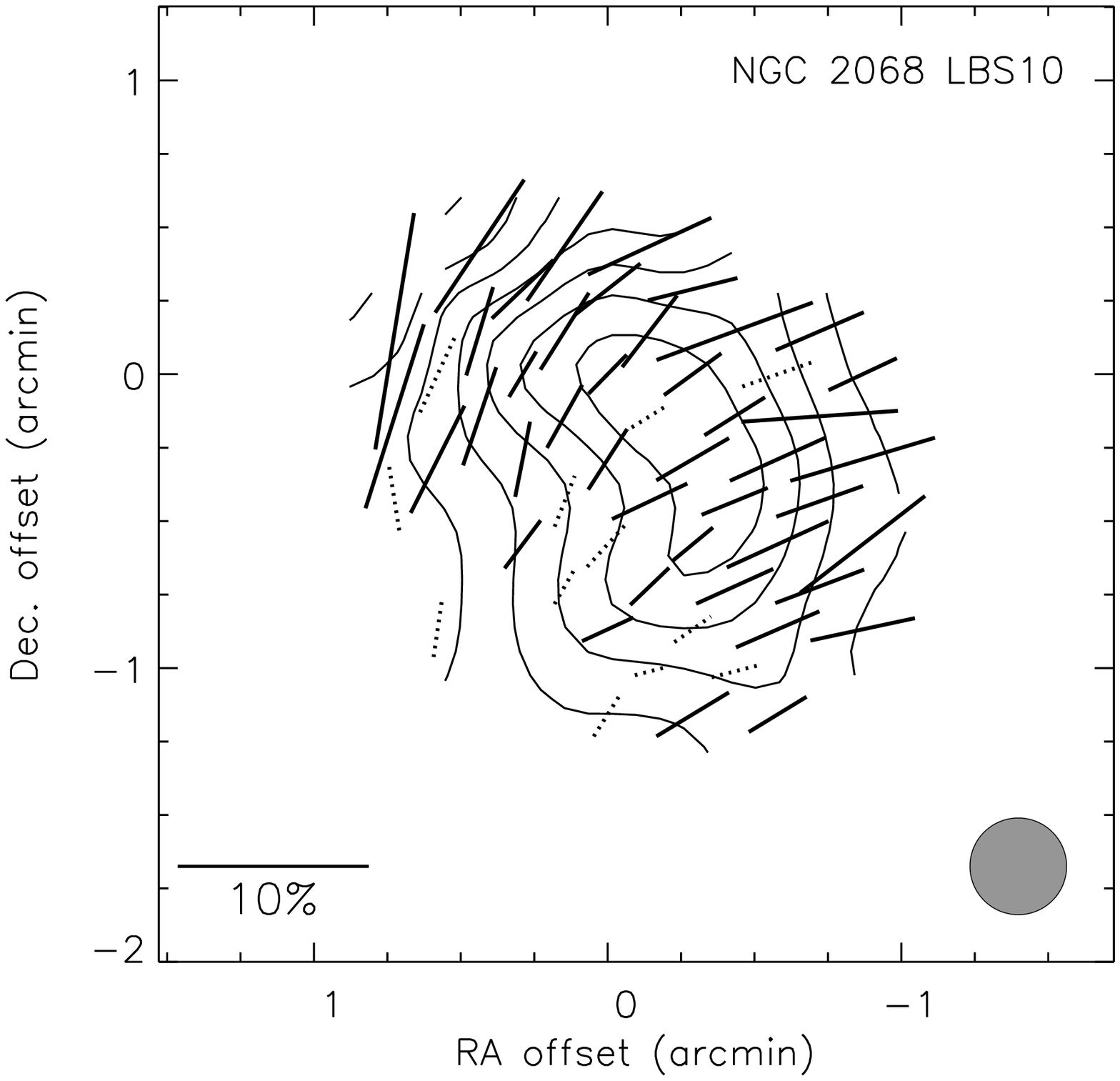}
\caption{NGC 2068 LBS10.
Offsets from $5\hour46\minute50\fs2$, $
0\arcdeg2\arcmin1\arcsec$ (J2000).
Contours at 30, 40, ..., 90\% of the peak flux of 28\,Jy.  The length of a 10\% vector is shown for scale.
\label{fig19}}
\end{figure*}

\begin{figure*}
\plotone{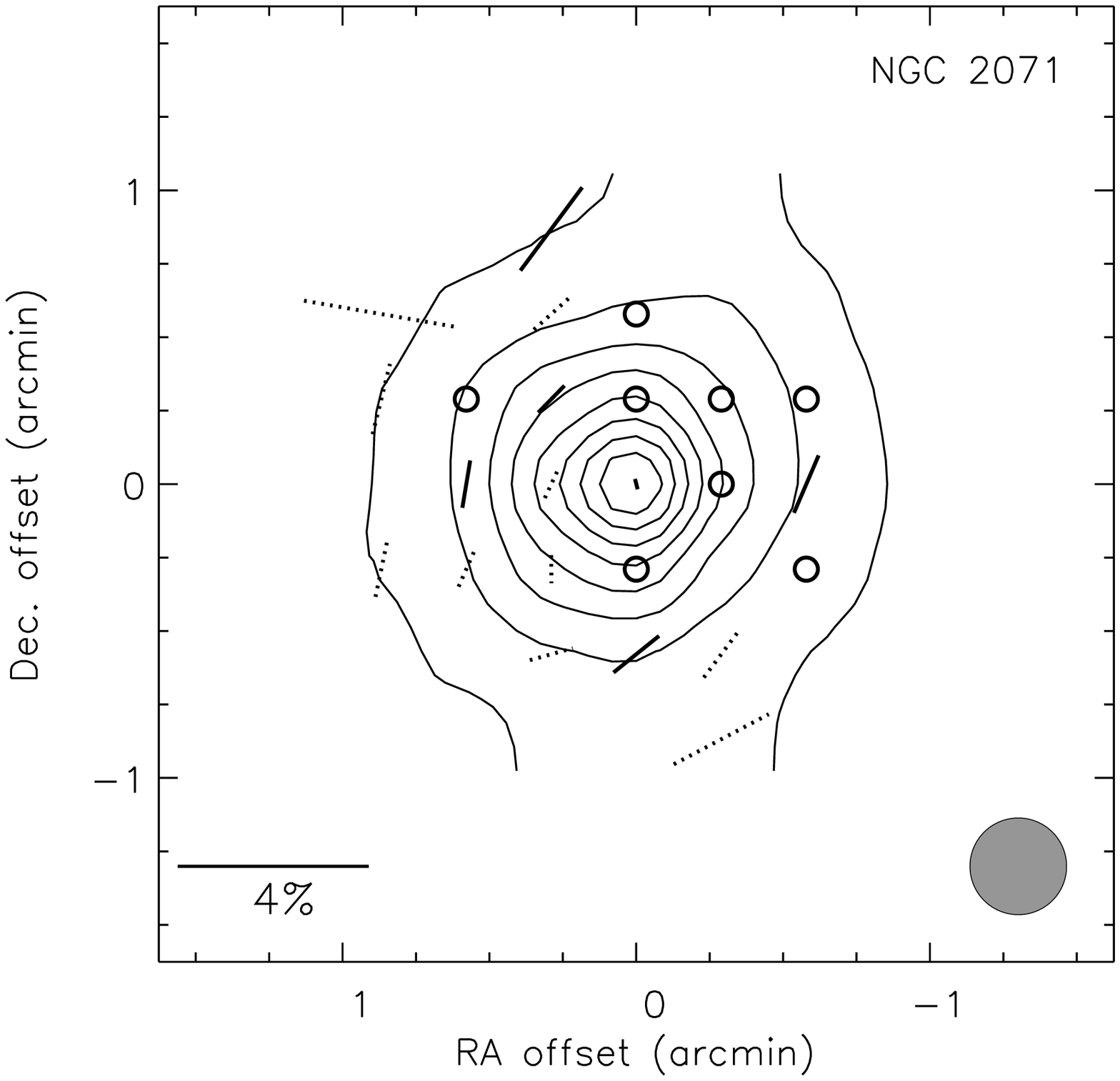}
\caption{NGC 2071.
Offsets from $5\hour47\minute4\fs9$, $
0\arcdeg21\arcmin47\arcsec$ (J2000).
Contours at 20, 30, ..., 90\% of the peak flux of 180\,Jy.
\label{fig20}}
\end{figure*}

\begin{figure*}
\plotone{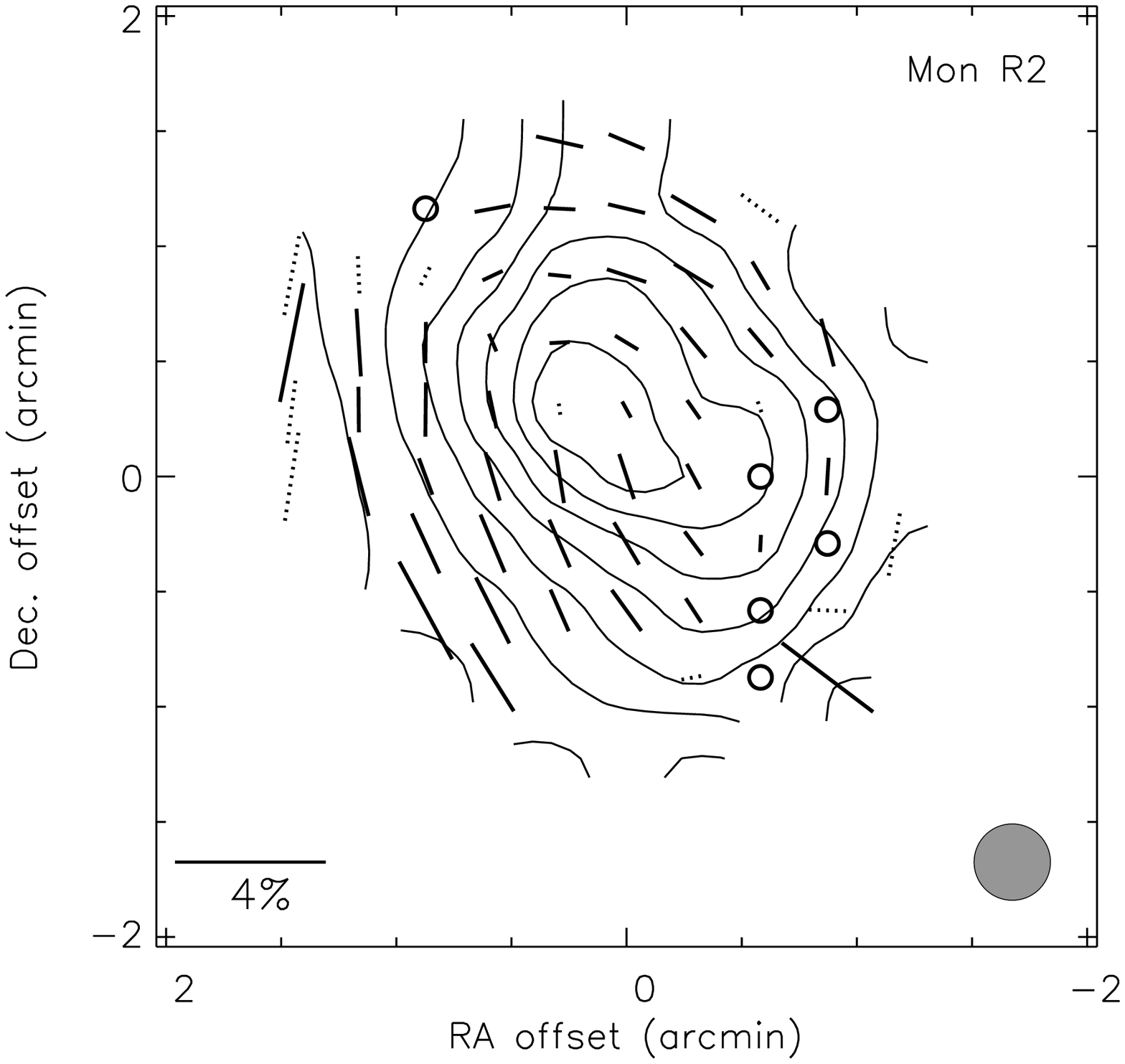}
\caption{Mon R2.
Offsets from $6\hour7\minute46\fs6$, $
-6\arcdeg23\arcmin16\arcsec$ (J2000).
Contours at 30, 40, ..., 90\% of the peak flux of 280\,Jy.
\label{fig21}}
\end{figure*}

\begin{figure*}
\plotone{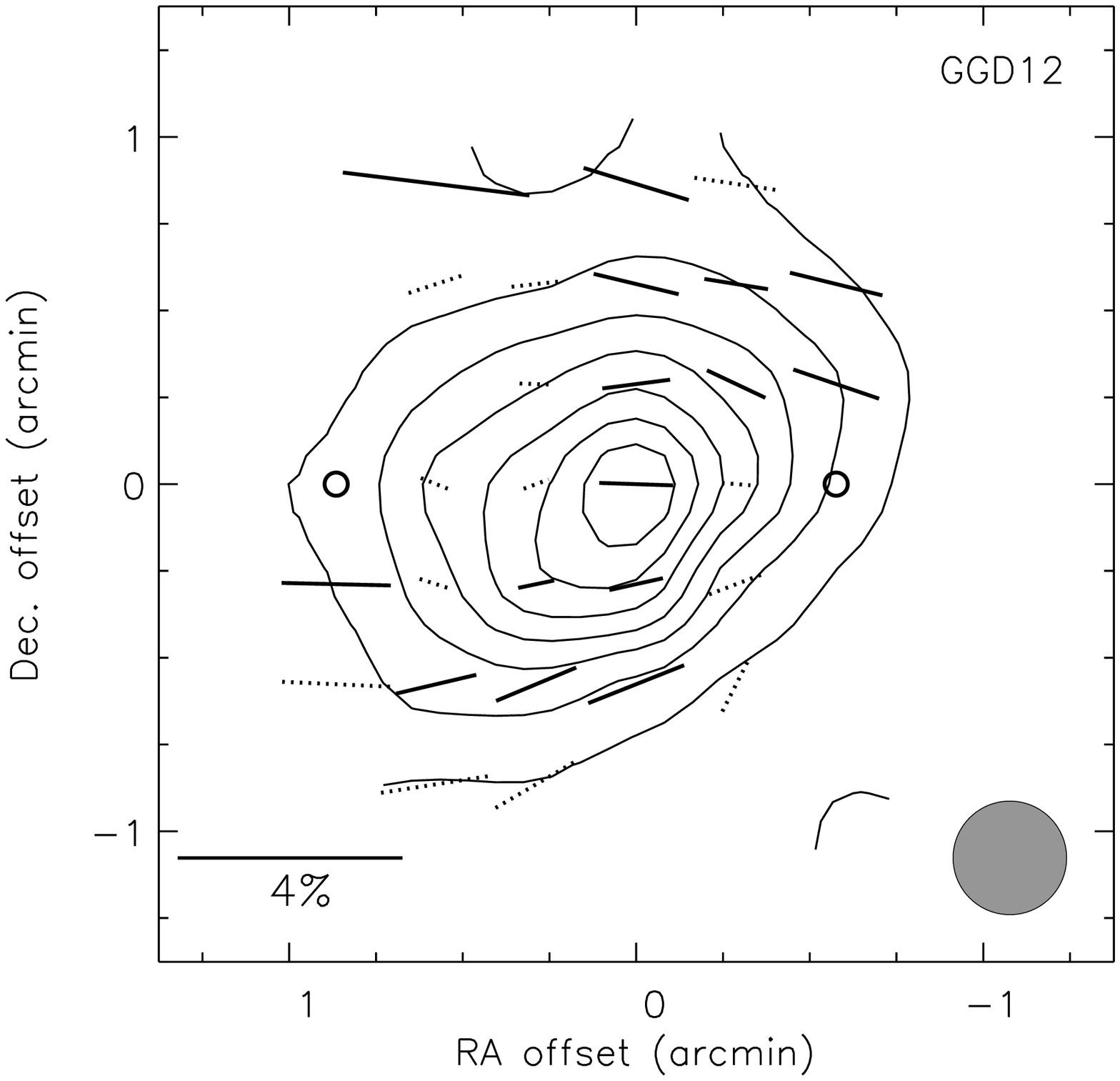}
\caption{GGD12.
Offsets are centered at $6\hour10\minute50\fs4$, $
-6\arcdeg11\arcmin46\arcsec$ (J2000).
Contours at 30, 40, ..., 90\% of the peak flux of 260\,Jy.
\label{fig22}}
\end{figure*}

\begin{figure*}
\plotone{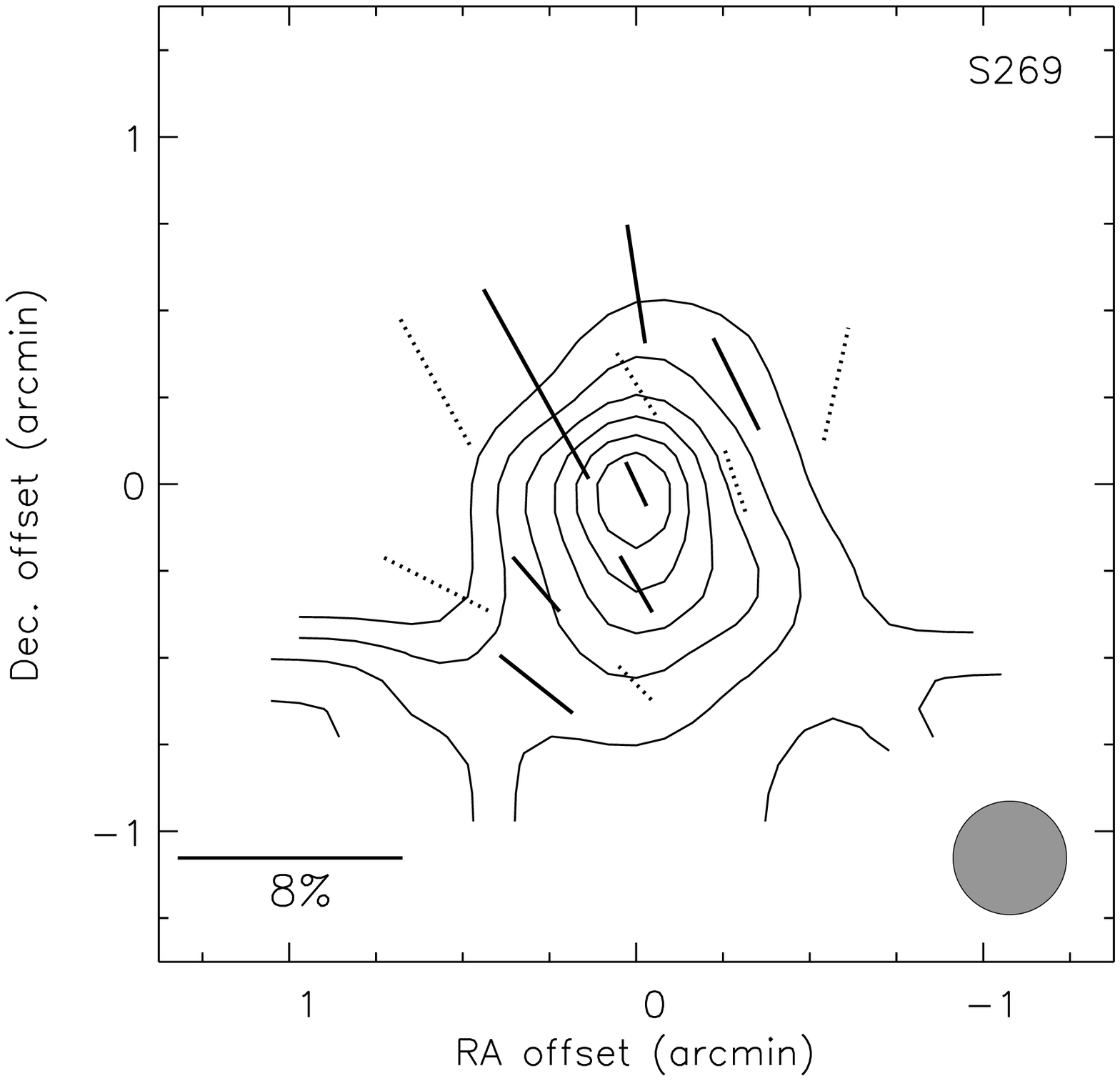}
\caption{S269.
Offsets from $6\hour14\minute36\fs6$, $
13\arcdeg49\arcmin35\arcsec$ (J2000).
Contours at 40, 50, ..., 90\% of the peak flux of 58\,Jy.
\label{fig23}}
\end{figure*}

\begin{figure*}
\plotone{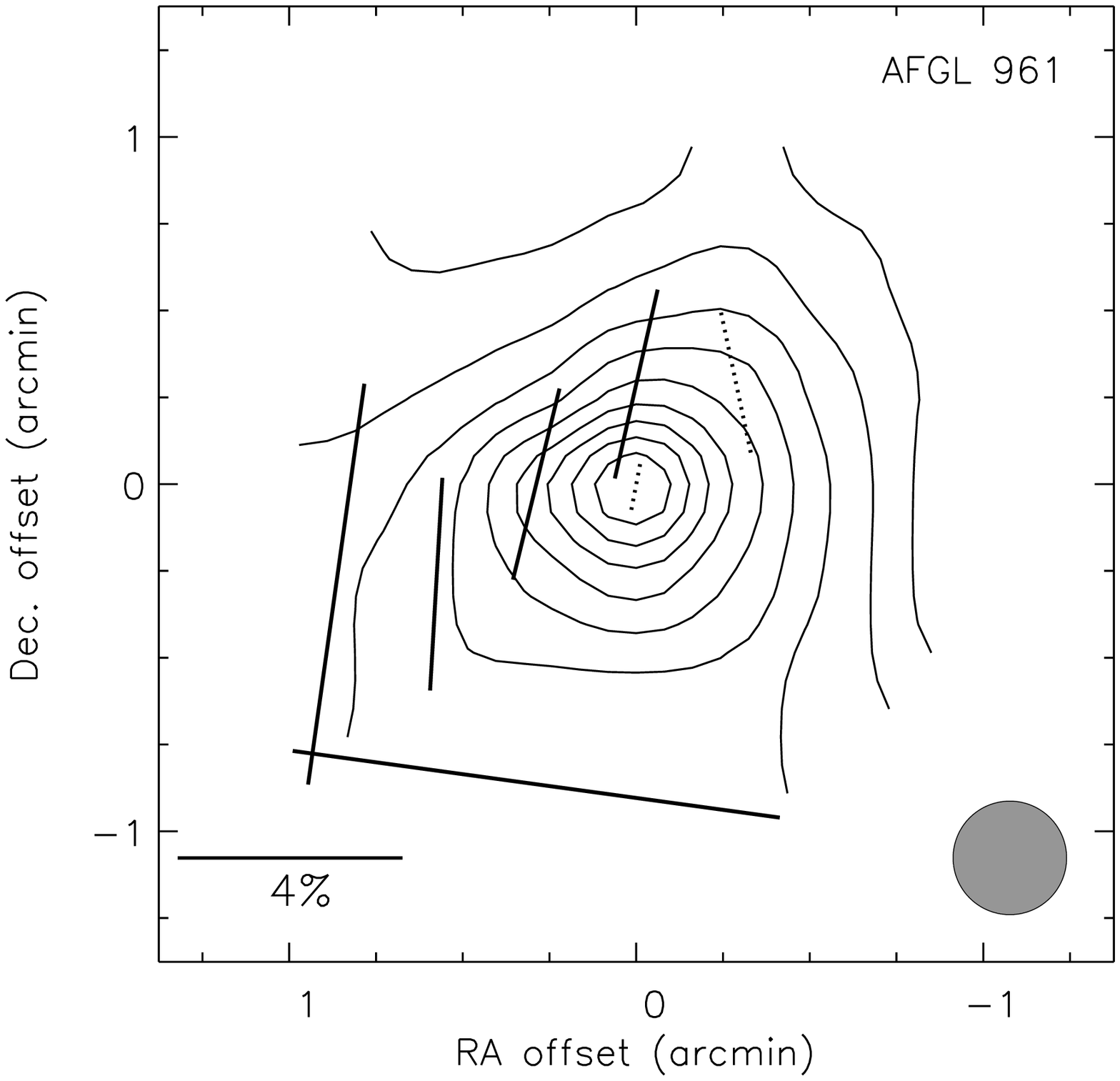}
\caption{AFGL 961.
Offsets from $6\hour34\minute37\fs7$, $
4\arcdeg12\arcmin44\arcsec$ (J2000).
Contours at 10, 20, ..., 90\% of the peak flux of 37\,Jy.
\label{fig24}}
\end{figure*}

\begin{figure*}
\plotone{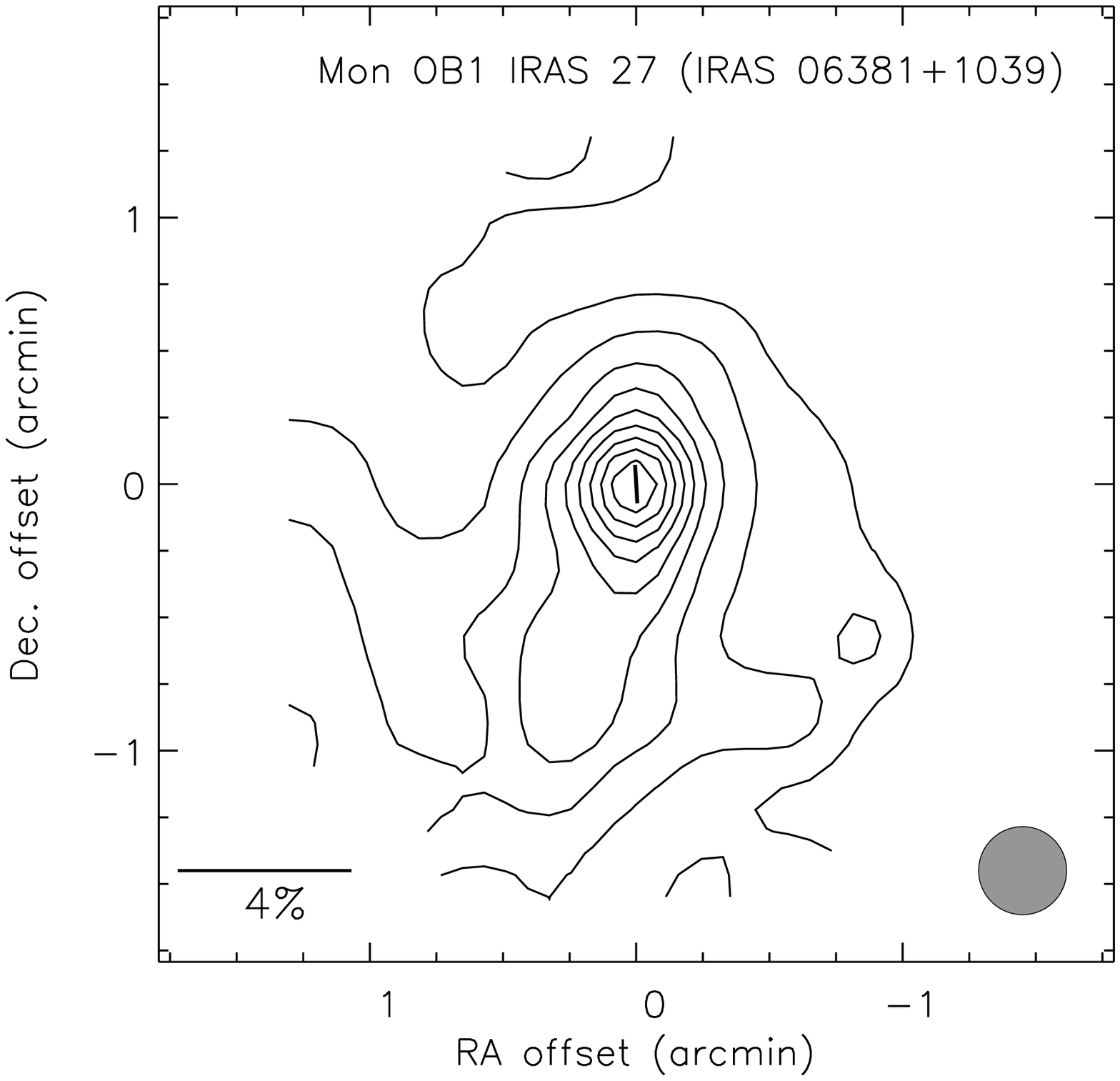}
\caption{Mon OB1 IRAS 27 (IRAS 06381+1039).
Offsets from $6\hour40\minute58\fs3$, $
10\arcdeg36\arcmin54\arcsec$ (J2000).
Contours at 10, 20, ..., 90\% of the peak flux of 29\,Jy. 
\label{fig25}}
\end{figure*}

\begin{figure*}
\plotone{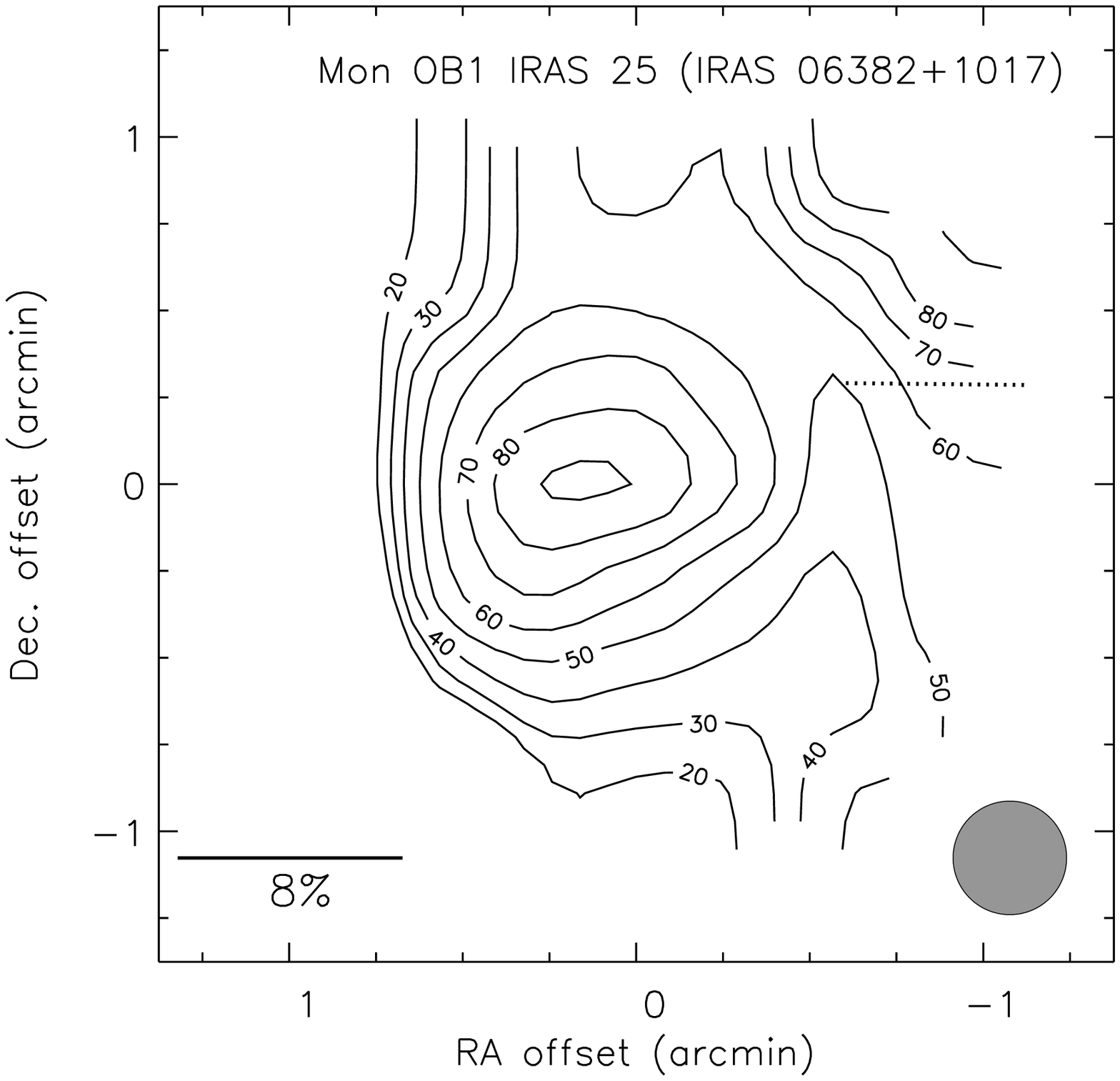}
\caption{Mon OB1 IRAS 25 (IRAS 06382+1017).
Offsets from $6\hour41\minute3\fs7$, $
10\arcdeg15\arcmin7\arcsec$ (J2000).
Contours at 20, 30, ..., 90\% of the peak flux of 25\,Jy.  The peak flux occurs in the northwest corner
of the mapped region.
\label{fig26}}
\end{figure*}

\begin{figure*}
\plotone{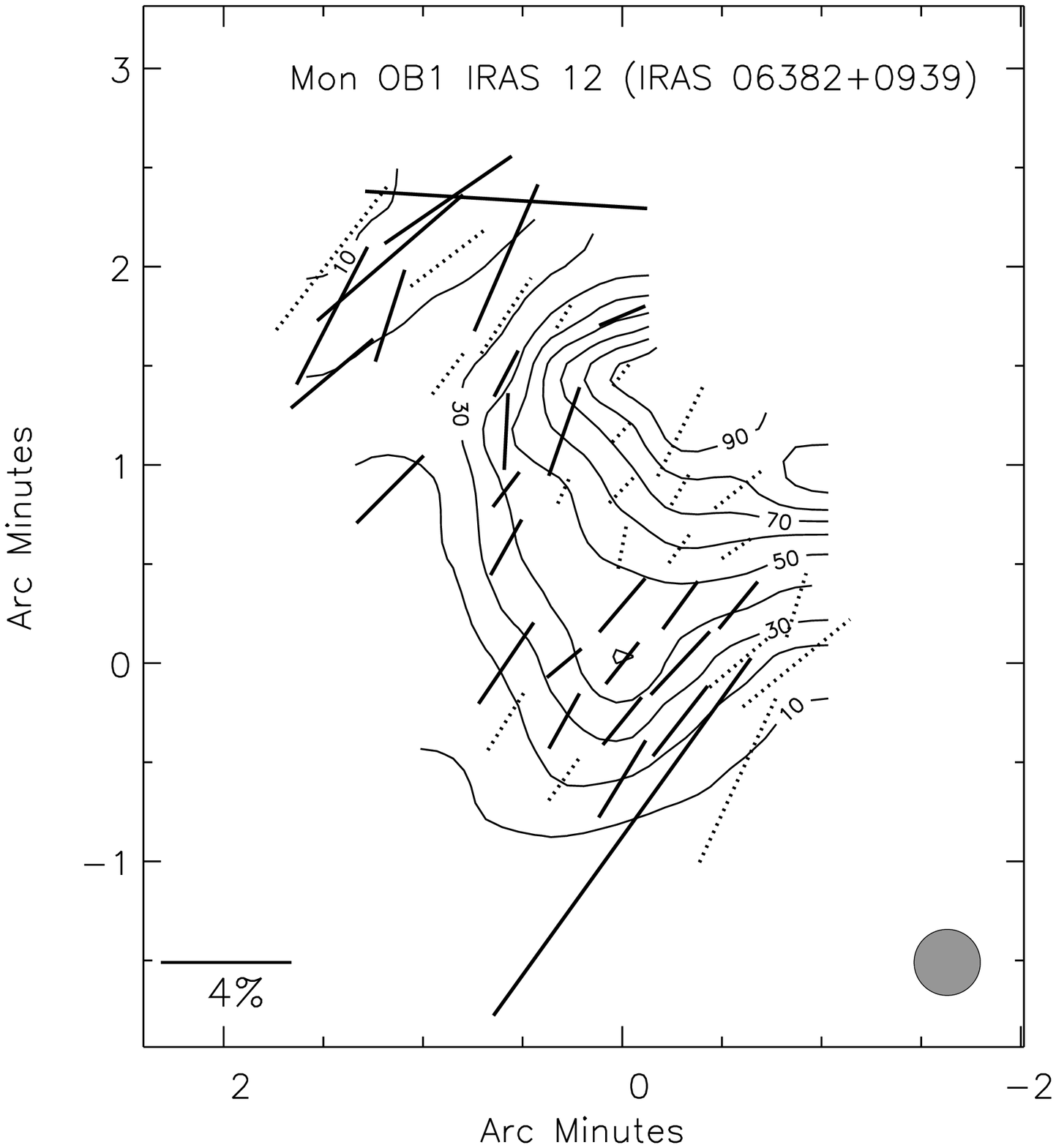}
\caption{Mon OB1 IRAS 12 (IRAS 06382+0939).
Offsets from $6\hour41\minute6\fs1$, $
9\arcdeg34\arcmin9\arcsec$ (J2000).
Contours at 10, 20, ..., 90\% of the peak flux of 62\,Jy.
\label{fig27}}
\end{figure*}

\begin{figure*}
\plotone{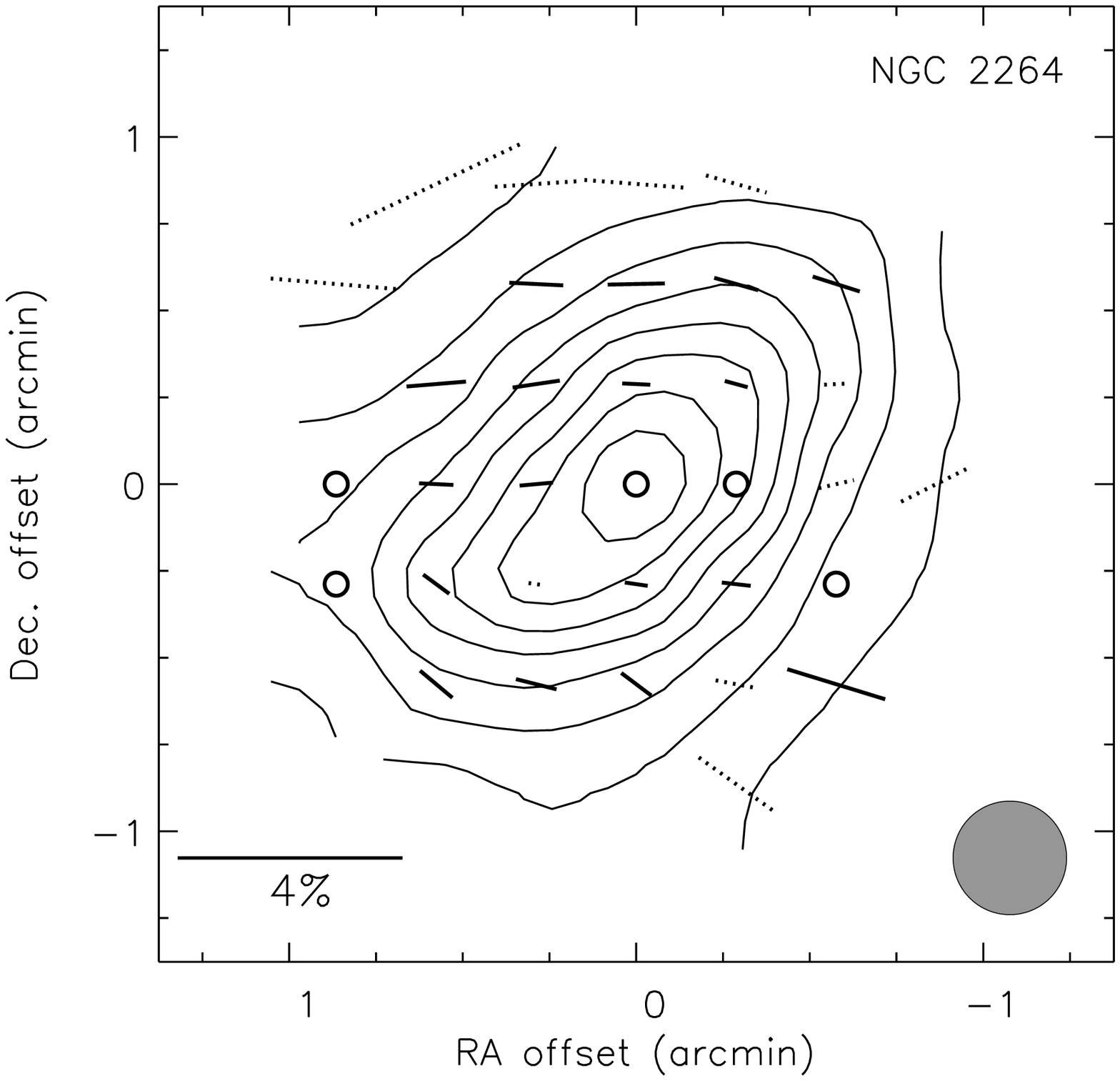}
\caption{NGC 2264.
Offsets from $6\hour41\minute10\fs3$, $
9\arcdeg29\arcmin27\arcsec$ (J2000).
Contours at 20, 30, ..., 90\% of the peak flux of 180\,Jy.
\label{fig28}}
\end{figure*}

\begin{figure*}
\plotone{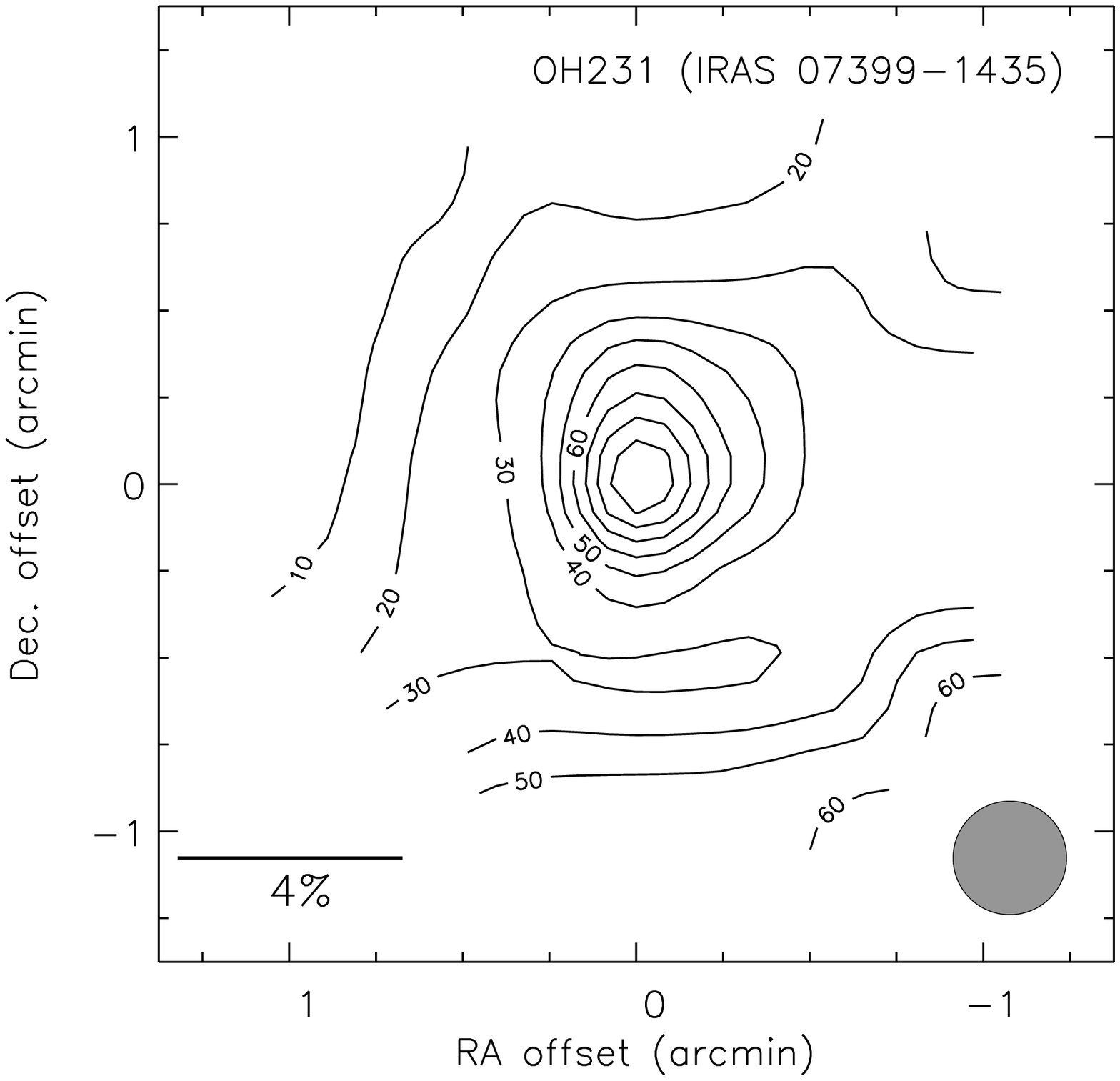}
\caption{OH231 (IRAS 07399-1435).
Offsets from $7\hour42\minute17\fs0$, $
-14\arcdeg42\arcmin49\arcsec$ (J2000).
Contours at 10, 20, ..., 90\% of the peak flux of 20\,Jy.
\label{fig29}}
\end{figure*}

\begin{figure*}
\plotone{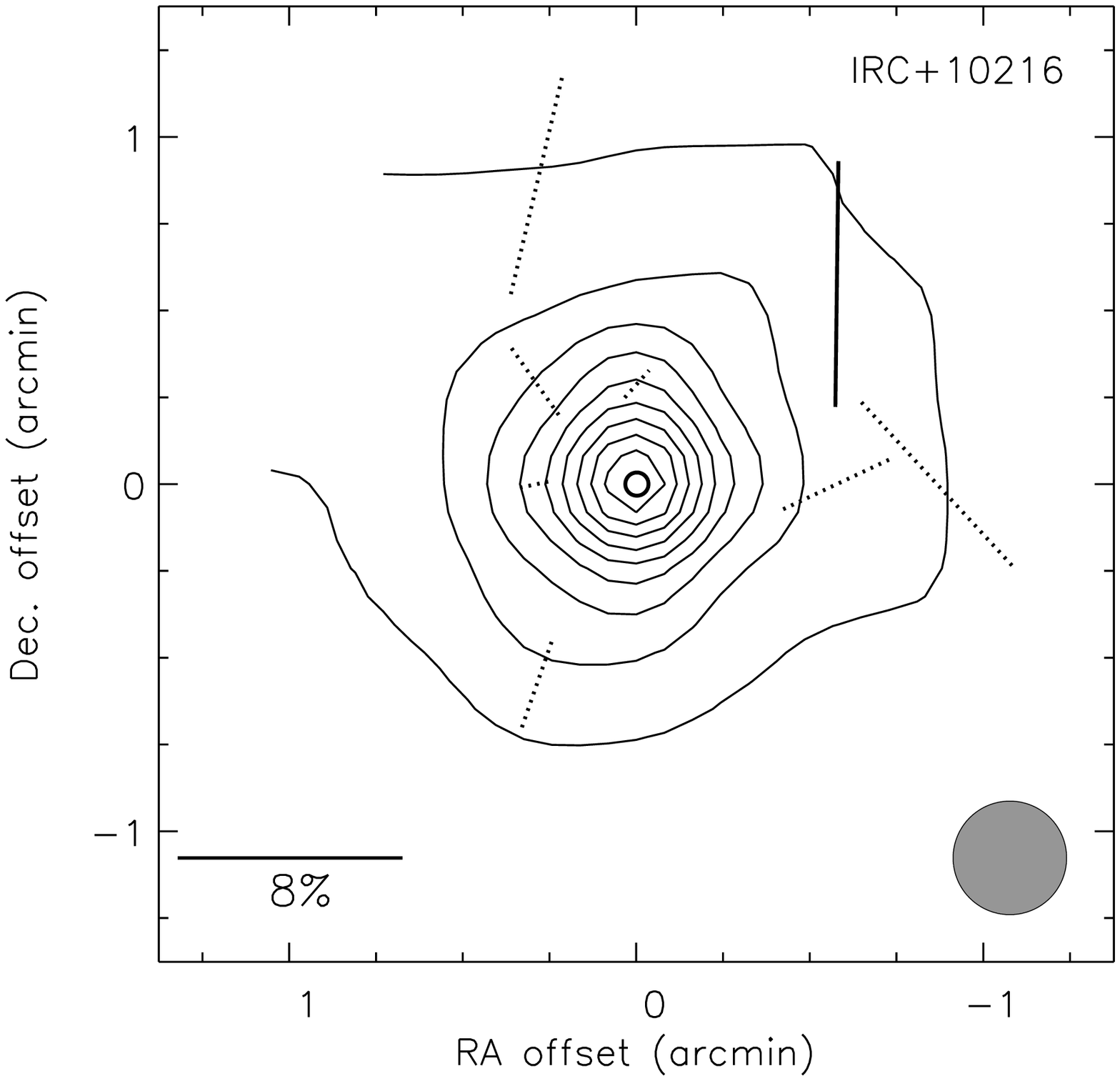}
\caption{IRC+10216.
Offsets from $9\hour47\minute57\fs3$, $
13\arcdeg16\arcmin43\arcsec$ (J2000).
Contours at 10, 20, ..., 90\% of the peak flux of 30\,Jy.
\label{fig30}}
\end{figure*}

\begin{figure*}
\plotone{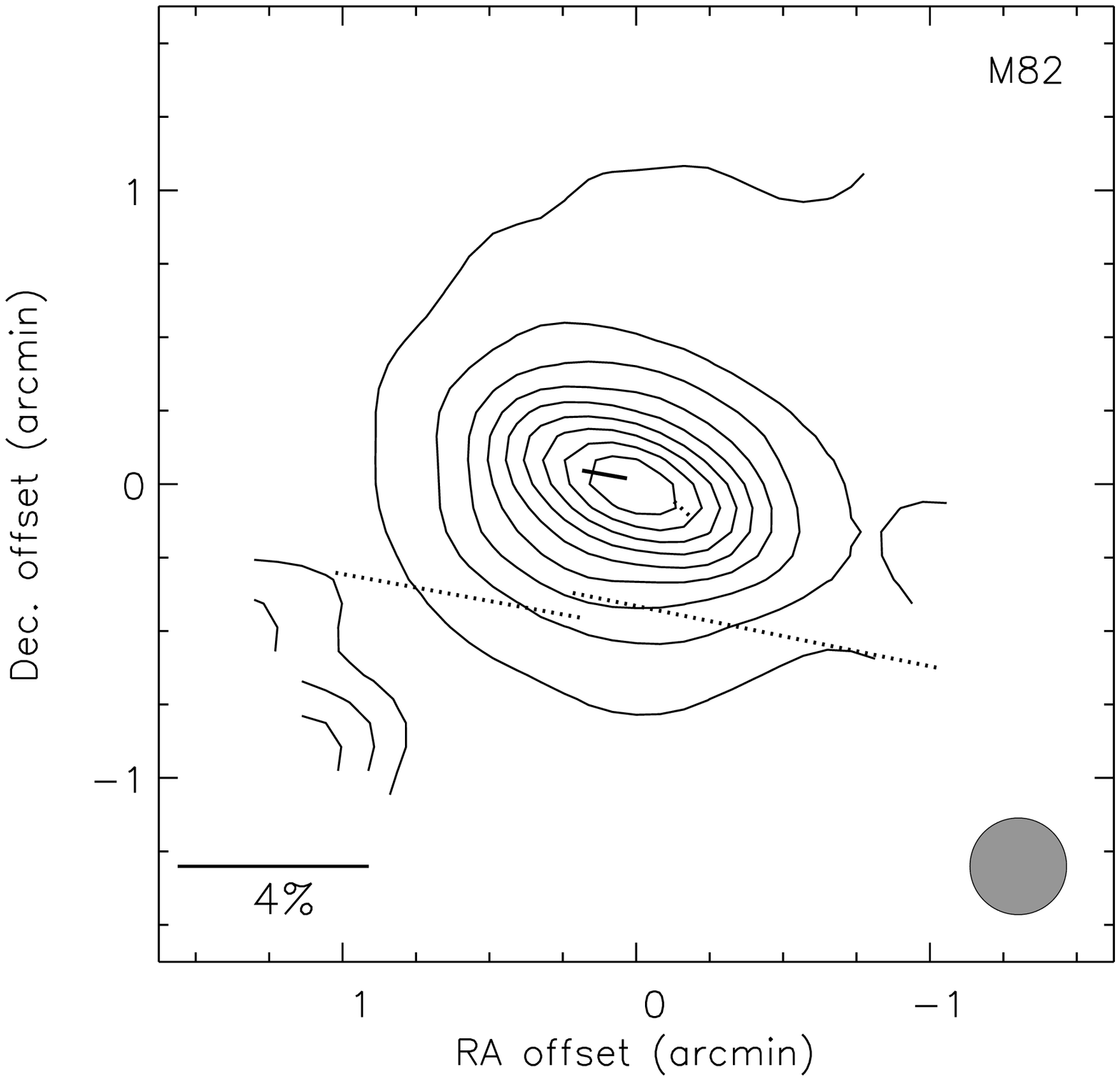}
\caption{M82.
Offsets from $9\hour55\minute52\fs2$, $
69\arcdeg40\arcmin46\arcsec$ (J2000).
Contours at 10, 20, ..., 90\% of the peak flux of 45\,Jy.
\label{fig31}}
\end{figure*}

\begin{figure*}
\plotone{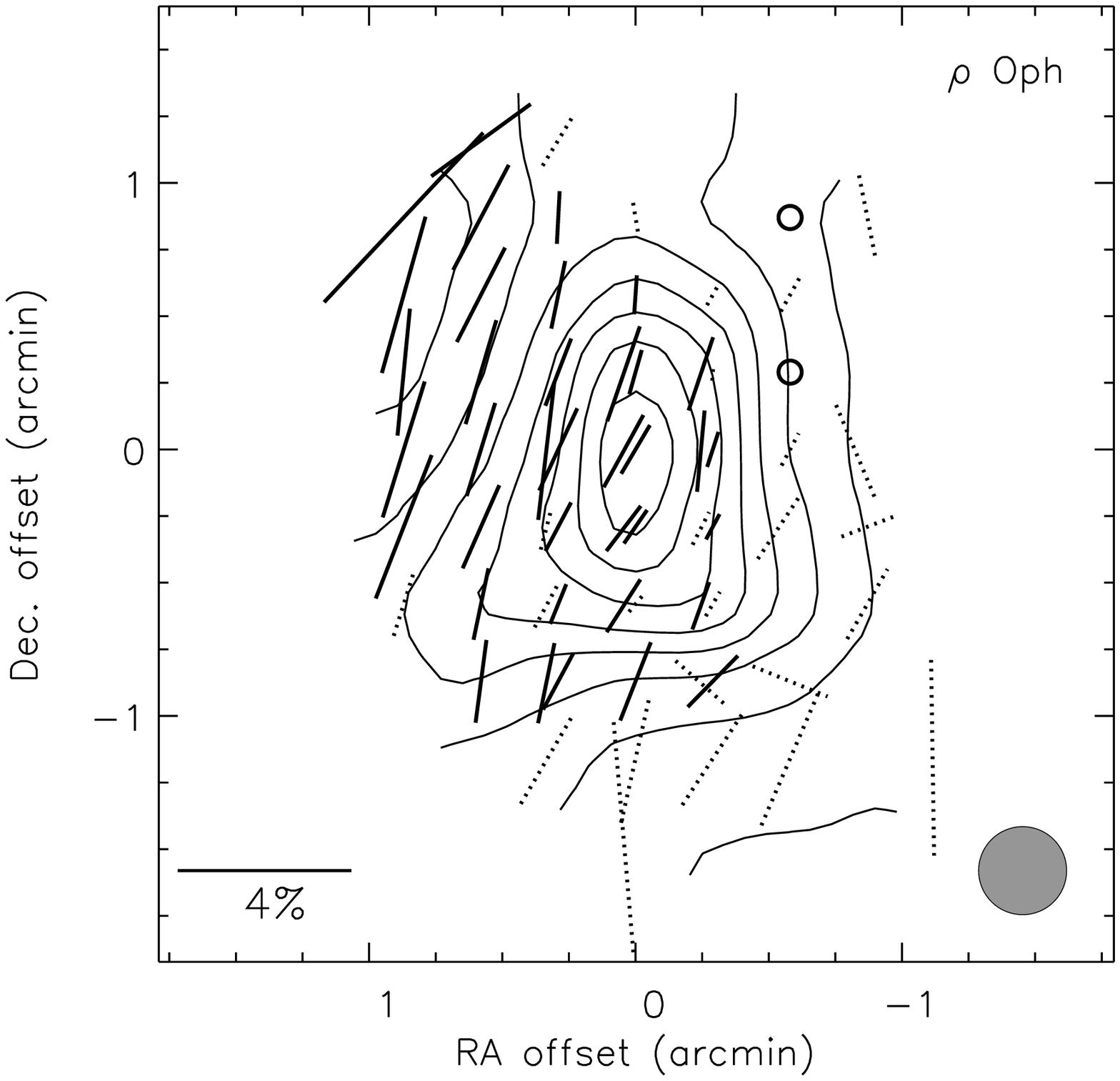}
\caption{$\rho$ Oph.
Offsets from $16\hour26\minute27\fs5$, $
-24\arcdeg23\arcmin54\arcsec$ (J2000).
Contours at 20, 30, ..., 90\% of the peak flux of 110\,Jy.
\label{fig32}}
\end{figure*}

\begin{figure*}
\plotone{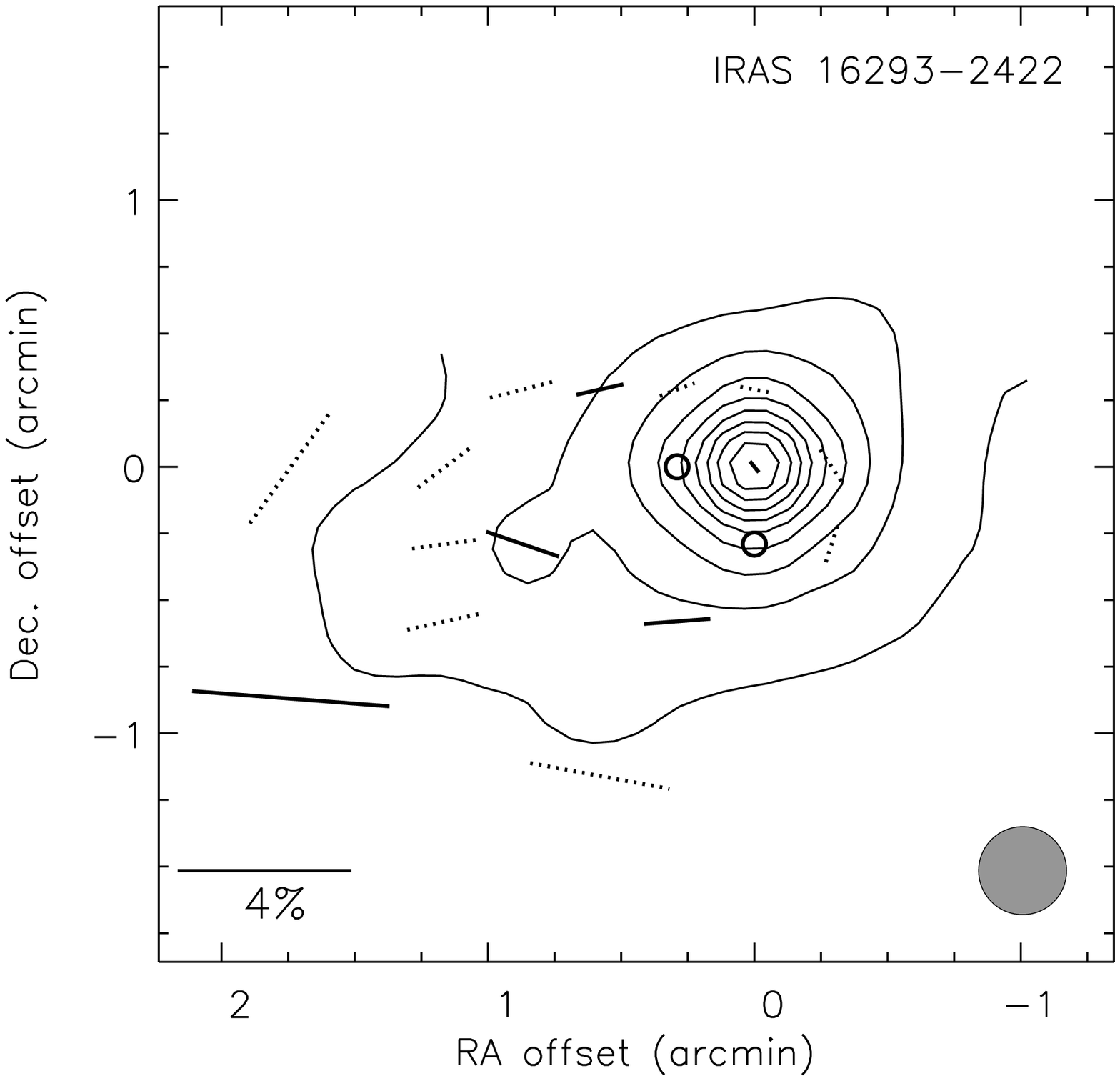}
\caption{IRAS 16293-2422.
Offsets from $16\hour32\minute22\fs9$, $
-24\arcdeg28\arcmin36\arcsec$ (J2000).
Contours at 10, 20, ..., 90\% of the peak flux of 240\,Jy.
\label{fig33}}
\end{figure*}

\begin{figure*}
\plotone{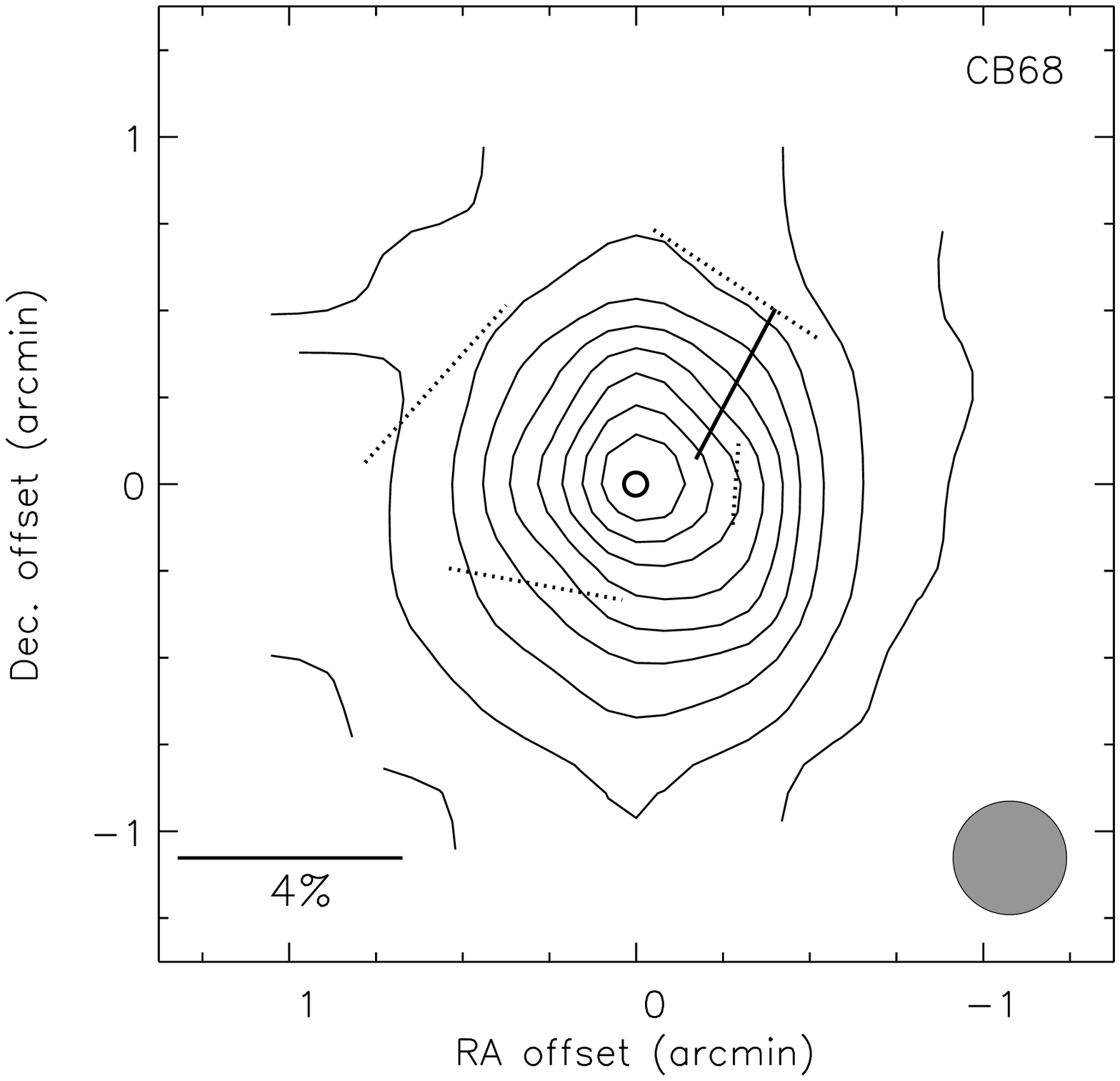}
\caption{CB68.
Offsets from $16\hour57\minute19\fs5$, $
-16\arcdeg9\arcmin21\arcsec$ (J2000).
Contours at 10, 20, ..., 90\% of the peak flux of 14\,Jy.
\label{fig34}}
\end{figure*}

\begin{figure*}
\plotone{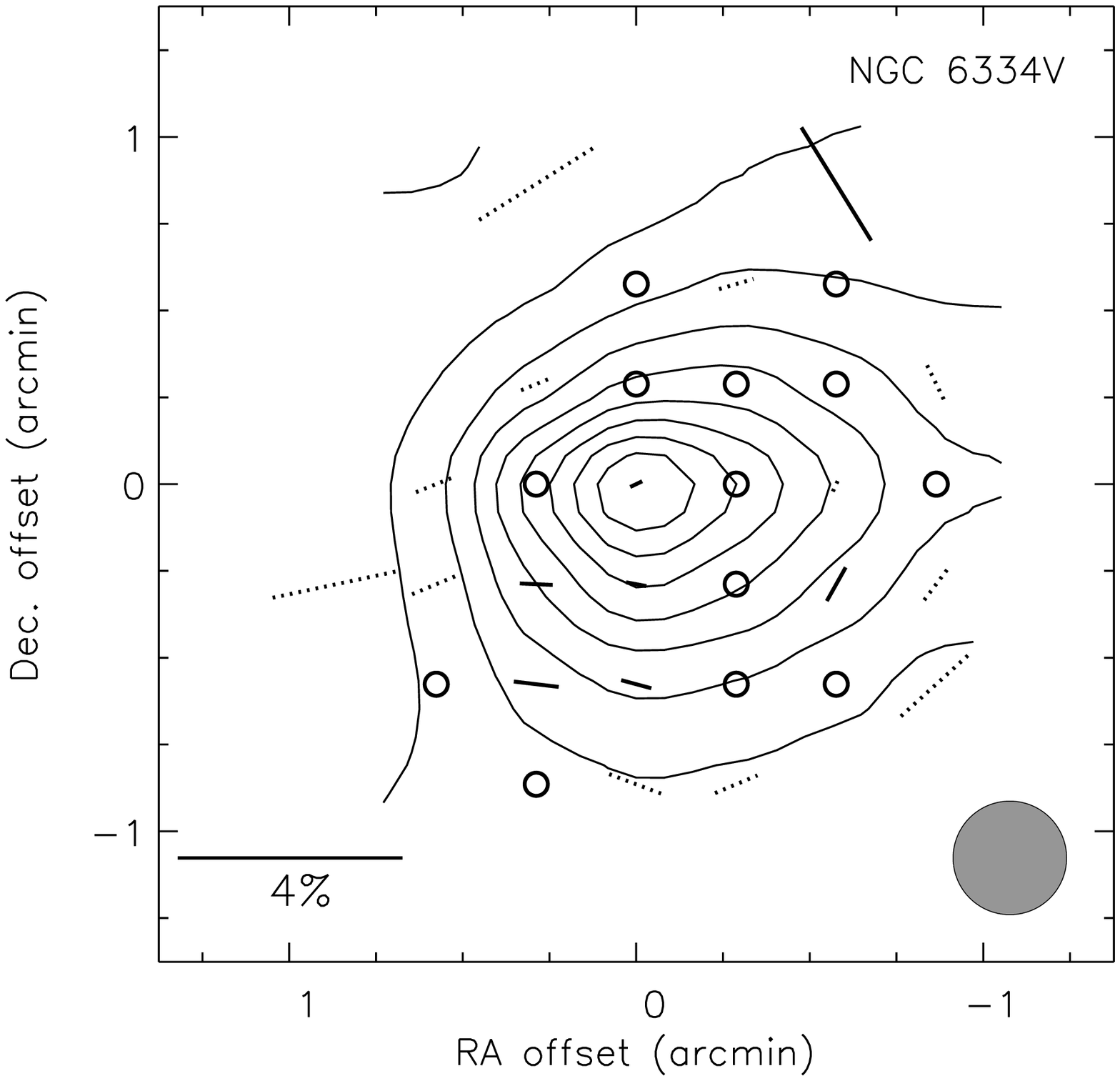}
\caption{NGC 6334V.
Offsets from $17\hour19\minute57\fs4$, $
-35\arcdeg57\arcmin46\arcsec$ (J2000).
Contours at 10, 20, ..., 90\% of the peak flux of 650\,Jy.
\label{fig35}}
\end{figure*}

\begin{figure*}
\plotone{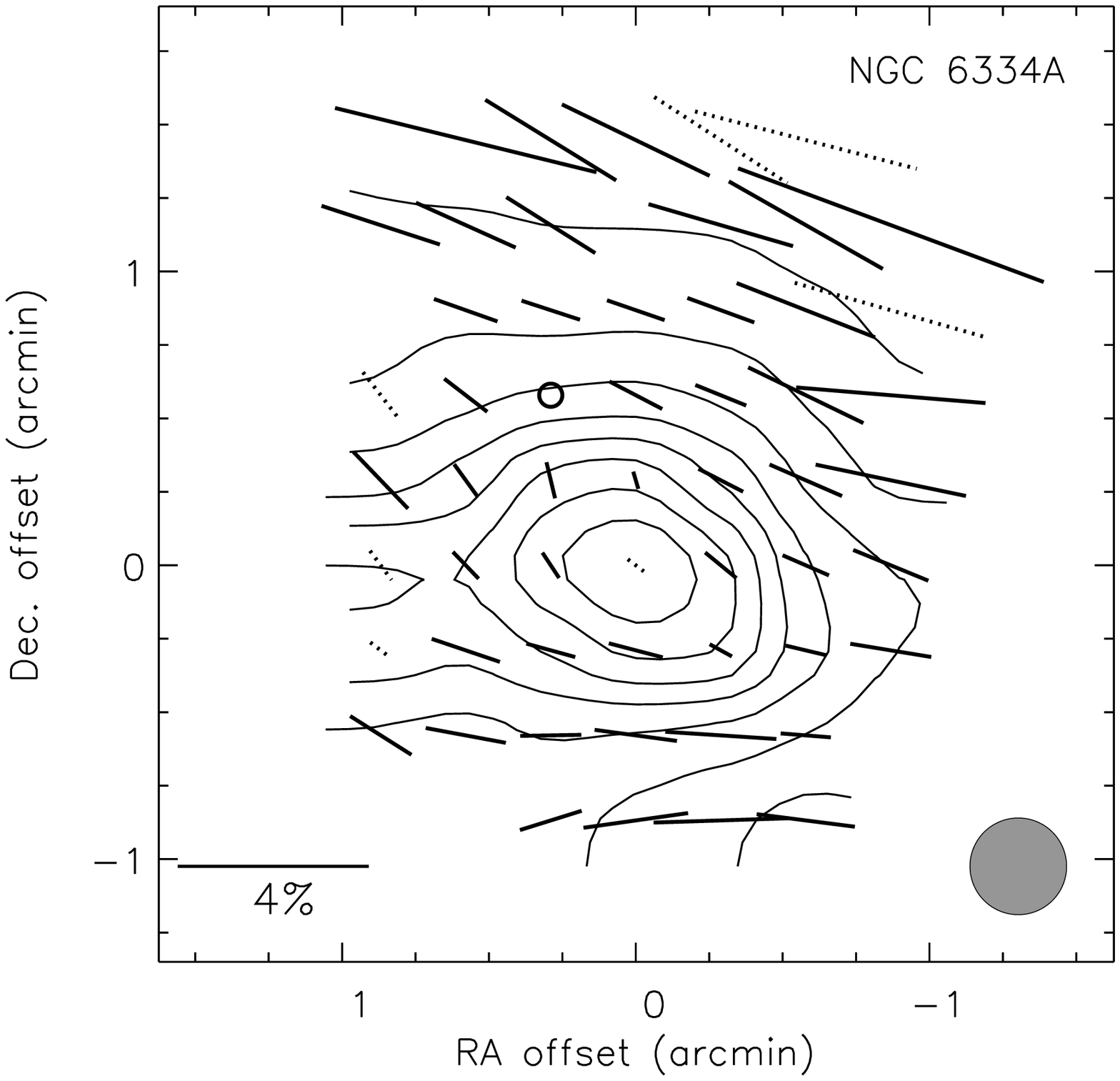}
\caption{NGC 6334A.
Offsets from $17\hour20\minute19\fs1$, $
-35\arcdeg54\arcmin45\arcsec$ (J2000).
Contours at 20, 30, ..., 90\% of the peak flux of 480\,Jy.
\label{fig36}}
\end{figure*}

\begin{figure*}
\plotone{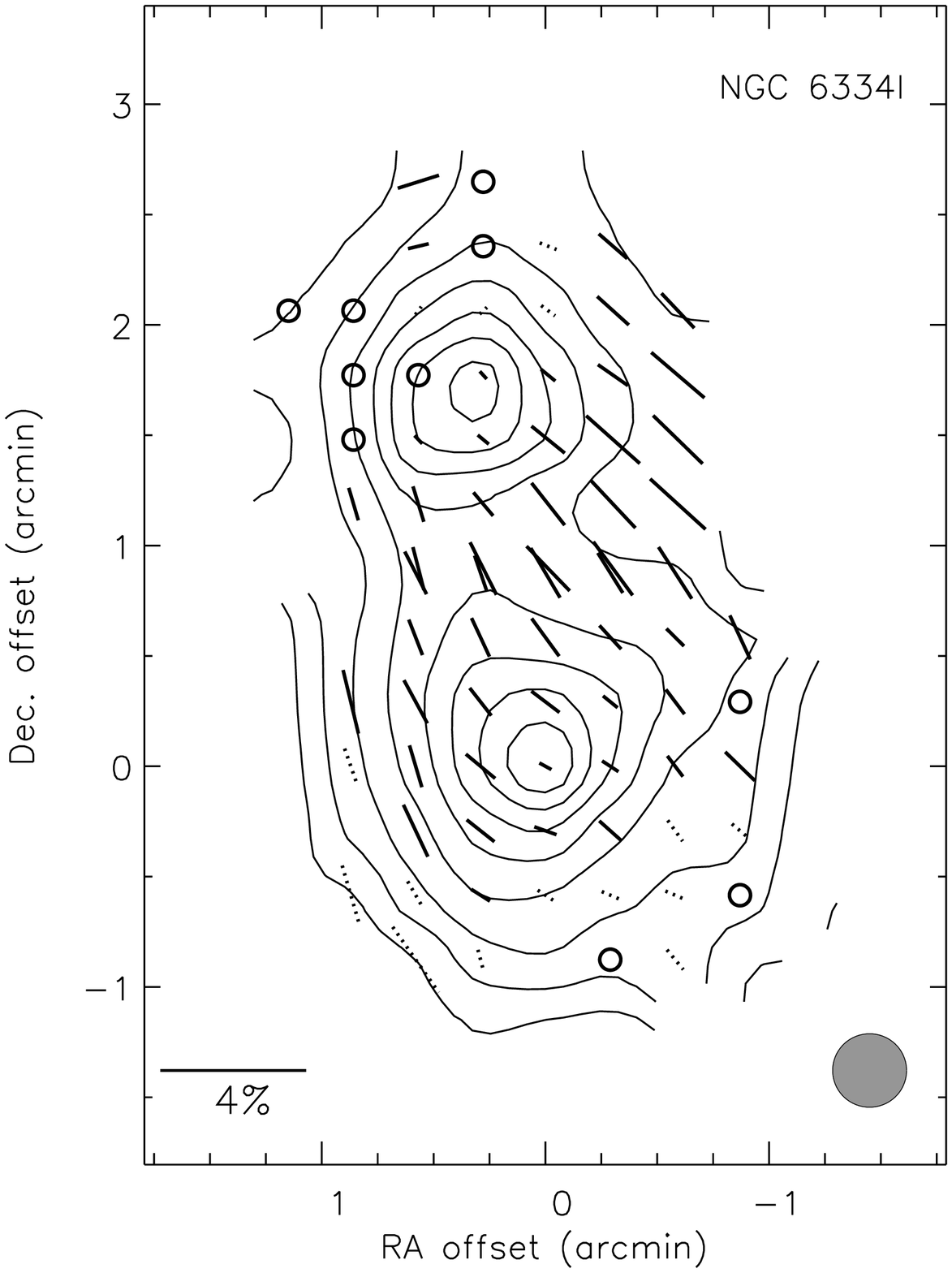}
\caption{NGC 6334I.
Offsets from $17\hour20\minute53\fs4$, $
-35\arcdeg47\arcmin0\arcsec$ (J2000).
Contours at 20, 30, ..., 90\% of the peak flux of 1200\,Jy.
\label{fig37}}
\end{figure*}

\begin{figure*}
\plotone{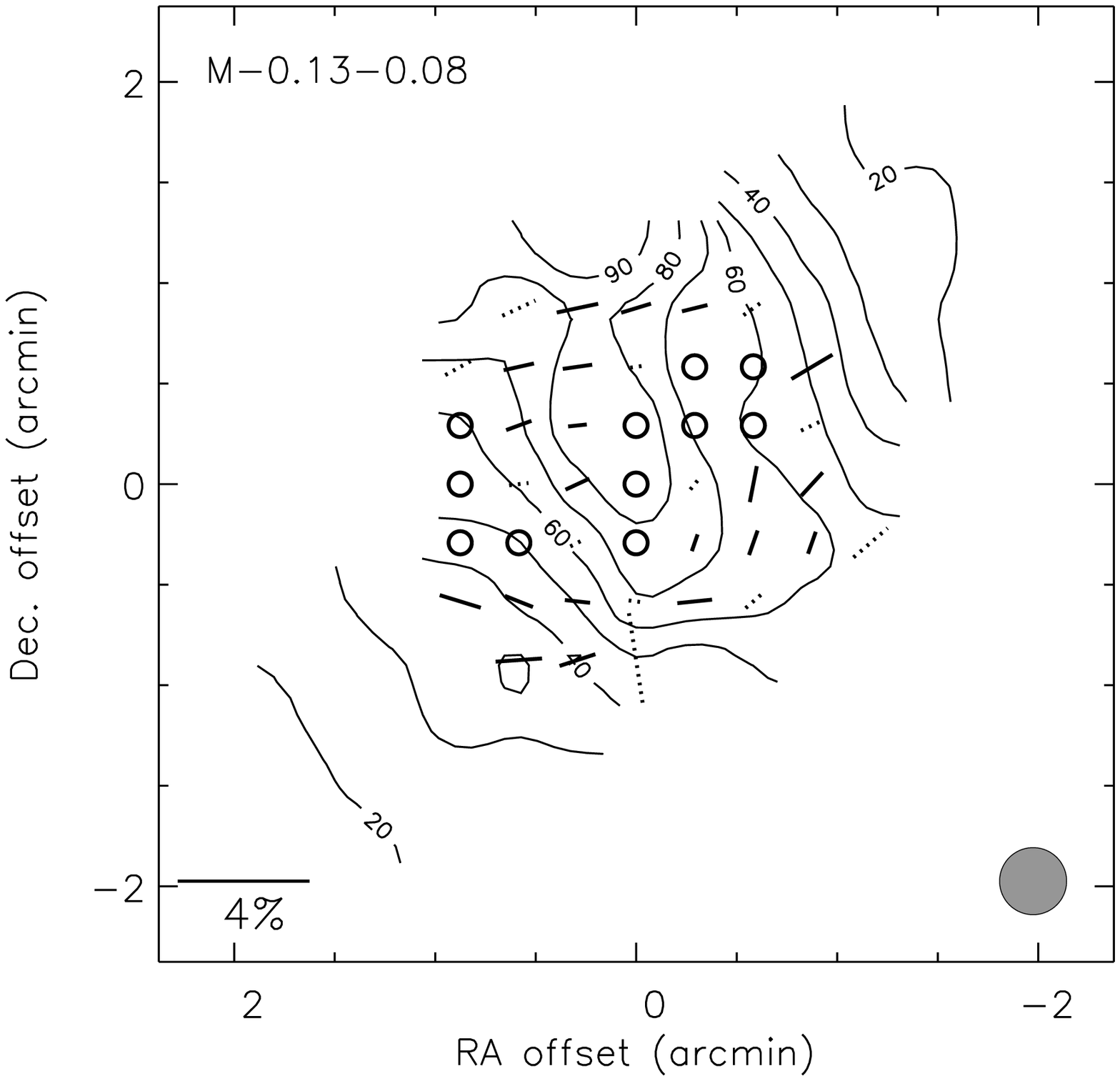}
\caption{M-0.13-0.08.
Offsets from $17\hour45\minute37\fs4$, 
$-29\arcdeg5\arcmin40\arcsec$ (J2000).
Contours at 20, ..., 90\% of the peak flux of 350\,Jy.
\label{fig38}}
\end{figure*}

\begin{figure*}
\plotone{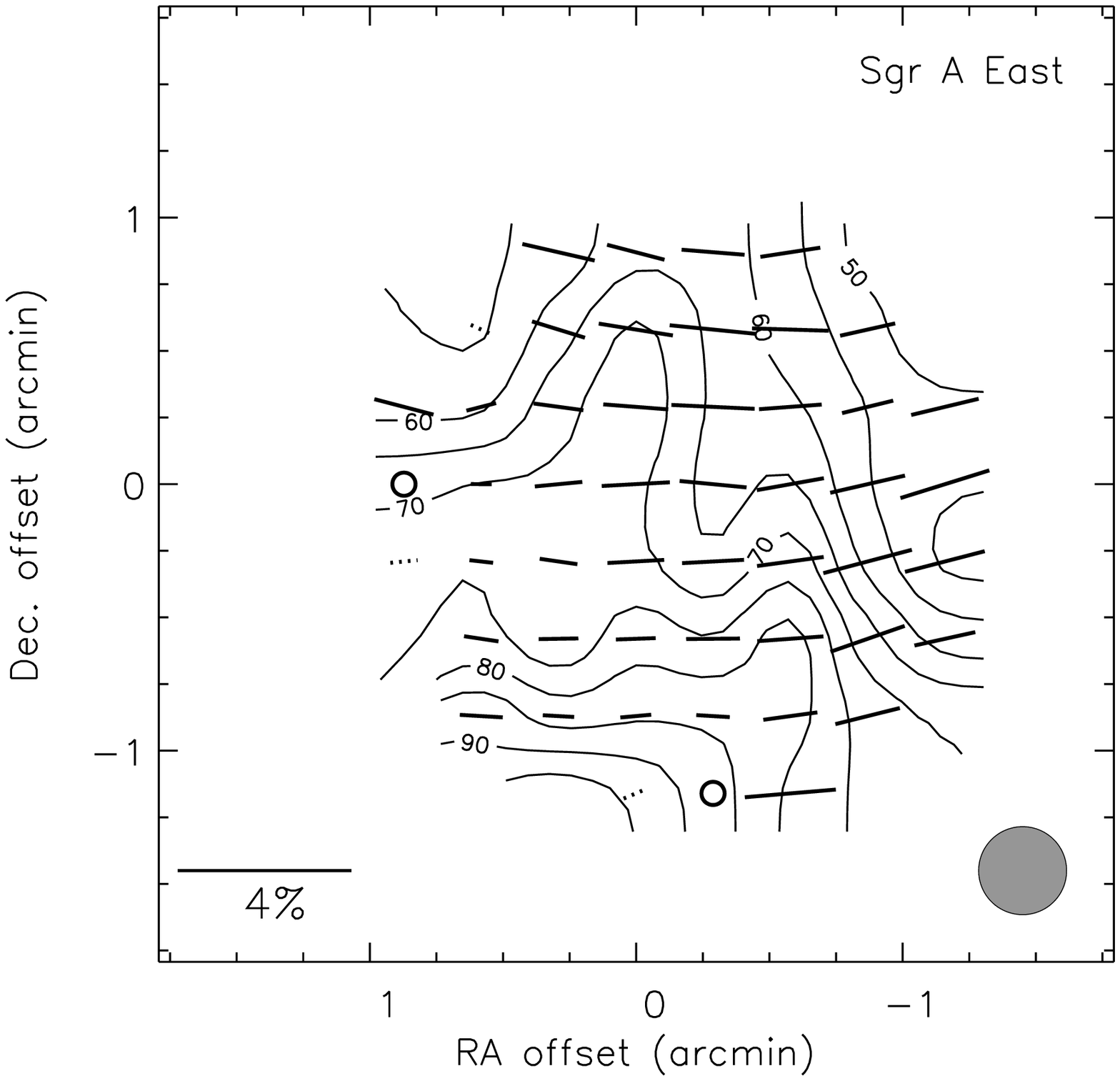}
\caption{Sgr A East.
Offsets from $17\hour45\minute41\fs5$, $
-29\arcdeg0\arcmin9\arcsec$ (J2000).
Contours at 50, 55, 60, 65, ..., 95\% of the peak flux of 200\,Jy.
\label{fig39}}
\end{figure*}

\begin{figure*}
\plotone{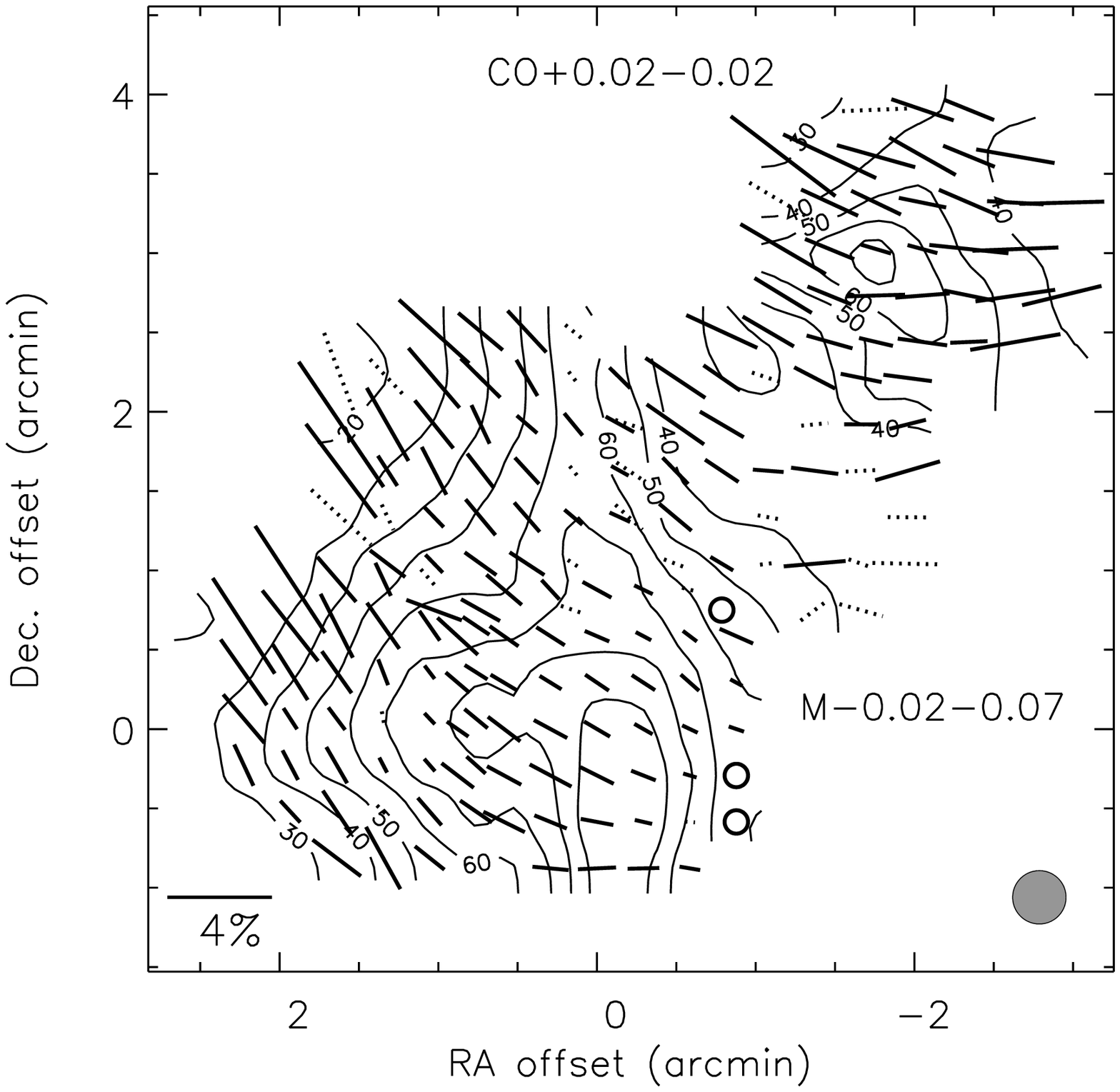}
\caption{CO+0.02-0.02 and M-0.02-0.07.
Offsets from $17\hour45\minute51\fs7$, $
-28\arcdeg59\arcmin9\arcsec$ (J2000).
Contours at 20, 30, ..., 90\% of the peak flux of 270\,Jy.
\label{fig40}}
\end{figure*}

\begin{figure*}
\plotone{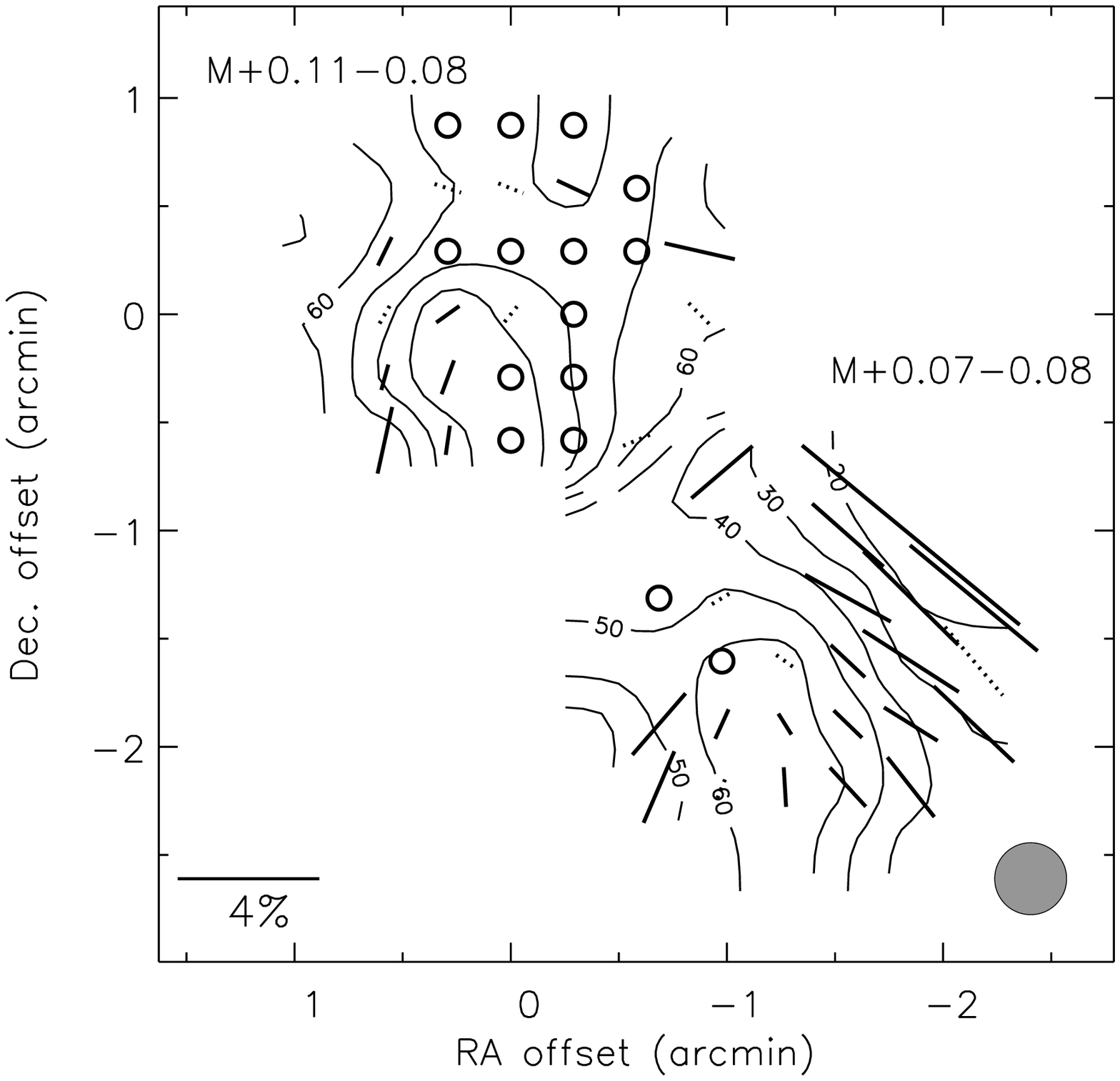}
\caption{M+0.11-0.08 and M+0.07-0.08.
Offsets from $17\hour46\minute10\fs3$, $
-28\arcdeg53\arcmin6\arcsec$ (J2000).
Contours at 20, 30, ..., 90\% of the peak flux of 210\,Jy.
\label{fig41}}
\end{figure*}

\begin{figure*}
\plotone{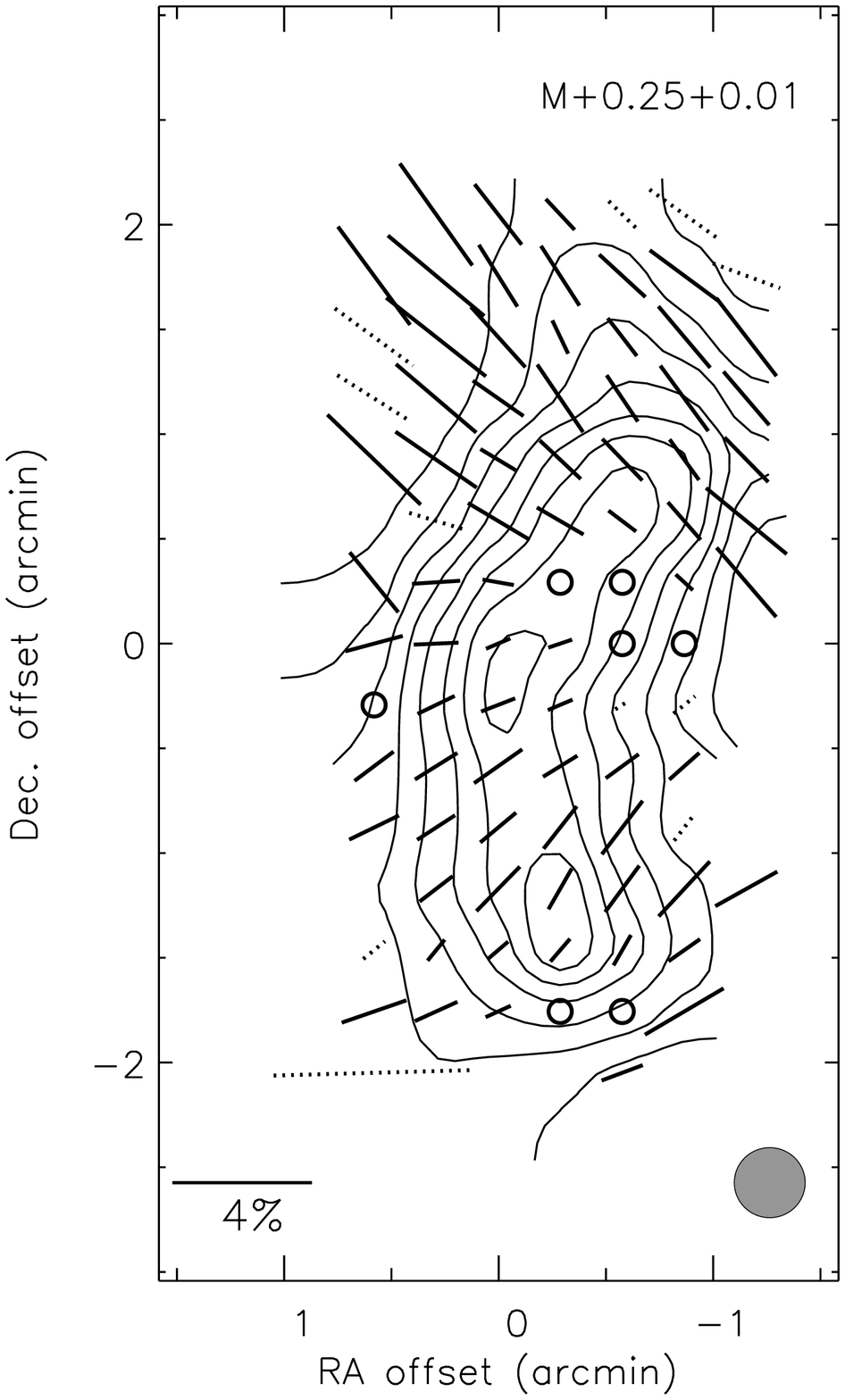}
\caption{M+0.25+0.01.
Offsets from $17\hour46\minute10\fs6$, $
-28\arcdeg42\arcmin17\arcsec$ (J2000).
Contours at 20, 30, ..., 90\% of the peak flux of 310\,Jy.
\label{fig42}}
\end{figure*}

\begin{figure*}
\plotone{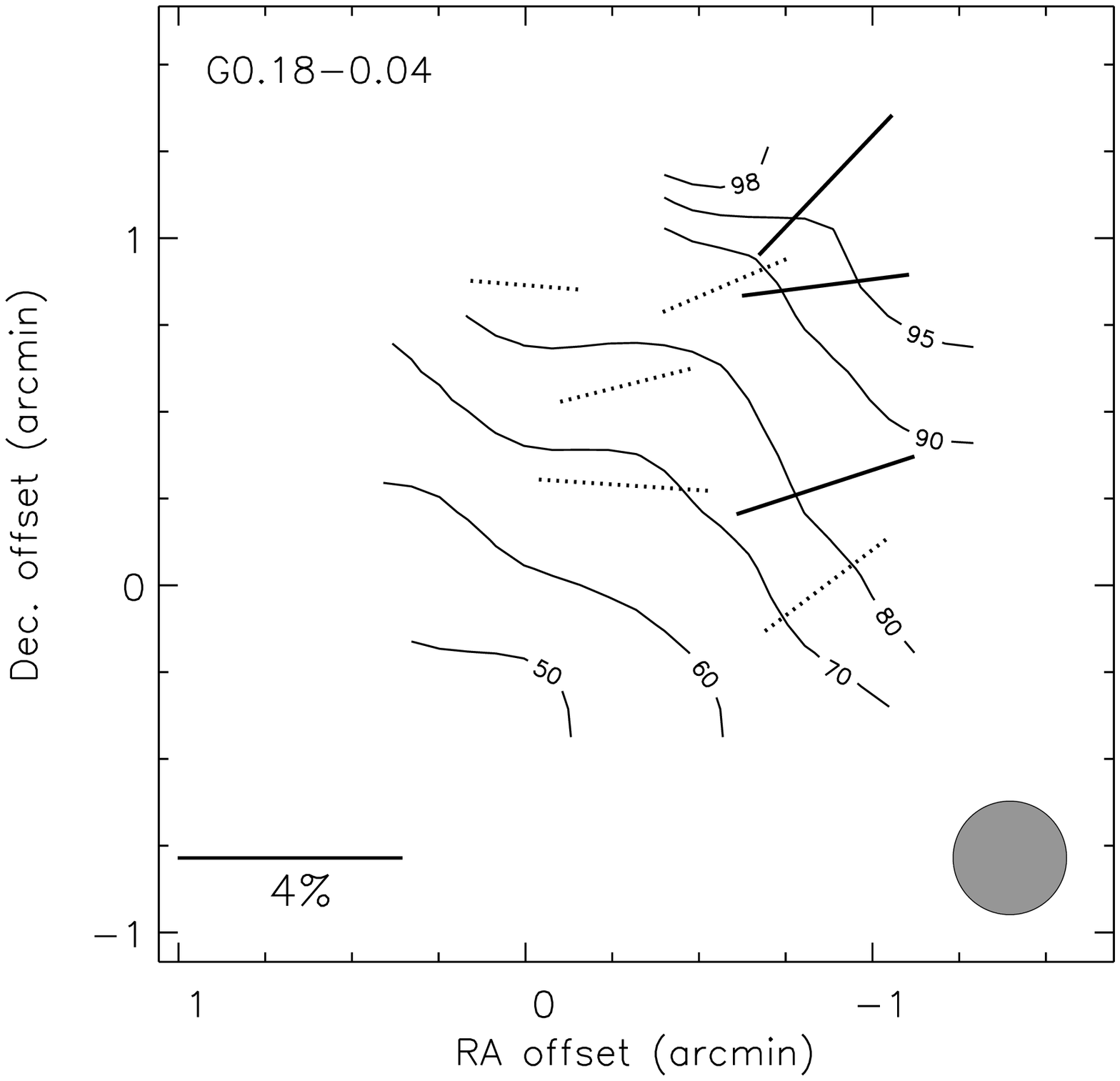}
\caption{Sickle (G0.18-0.04).
Offsets from $17\hour46\minute14\fs9$, $
-28\arcdeg48\arcmin3\arcsec$ (J2000).
Contours labeled in percent of the peak flux of 150\,Jy.
\label{fig43}}
\end{figure*}

\begin{figure*}
\plotone{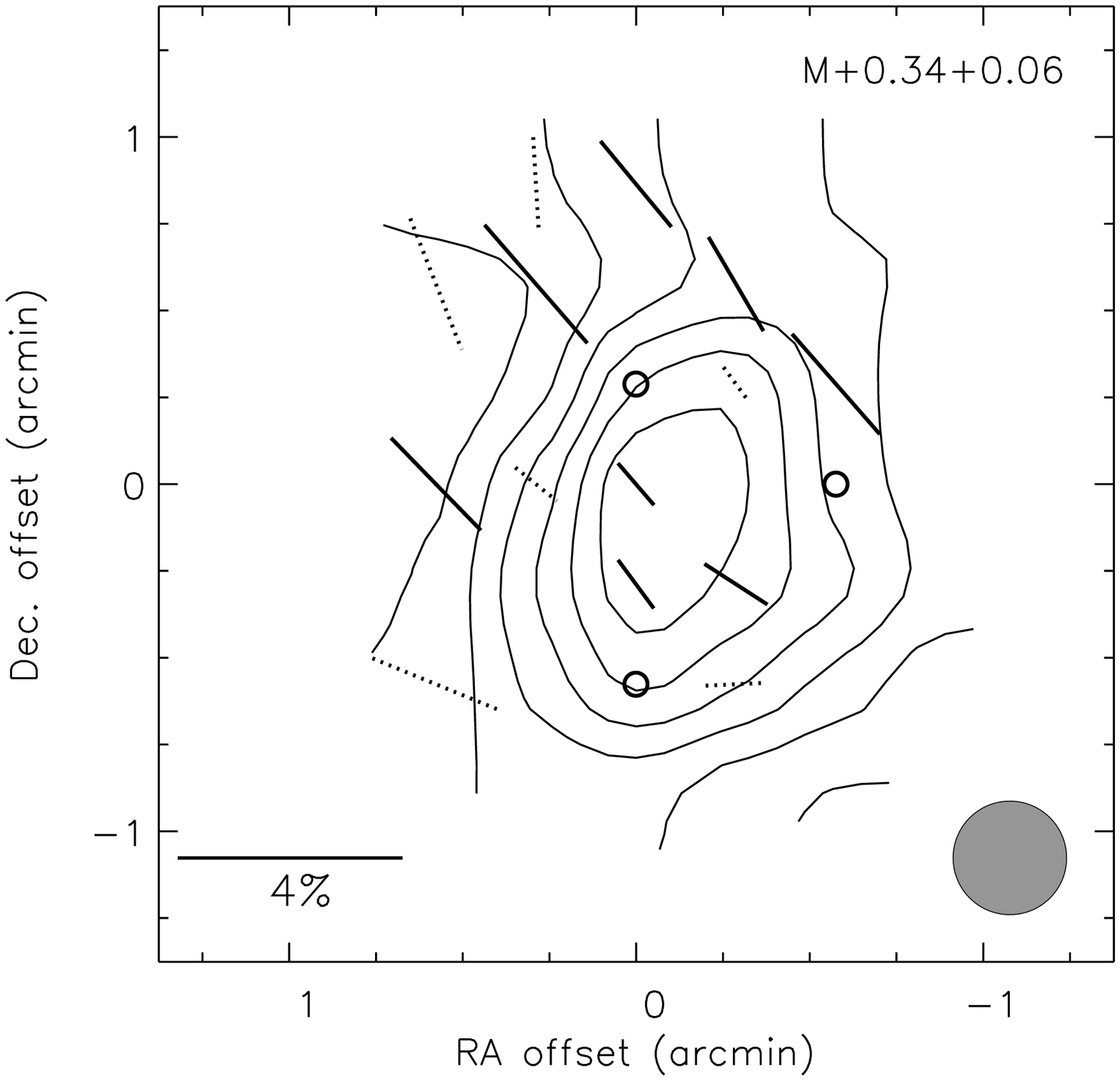}
\caption{M+0.34+0.06.
Offsets from $17\hour46\minute13\fs2$, $
-28\arcdeg36\arcmin53\arcsec$ (J2000).
Contours at 40, 50, ..., 90\% of the peak flux of 160\,Jy.
\label{fig44}}
\end{figure*}

\begin{figure*}
\plotone{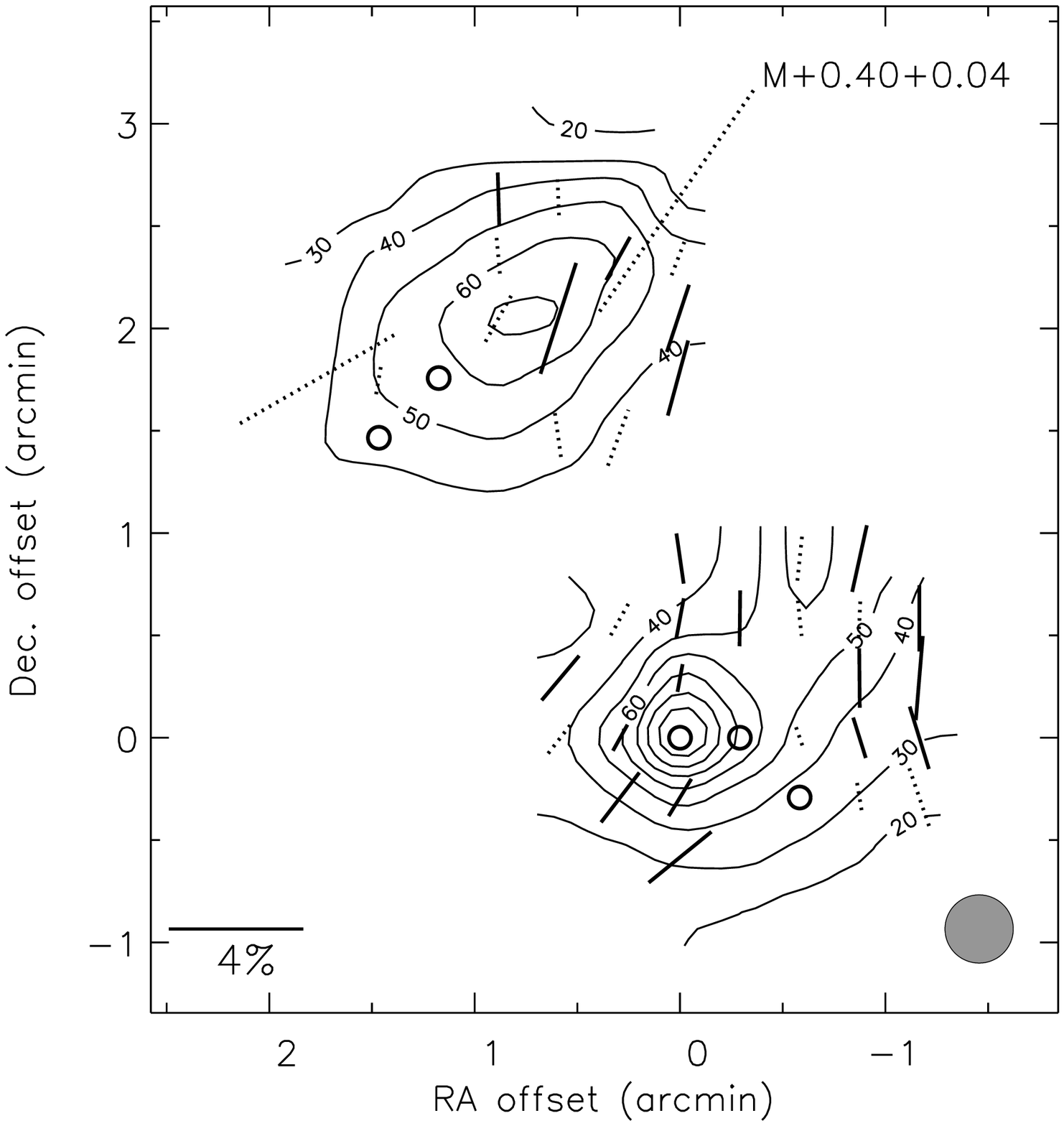}
\caption{M+0.40+0.04.
Offsets from $17\hour46\minute21\fs4$, $
-28\arcdeg35\arcmin41\arcsec$ (J2000).
Contours at 20, 30, ..., 90\% of the peak flux of 220\,Jy.
\label{fig45}}
\end{figure*}

\begin{figure*}
\plotone{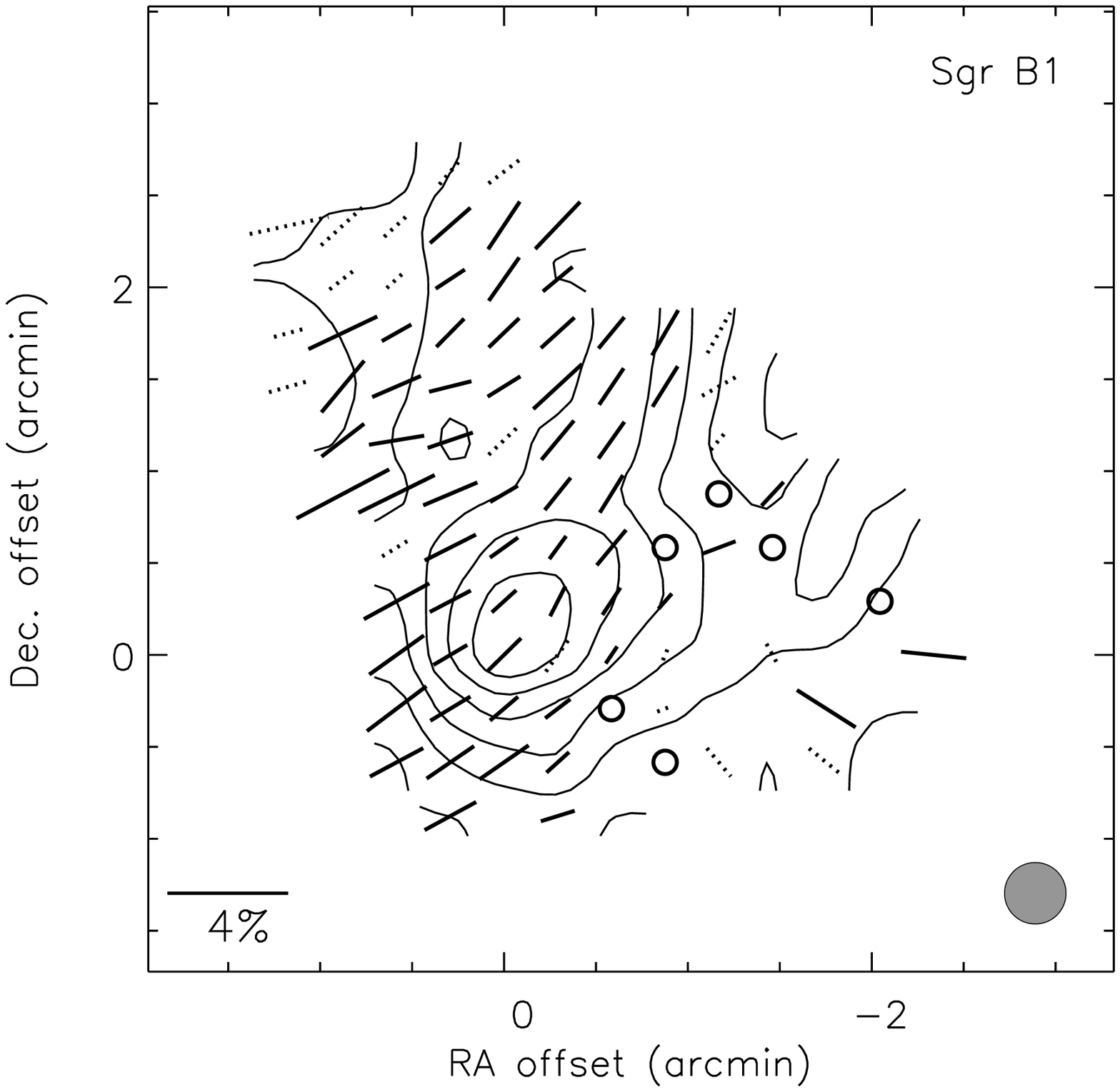}
\caption{Sgr B1.
Offsets from $17\hour46\minute47\fs2$, $
-28\arcdeg32\arcmin0\arcsec$ (J2000).
Contours at 40, 50, ..., 90\% of the peak flux of 310\,Jy.
\label{fig46}}
\end{figure*}

\begin{figure*}
\plotone{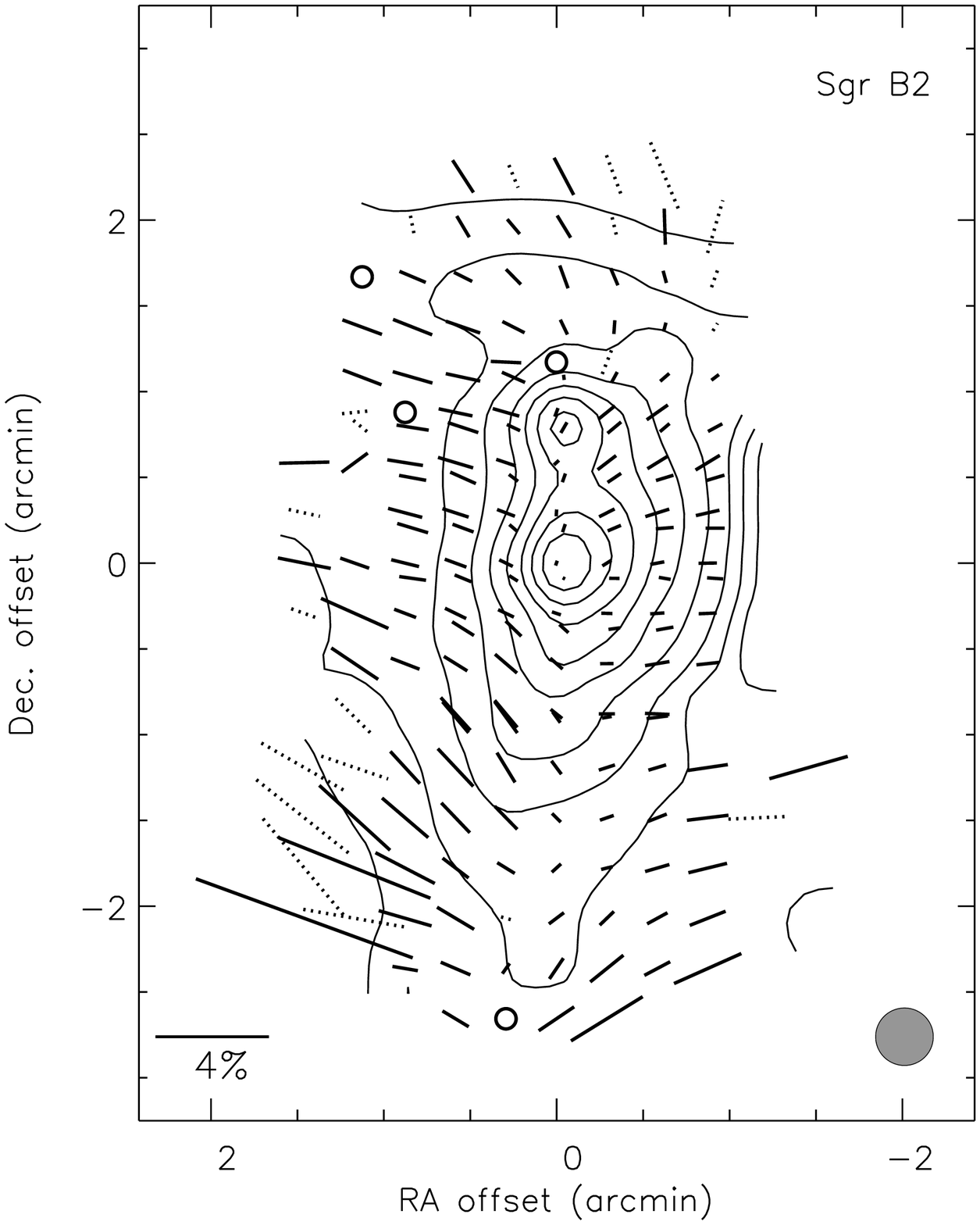}
\caption{Sgr B2.
Offsets from $17\hour47\minute20\fs2$, $
-28\arcdeg23\arcmin6\arcsec$ (J2000).
Contours at 10, 20, ..., 90\% of the peak flux of 3300\,Jy.
\label{fig47}}
\end{figure*}

\begin{figure*}
\plotone{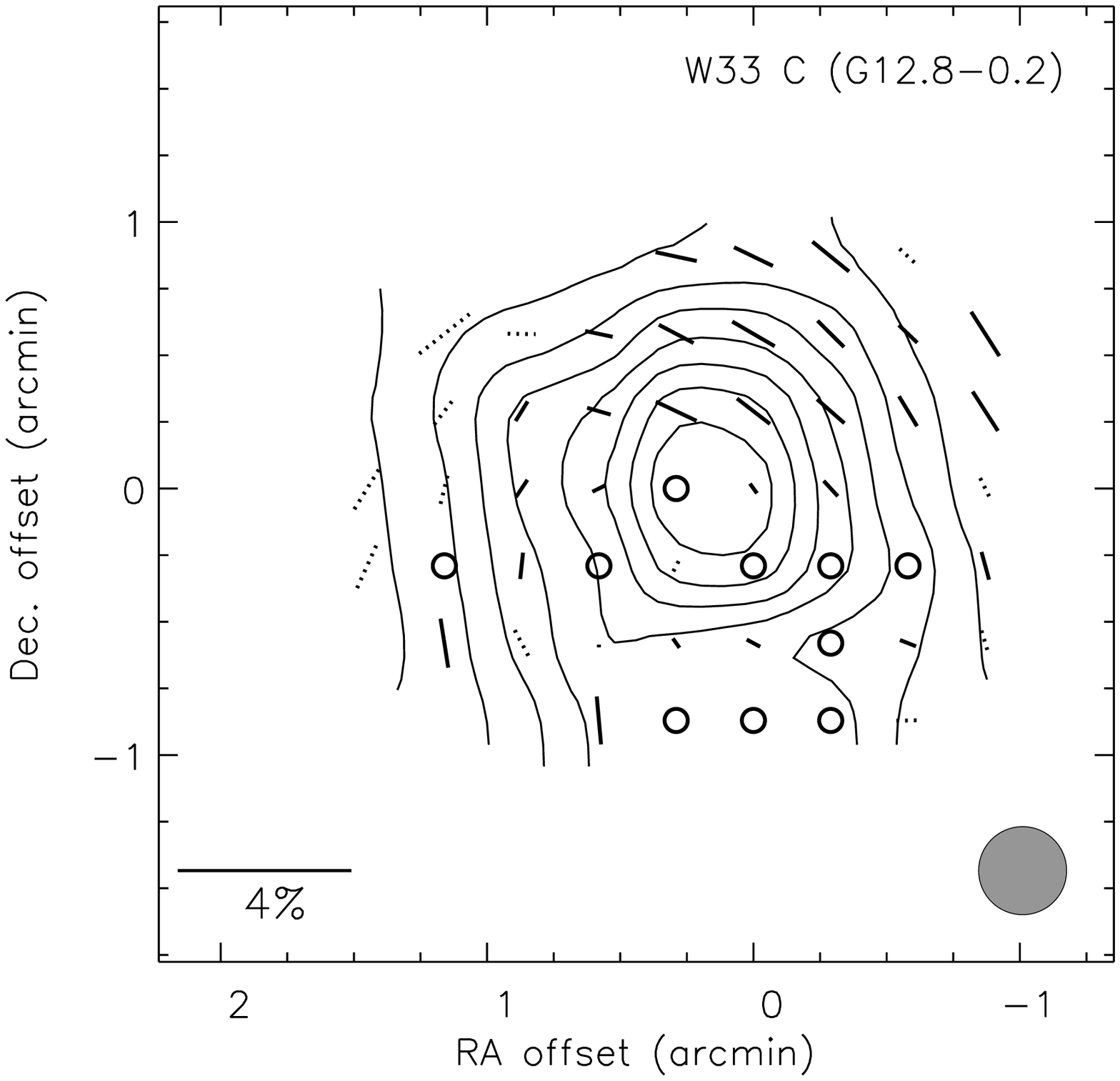}
\caption{W33 C (G12.8-0.2).
Offsets from $18\hour14\minute13\fs5$, $
-17\arcdeg55\arcmin32\arcsec$ (J2000).
Contours at 20, 30, ..., 90\% of the peak flux of 480\,Jy.
\label{fig48}}
\end{figure*}

\begin{figure*}
\plotone{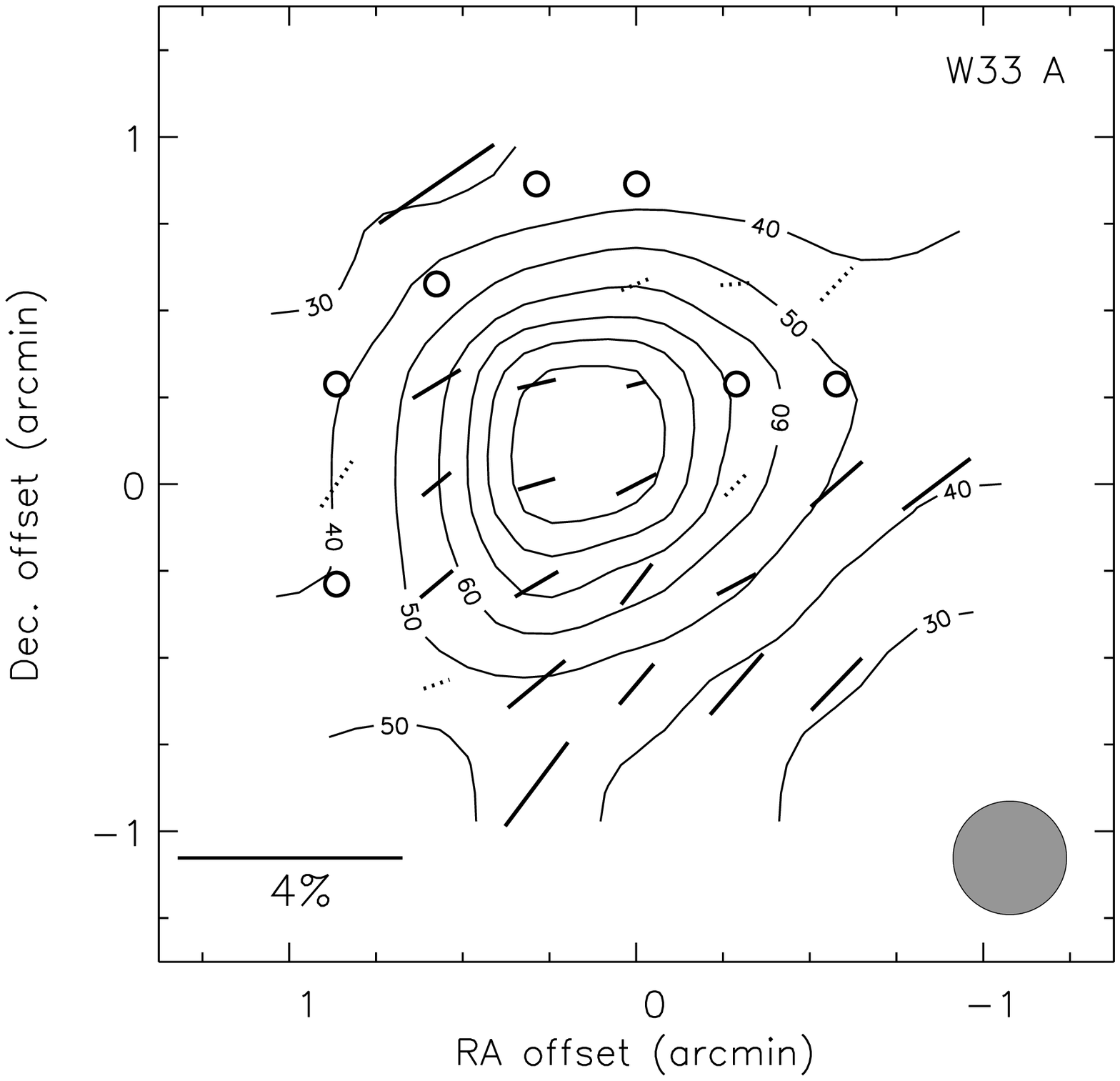}
\caption{W33 A.
Offsets from $18\hour14\minute39\fs0$, $
-17\arcdeg52\arcmin4\arcsec$ (J2000).
Contours at 30, 40, ..., 90\% of the peak flux of 170\,Jy.
\label{fig49}}
\end{figure*}

\begin{figure*}
\plotone{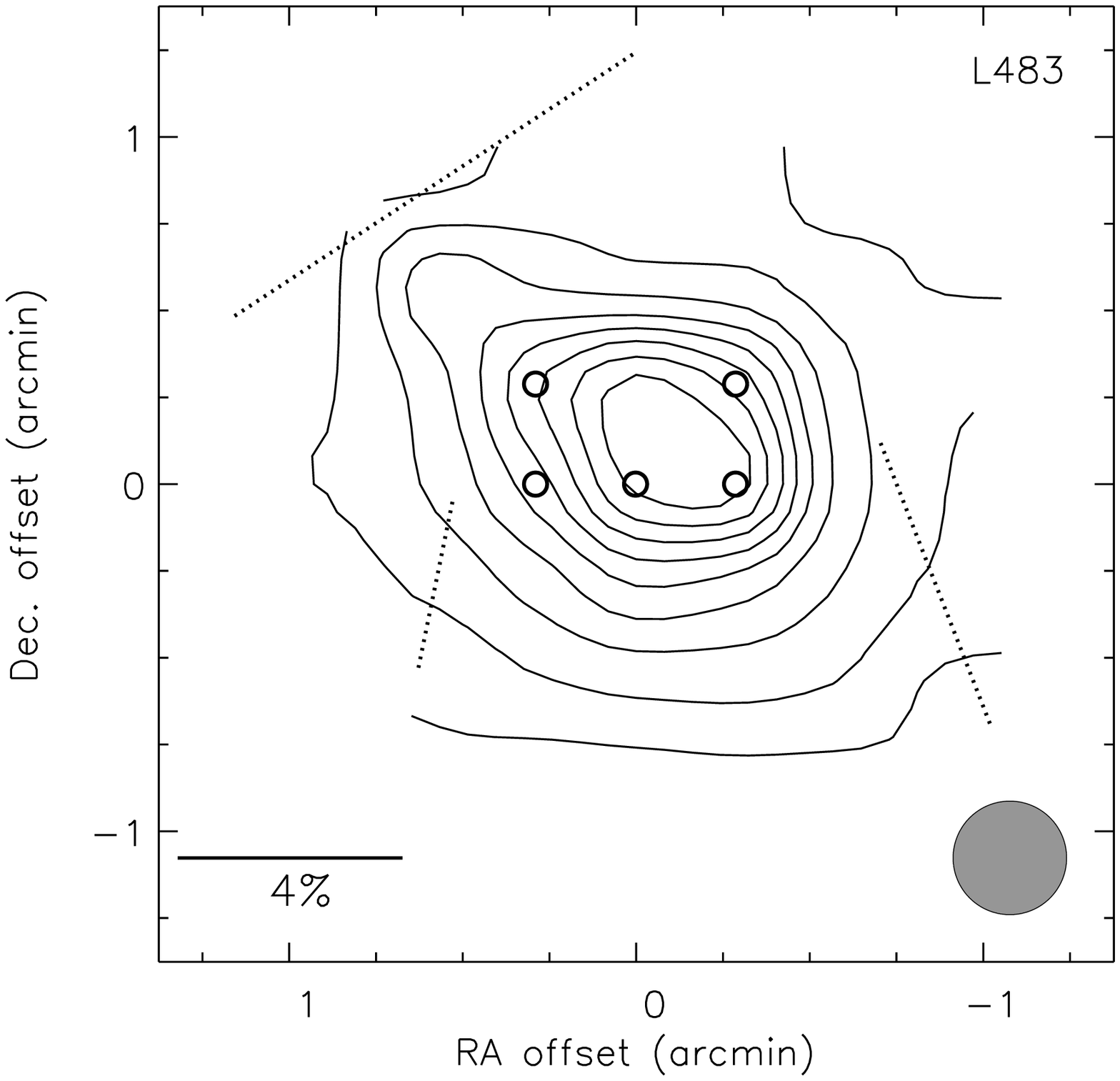}
\caption{L483.
Offsets from $18\hour17\minute29\fs8$, $
-4\arcdeg39\arcmin38\arcsec$ (J2000).
Contours at 10, 20, ..., 90\% of the peak flux of 31\,Jy.
\label{fig50}}
\end{figure*}

\begin{figure*}
\plotone{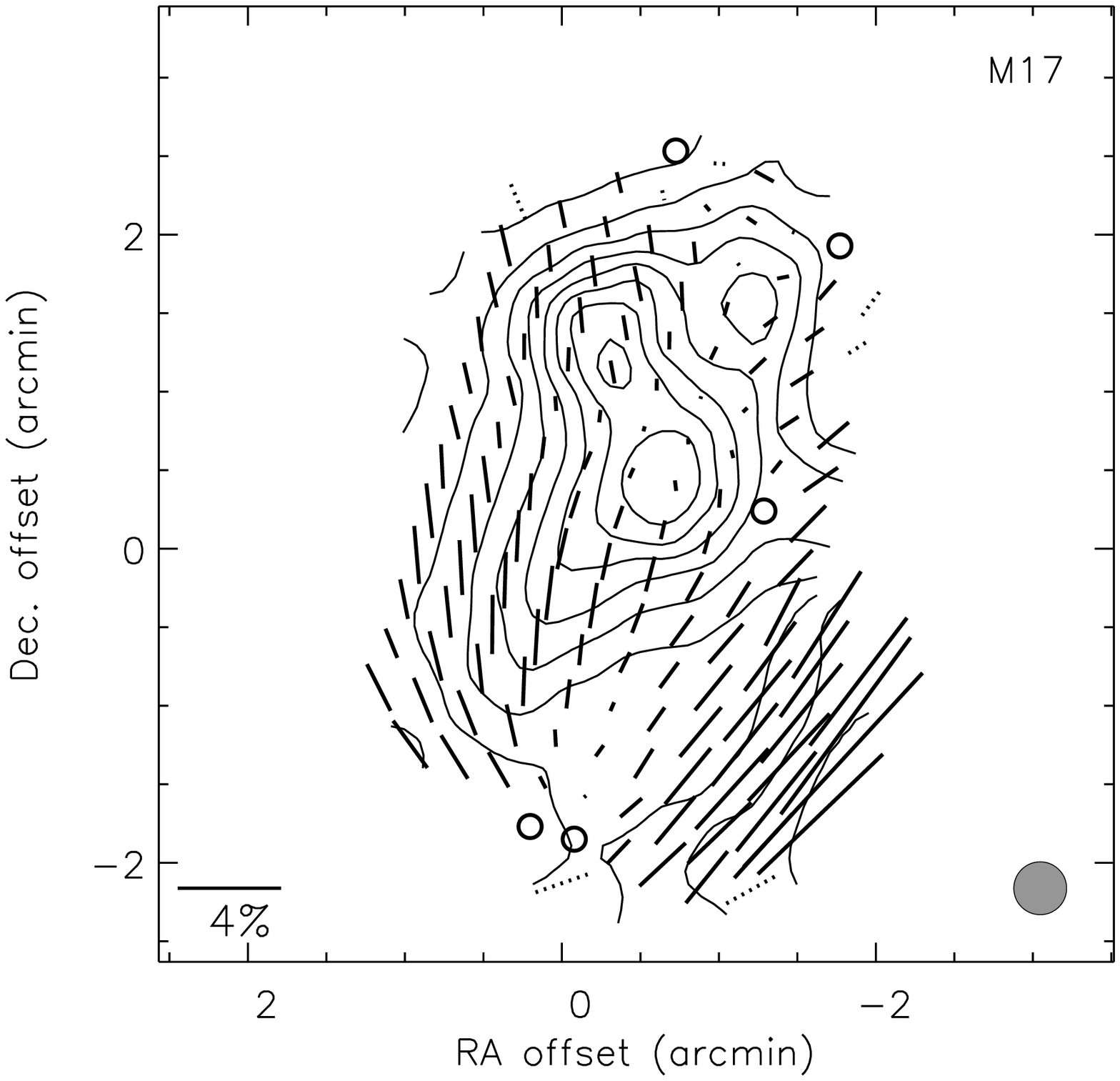}
\caption{M17.
Offsets from $18\hour20\minute24\fs6$, $
-16\arcdeg13\arcmin2\arcsec$ (J2000).
Contours at 10, 20, ..., 90\% of the peak flux of 700\,Jy.
\label{fig51}}
\end{figure*}

\begin{figure*}
\plotone{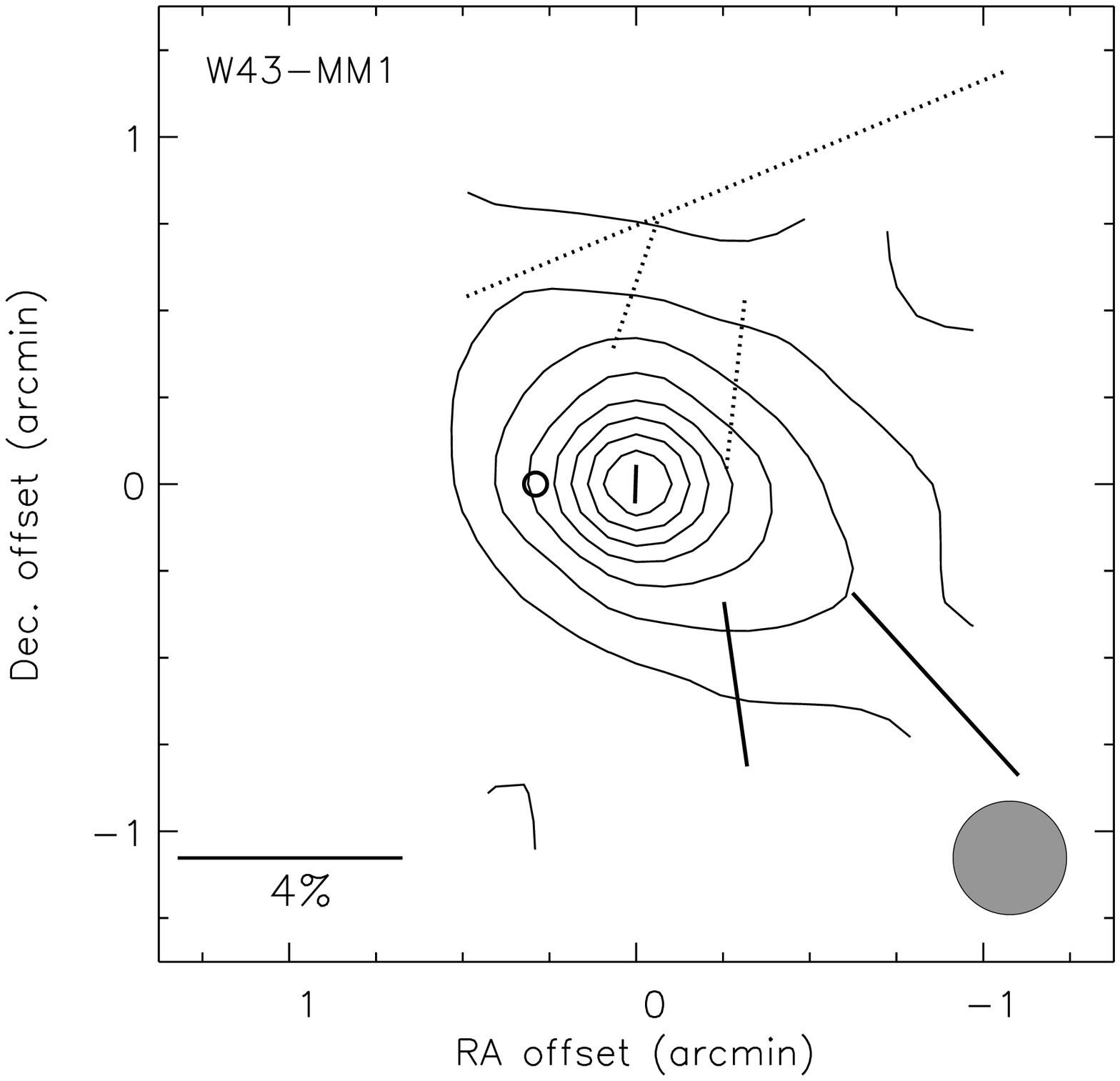}
\caption{W43-MM1.
Offsets from $18\hour47\minute47\fs0$, $
-1\arcdeg54\arcmin29\arcsec$ (J2000).
Contours at 20, 30, ..., 90\% of the peak flux of 590\,Jy.
\label{fig52}}
\end{figure*}

\begin{figure*}
\plotone{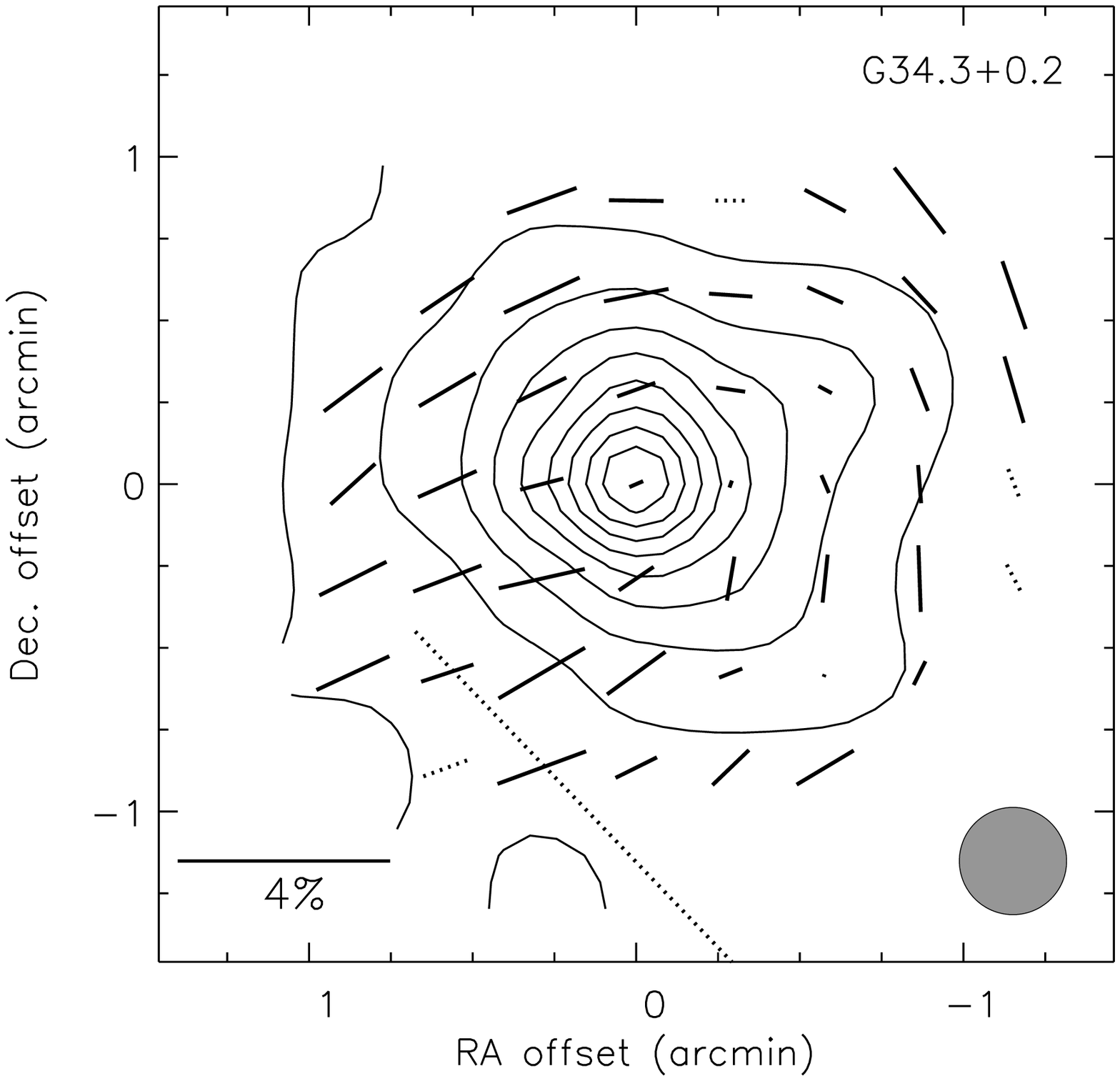}
\caption{G34.3+0.2.
Offsets from $18\hour53\minute18\fs6$, $
1\arcdeg14\arcmin59\arcsec$ (J2000).
Contours at 10, 20, ..., 90\% of the peak flux of 1100\,Jy.
\label{fig53}}
\end{figure*}

\begin{figure*}
\plotone{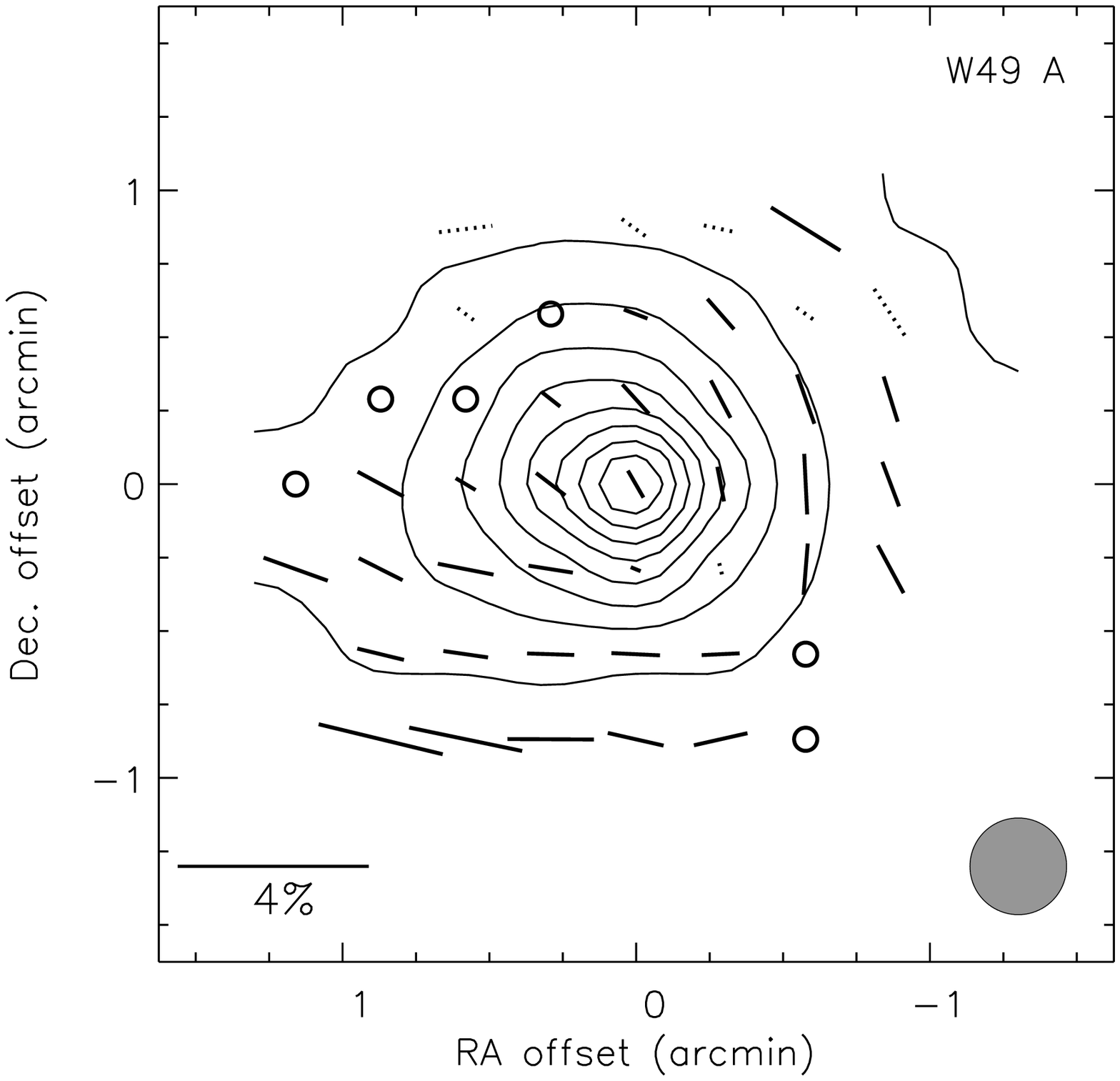}
\caption{W49 A.
Offsets from $19\hour10\minute13\fs6$, $
9\arcdeg6\arcmin17\arcsec$ (J2000).
Contours at 10, 20, ..., 90\% of the peak flux of 730\,Jy.
\label{fig54}}
\end{figure*}

\begin{figure*}
\plotone{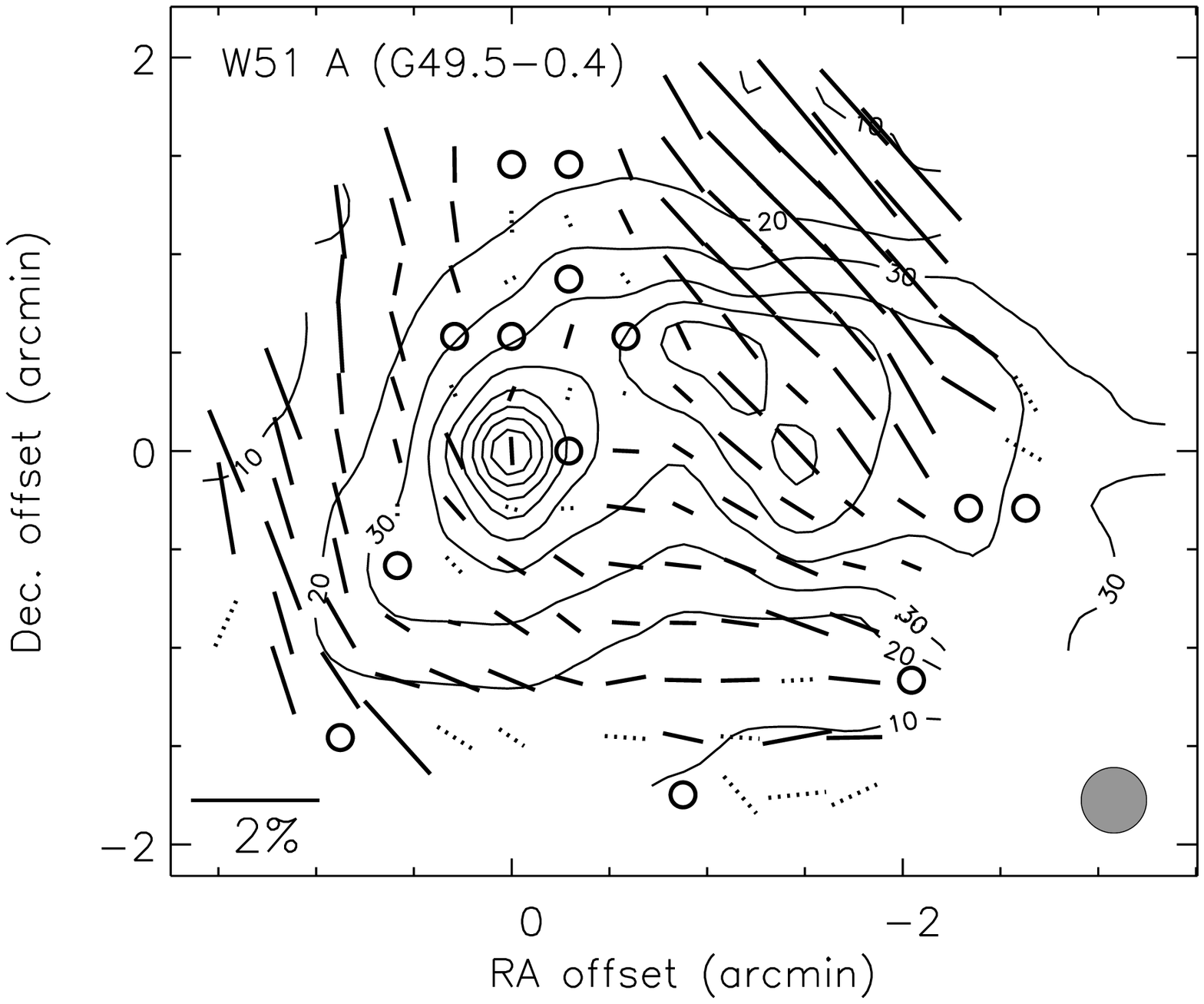}
\caption{W51 A (G49.5-0.4).
Offsets from $19\hour23\minute44\fs1$, $
14\arcdeg30\arcmin32\arcsec$ (J2000).
Contours at 10, 20, ..., 90\% of the peak flux of 1200\,Jy.
\label{fig55}}
\end{figure*}

The results are tabulated in Table~2 and shown in Figures \ref{fig5}
-- \ref{fig59}.  The plotted vectors have all been debiased (see
Section~2.3).  The results listed in Table~2 have not been debiased.
All maps include a reference vector and a shaded circle indicating the
Hertz beam size of 20\arcsec.  Coordinates in right ascension and
declination are offset from the coordinates given in the caption and
Table~\ref{tbl-object}.  Solid vectors denote measurements where
$P/\sigma_P \geq 3$; dotted vectors denote $2 \leq P/\sigma_P < 3$; and open
circles denote $P + 2\sigma_P \leq 1\%$ .  Note that any vector which
meets the criterion $P/\sigma_P > 3$ has $\Delta\phi < 10\arcdeg$ and
any vector with $P/\sigma_P > 2$ has $\Delta\phi < 15\arcdeg$.  Flux
contours were generated by smoothing the data with a $20\arcsec$
gaussian, choosen to mimic the Hertz beam.  No smoothing was applied
to the polarizations and fluxes tabulated in Table~2.

\epsscale{0.74}
\epsscale{0.90}
\epsscale{0.90}
\epsscale{0.70}
\epsscale{1}
\epsscale{0.74}
\epsscale{0.73}
\clearpage
\epsscale{0.8}
\epsscale{0.73}
\epsscale{0.8}
\clearpage
\epsscale{0.74}
\epsscale{1}
\epsscale{.74}
\epsscale{0.90}
\epsscale{0.74}
\epsscale{0.90}
\epsscale{0.74}
\clearpage
\epsscale{1}
\epsscale{0.74}
\begin{figure*}
\plotone{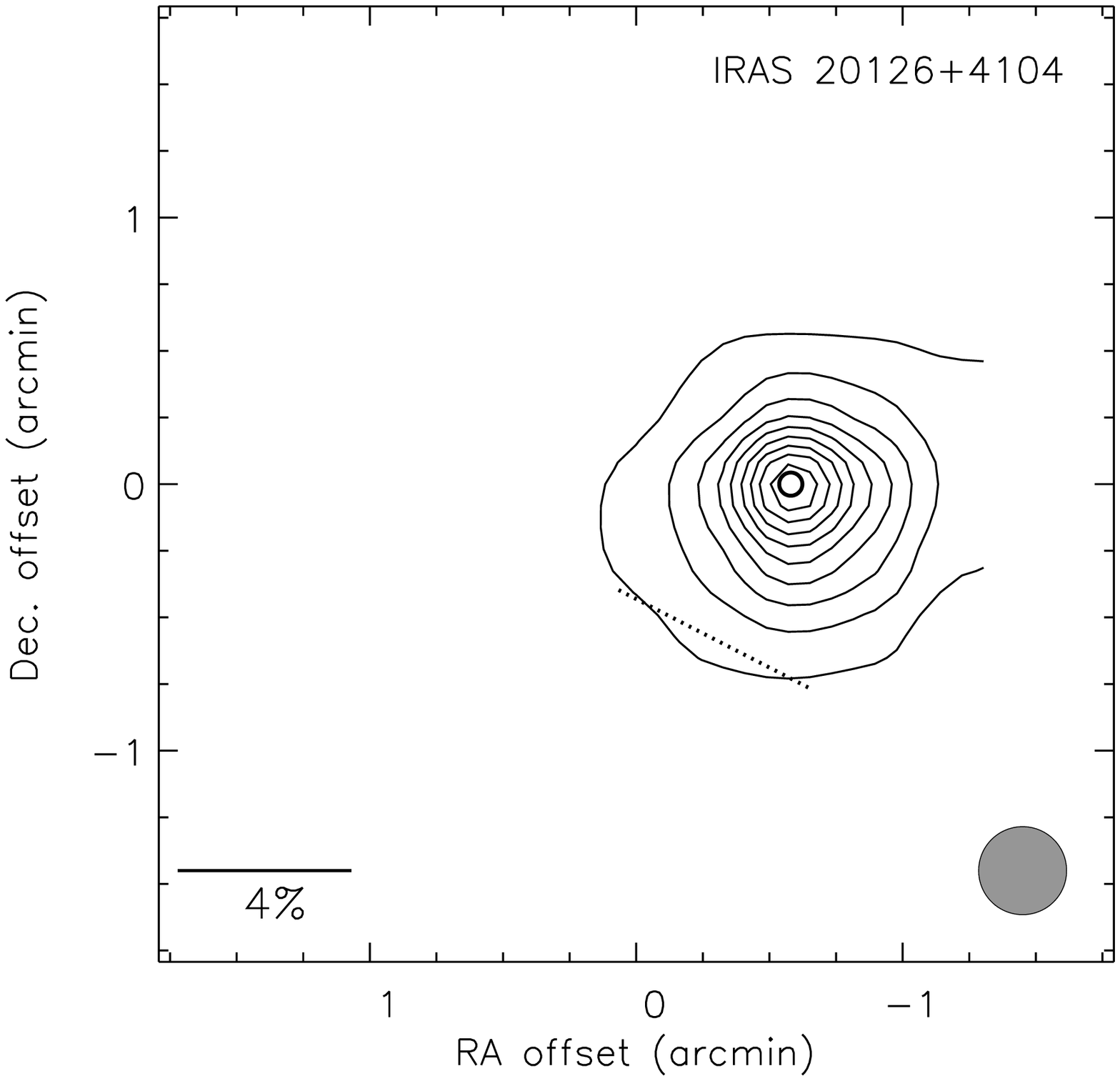}
\caption{IRAS 20126+4104.
Offsets from $20\hour14\minute29\fs4$, $
41\arcdeg13\arcmin34\arcsec$ (J2000).
Contours at 10, 20, ..., 90\% of the peak flux of 110\,Jy.
\label{fig56}}
\end{figure*}
\begin{figure*}
\plotone{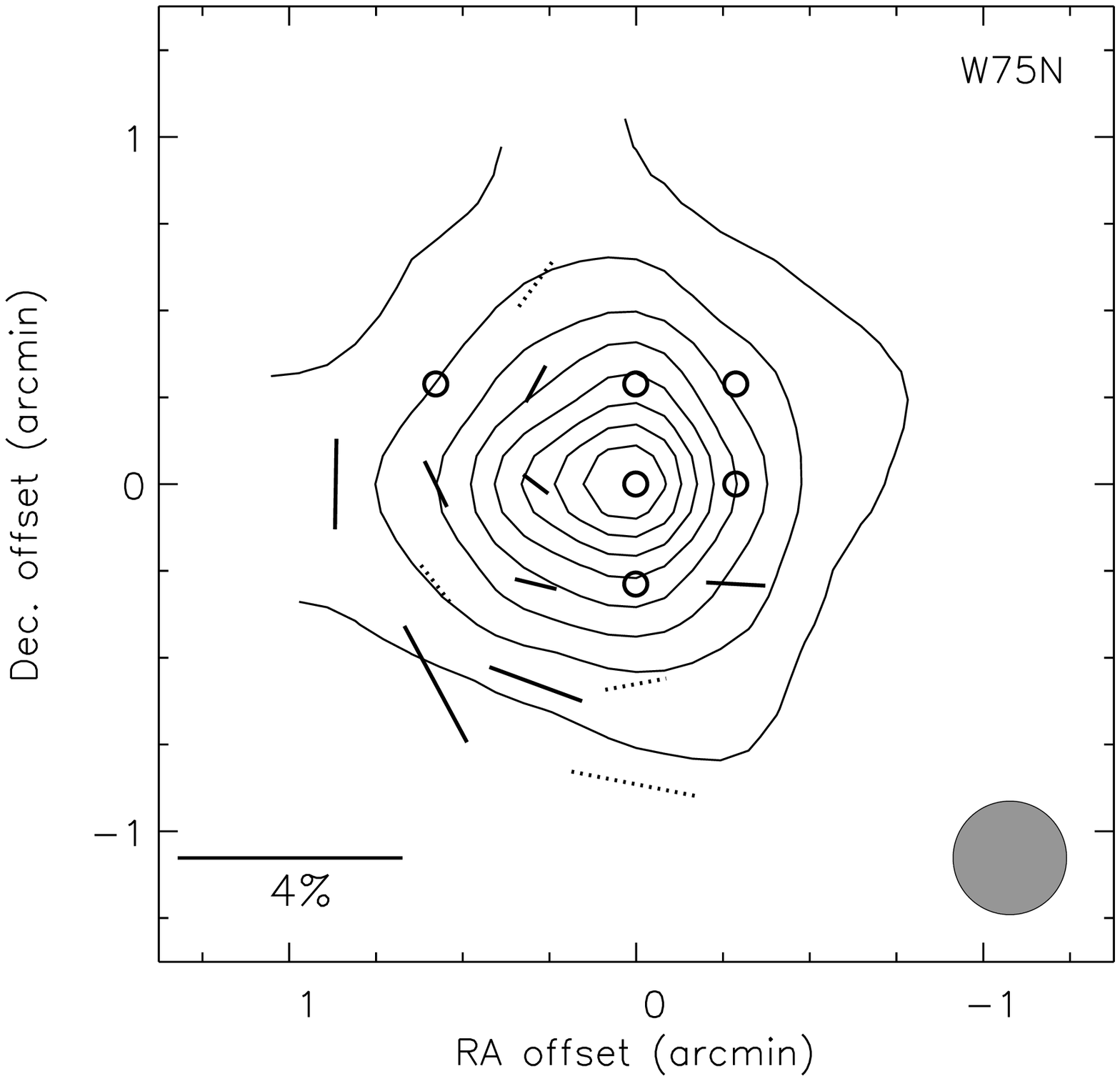}
\caption{W75N.
Offsets from $20\hour38\minute36\fs4$, $
42\arcdeg37\arcmin34\arcsec$ (J2000).
Contours at 20, 30, ..., 90\% of the peak flux of 670\,Jy.
\label{fig57}}
\end{figure*}
\begin{figure*}
\plotone{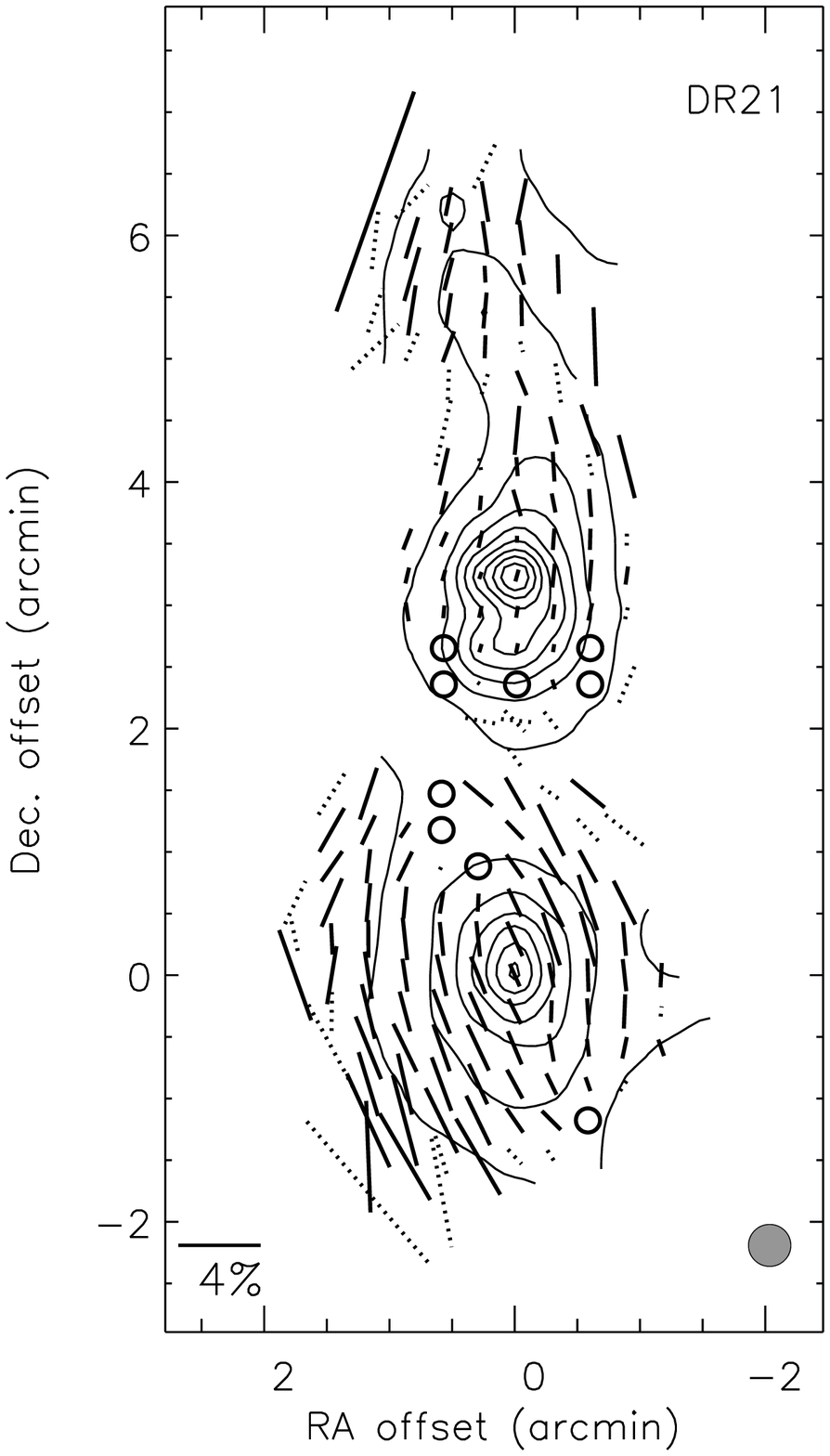}
\caption{DR21.
Offsets from $20\hour39\minute1\fs1$, $
42\arcdeg19\arcmin31\arcsec$ (J2000).
Contours at 10, 20, ..., 90\% of the peak flux of 820\,Jy (at ($\Delta\alpha, \Delta\delta$) = (0\arcmin, 3\farcm3)).
\label{fig58}}
\end{figure*}
\begin{figure*}
\plotone{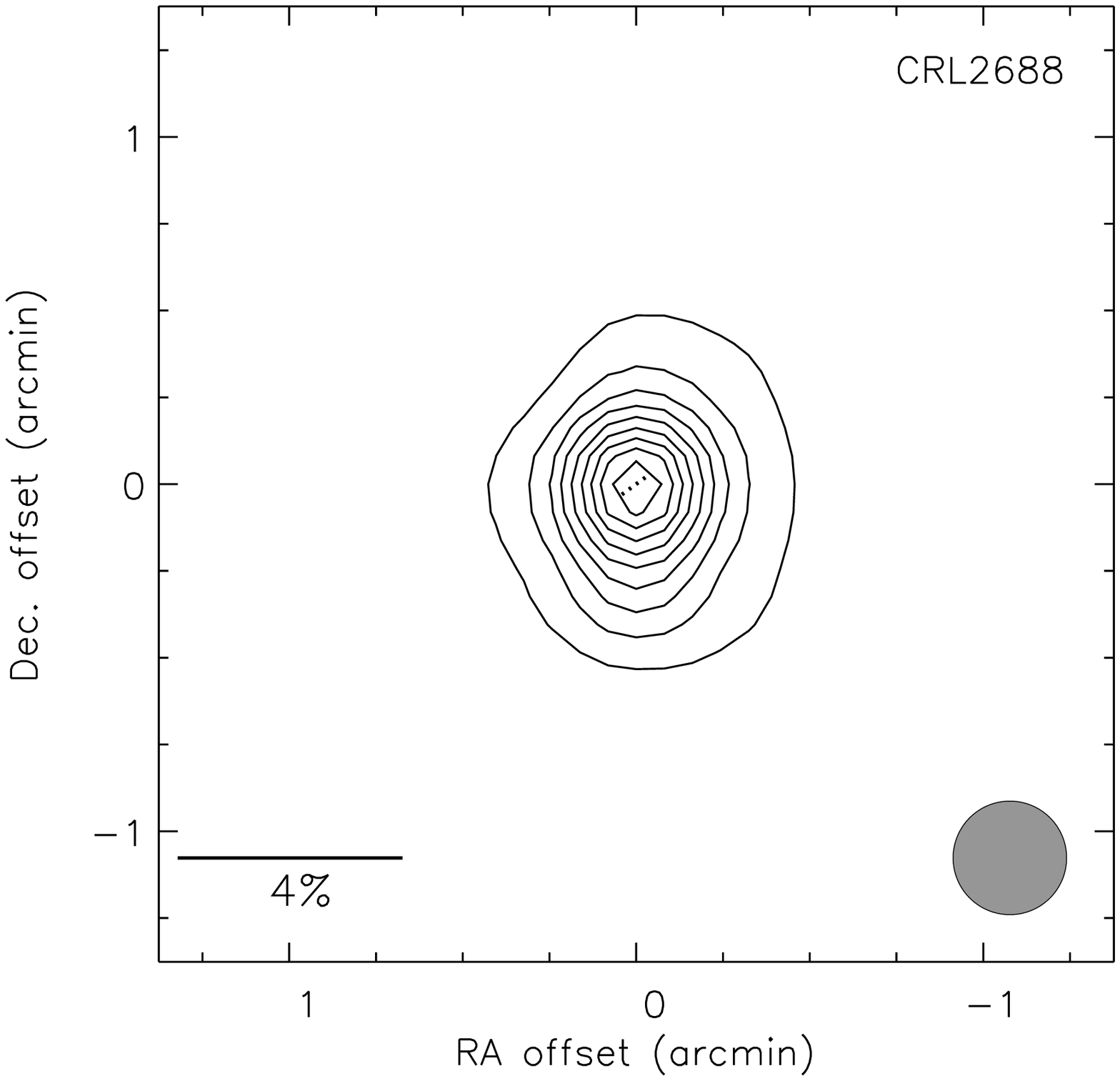}
\caption{CRL2688.
Offsets from $21\hour2\minute18\fs7$, $
36\arcdeg41\arcmin38\arcsec$ (J2000).
Contours at 10, 20, ..., 90\% of the peak flux of 46\,Jy.
\label{fig59}}
\end{figure*}

\acknowledgments We are grateful to the staff of the Caltech
Submillimeter Observatory for their invaluable assistance.  We would
also like to thank Giles Novak, Dave Chuss, Martin Houde, and David
Schleuning for assistance with the observations and many invaluable
conversations.  This work was supported by NSF grant AST 0505124 and
by Vaillancourt's NASA Graduate Student Research Program (NGT
5-63). The CSO has been supported by NSF grants AST 05-40882 and
08-38261.

{\it Facilities:} \facility{CSO (Hertz)}

\clearpage

\hsize 11in
\vsize8.5in

\LongTables
\begin{landscape}
  \tabletypesize{\scriptsize}
  \begin{deluxetable*}{llclllllccc}

\tablewidth{0pt}
\tablecaption{Object List \label{tbl-object}}
\tablehead{ \colhead{Source} & \colhead{Runs\tablenotemark{a}} & \colhead{$\alpha$ (2000)} & \colhead{$\delta$ (2000)}
 & \colhead{$l$} & \colhead{$b$} & \colhead{Chop Throw} & \colhead{Chop Angle} & \colhead{Peak Flux} & \colhead{Flux} & \colhead{Previously}\\
\colhead{} & \colhead{} & \colhead{(hh:mm:ss.s)} & \colhead{(dd:mm:ss)}
 & \colhead{(deg.)} & \colhead{(deg.)} & \colhead{(arcsec)} & \colhead{(degrees E of N)} & \colhead{(Jy/beam\tablenotemark{b})} & \colhead{Reference\tablenotemark{c}} & \colhead{Published\tablenotemark{d}}}

\startdata
\object{NGC 253} 	&	7	&	      00:47:33.1 	&	                 $-$25:17:15 	&	     \phn$97.37$ 	&	       $-87.96$ 	&	188	&	89 -- 111	&	    \phn110\phd\phn 	&	          \\
W3 	&	 2,5 	&	      02:25:40.7 	&	          \phs62:05:52 	&	        $133.71$ 	&	 \phs\phn$1.22$ 	&	347 -- 386	&	42 -- 123	&	    \phn480\phd\phn 	&	& 8,17         \\
\object{NGC 1333} 	&	 7,12 	&	  03:29:03.8 	&	          \phs31:16:03 	&	        $158.35$ 	&	       $-20.56$ 	&	301 -- 381	&	12 -- 160	&	 \phn\phn75\phd\phn 	&	          \\
L1551 	&	7	&	      04:31:34.2 	&	      \phs18:08:05 	&	        $178.93$ 	&	       $-20.05$ 	&	378 -- 380	&	 10, 171 -- 172 	&	 \phn\phn95\phd\phn 	&	        1 \\
\object[Ced 55e]{IRAS 05327-0457} 	&	 9,12 	&	      05:35:14.4 	&	              $-$04:55:45 	&	        $208.57$ 	&	       $-19.18$ 	&	246 -- 507	&	32 -- 62	&	 \phn\phn25\phd\phn 	&	  NGC 2071\\
\object[OMC 1]{OMC-1} 	&	 3,5,6,7,9,10,12 	&	      05:35:14.5 	&	              $-$05:22:32 	&	        $208.99$ 	&	       $-19.38$ 	&	297 -- 502	&	32 -- 151	&	       2100\phd\phn 	&	& 5,6,7,8,9          \\
\object[OMC 2]{OMC-2} 	&	 2,3,5,12 	&	      05:35:26.7 	&	          $-$05:10:00 	&	        $208.82$ 	&	       $-19.25$ 	&	306 -- 386	&	 36 -- 79, 99 -- 148 	&	    \phn200\phd\phn 	&	& 5,6,7,8,9          \\
\object[OMC 3]{OMC-3} 	&	 2,3,5,12 	&	      05:35:23.5 	&	          $-$05:01:32 	&	        $208.68$ 	&	       $-19.19$ 	&	246 -- 386	&	37 -- 146	&	    \phn110\phd\phn 	&	& 5,6,7,8,9          \\
\object[OMC 4]{OMC-4} 	&	 7,10,12 	&	  05:35:08.2 	&	              $-$05:35:56 	&	        $209.19$ 	&	       $-19.51$ 	&	306 -- 385	&	 101, 137 -- 152 	&	 \phn\phn37\phd\phn 	&	& 5,6,7,8,9          \\
\object[LDN 1641N]{L1641N} 	&	 3,6 	&	      05:36:18.8 	&	              $-$06:22:11 	&	        $210.06$ 	&	       $-19.59$ 	&	372 -- 387	&	 108 -- 126, 143 -- 152 	&	    \phn\phn89\phd\phn 	&	          \\
\object{NGC 2023} 	&	6	&	      05:41:25.4 	&	          $-$02:18:06 	&	        $206.86$ 	&	       $-16.60$ 	&	363 -- 365	&	132 -- 141	&	 \phn\phn43\phd\phn 	&	          \\
\object{NGC 2024} 	&	 2,3 	&	      05:41:43.0 	&	              $-$01:54:22 	&	        $206.53$ 	&	       $-16.36$ 	&	350 -- 387	&	32 -- 126	&	    \phn470\phd\phn 	&	          \\
HH24MMS 	&	10	&	  05:46:08.4 	&	              $-$00:10:43 	&	        $205.49$ 	&	       $-14.57$ 	&	370 -- 374	&	147 -- 154	&	 \phn\phn26\phd\phn 	&	          \\
\object[M78]{NGC 2068 LBS17} 	&	12	&	      05:46:28.0 	&	          $-$00:00:54 	&	        $205.38$ 	&	       $-14.42$ 	&	300 -- 302	&	131 -- 150	&	 \phn\phn17\phd\phn 	&	  NGC 2071\\
\object[M78]{NGC 2068 LBS10} 	&	12	&	      05:46:50.2 	&	  \phs00:02:01 	&	        $205.38$ 	&	       $-14.32$ 	&	300 -- 306	&	95 -- 154	&	 \phn\phn28\phd\phn 	&	  NGC 2071\\
\object{NGC 2071} 	&	 1,3,5 	&	  05:47:04.9 	&	          \phs00:21:47 	&	        $205.11$ 	&	       $-14.11$ 	&	250 -- 387	&	65 -- 162	&	    \phn180\phd\phn 	&	        1 \\
\object[GAL 213.71-12.60]{Mon R2} 	&	 3,10 	&	  06:07:46.6 	&	              $-$06:23:16 	&	        $213.71$ 	&	       $-12.60$ 	&	370 -- 387	&	128 -- 154	&	    \phn280\phd\phn 	&	          \\
\object[GGD 12]{GGD12} 	&	3	&	      06:10:50.4 	&	              $-$06:11:46 	&	        $213.88$ 	&	       $-11.84$ 	&	393	&	118 -- 134	&	    \phn260\phd\phn 	&	          \\
S269 	&	 6,7 	&	      06:14:36.6 	&	              \phs13:49:35 	&	        $196.45$ 	&	    \phn$-1.68$ 	&	363 -- 410	&	163 -- 166	&	 \phn\phn58\phd\phn 	&	          \\
\object[IRAS 06319+0415]{AFGL 961} 	&	6	&	      06:34:37.7 	&	          \phs04:12:44 	&	        $207.27$ 	&	    \phn$-1.81$ 	&	363 -- 367	&	 84 -- 98, 153 -- 160 	&	 \phn\phn37\phd\phn 	&	          \\
\object[IRAS 06381+1039]{Mon OB1 27 (IRAS 06381+1039)} 	&	9	&	     06:40:58.3 	&	              \phs10:36:54 	&	        $202.30$ 	&	 \phs\phn$2.53$ 	&	175 -- 196	&	107 -- 163	&	 \phn\phn29\phd\phn 	&	          \\
 &	10	&		&		&		&		&	372 -- 375	&	158 -- 162	&		&	          \\
\object[IRAS 06382+1017]{Mon OB1 25 (IRAS 06382+1017)} 	&	10	&	 06:41:03.7 	&	          \phs10:15:07 	&	        $202.63$ 	&	 \phs\phn$2.38$ 	&	372 -- 375	&	157 -- 163	&	 \phn\phn25\phd\phn 	&	          \\
\object[IRAS 06382+0939]{Mon OB1 12 (IRAS 06382+0939)}	&	 9,12 	&	  06:41:06.1 	&	      \phs09:34:09 	&	        $203.24$ 	&	 \phs\phn$2.08$ 	&	167 -- 182	&	20 -- 70	&	 \phn\phn62\phd\phn 	&	          \\
 &	12	&		&		&		&		&	240 -- 249	&	157 -- 162	&		&	          \\
\object{NGC 2264} 	&	 1,3,5 	&	      06:41:10.3 	&	          \phs09:29:27 	&	        $203.32$ 	&	 \phs\phn$2.06$ 	&	 249, 379 -- 393 	&	112 -- 162	&	    \phn180\phd\phn 	&	          \\
\object[NAME ROTTEN EGG NEBULA]{OH231 (IRAS 07399-1435)} 	&	10	&	      07:42:17.0 	&	                 $-$14:42:49 	&	        $231.83$ 	&	 \phs\phn$4.22$ 	&	370 -- 374	&	135 -- 142	&	 \phn\phn20\phd\phn 	&	        1 \\
\object[NAME PEANUT NEBULA]{IRC+10216} 	&	 1,3,6,10 	&	      09:47:57.3 	&	              \phs13:16:43 	&	        $221.45$ 	&	    \phs$45.06$ 	&	120 -- 387	&	83 -- 165	&	 \phn\phn30\phd\phn 	&	        1 \\
\object{M 82}	&	 1,3,5,6,10 	&	      09:55:52.2 	&	              \phs69:40:46 	&	        $141.41$ 	&	    \phs$40.57$ 	&	\phn90 -- 387	&	31 -- 109	&	 \phn\phn45\phd\phn 	&	          \\
\object[* rho Oph]{$\rho$ Oph} 	&	 1,3,5 	&	         16:26:27.5 	&	                 $-$24:23:54 	&	        $353.08$ 	&	    \phs$16.91$ 	&	341 -- 506	&	42 -- 94	&	    \phn110\phd\phn 	&	        2 \\
\object[NAME IRAS 1629A]{IRAS 16293-2422} 	&	 1,4,6,9,12 	&	         16:32:22.9 	&	                 $-$24:28:36 	&	        $353.94$ 	&	    \phs$15.84$ 	&	120 -- 249	&	42 -- 92	&	    \phn240\phd\phn 	&	        1 \\
 &	 3,5 	&		&		&		&		&	364 -- 378	&	138 -- 146	&		&	          \\
 &	 1,3 	&		&		&		&		&	378 -- 506	&	45 -- 91	&		&	          \\
CB68 	&	 9,10 	&	         16:57:19.5 	&	             $-$16:09:21 	&	  \phn\phn$4.50$ 	&	    \phs$16.34$ 	&	118 -- 382	&	35 -- 155	&	 \phn\phn14\phd\phn 	&	          \\
NGC 6334V 	&	3	&	         17:19:57.4 	&	                 $-$35:57:46 	&	        $351.16$ 	&	 \phs\phn$0.70$ 	&	377	&	145 -- 167	&	    \phn650\phd\phn 	&	 NGC 6334I\\
NGC 6334A 	&	 3,5 	&	         17:20:19.1 	&	                 $-$35:54:45 	&	        $351.25$ 	&	 \phs\phn$0.67$ 	&	364 -- 378	&	142 -- 178	&	    \phn480\phd\phn 	&	 NGC 6334I\\
\object[BDB2003 G351.41+00.64]{NGC 6334I} 	&	 3,5 	&	         17:20:53.4 	&	             $-$35:47:00 	&	        $351.42$ 	&	 \phs\phn$0.65$ 	&	367 -- 378	&	143 -- 168	&	       1200\phd\phn 	&	        1 \\
 &	6	&		&		&		&		&	367 -- 510	&	100 -- 129	&		&	          \\
\object[NAME SGR A-G]{M-0.13-0.08} 	&	 1,10 	&	         17:45:37.4 	&	             $-$29:05:40 	&	        $359.87$ 	&	    \phn$-0.08$ 	&	380 -- 500	&	 80 -- 86, 95 -- 117 	&	       350\phd\phn 	&	          \\
\object[SNR 000.0+00.0]{Sgr A East}	&	1	&	         17:45:41.5 	&	         $-$29:00:09 	&	        $359.95$ 	&	    \phn$-0.05$ 	&	376	&	80 -- 118	&	    \phn200\phd\phn 	&	& 14,15,16          \\
\object[CO 000.02 -00.02]{CO+0.02-0.02} 	&	10	&	         17:45:42.1 	&	             $-$28:56:05 	&	  $\phn\phn0.01$ 	&	    \phn$-0.01$ 	&	375 -- 392	&	92 -- 110	&	    \phn210\phd\phn 	&	          \\
\object[GCM -0.02 -0.07]{M-0.02-0.07} 	&	 1,10 	&	         17:45:51.7 	&	             $-$28:59:09 	&	        $359.99$ 	&	    \phn$-0.07$ 	&	375 -- 500	&	78 -- 128	&	    \phn270\phd\phn 	&	          \\
\object[GCM +0.07 -0.07]{M+0.07-0.08} 	&	10	&	      17:46:04.4 	&	                 $-$28:54:45 	&	  $\phn\phn0.07$ 	&	    \phn$-0.07$ 	&	387	&	109 -- 116	&	    \phn140\phd\phn 	&	          \\
M+0.11-0.08 	&	10	&	         17:46:10.3 	&	             $-$28:53:06 	&	  $\phn\phn0.11$ 	&	    \phn$-0.08$ 	&	387	&	86 -- 94	&	    \phn210\phd\phn 	&	          \\
\object[VB99 M0.25+0.01 P3]{M+0.25+0.01} 	&	 6,10,11,12 	&	         17:46:10.6 	&	                 $-$28:42:17 	&	  \phn\phn$0.26$ 	&	 \phs\phn$0.02$ 	&	360 -- 508	&	83 -- 141	&	    \phn310\phd\phn 	&	          \\
\object[GAL 000.18-00.04]{Sickle (G0.18-0.04) }	&	2	&	         17:46:14.9 	&	             $-$28:48:03 	&	  \phn\phn$0.19$ 	&	    \phn$-0.05$ 	&	480	&	91 -- 105	&	    \phn150\phd\phn 	&	          \\
M+0.34+0.06 	&	10	&	         17:46:13.2 	&	                 $-$28:36:53 	&	  \phn\phn$0.34$ 	&	 \phs\phn$0.05$ 	&	389 -- 403	&	 88 -- 91, 105 -- 108 	&	    \phn160\phd\phn 	&	          \\
\object[GAL 000.38+00.04]{M+0.40+0.04} 	&	 6,9,10 	&	         17:46:21.4 	&	                 $-$28:35:41 	&	  \phn\phn$0.38$ 	&	 \phs\phn$0.04$ 	&	383 -- 501	&	82 -- 132	&	    \phn220\phd\phn 	&	   Sgr B2 \\
\object[NAME SGR B1]{Sgr B1} 	&	6	&	         17:46:47.2 	&	             $-$28:32:00 	&	  \phn\phn$0.48$ 	&	    \phn$-0.01$ 	&	368 -- 385	&	128 -- 139	&	    \phn310\phd\phn 	&	          \\
 &	12	&		&		&		&		&	505	&	109 -- 120	&		&	\\
\object[SNR 000.7+00.0]{Sgr B2} 	&	 1,8,9,10,11,12 	&	         17:47:20.2 	&	             $-$28:23:06 	&	  \phn\phn$0.67$ 	&	    \phn$-0.04$ 	&	360 -- 508	&	50 -- 141	&	       3300\phd\phn 	&	        3    &    10,11,12\\
\object[W 33C]{W33 C (G12.8-0.2)} 	&	 6,10 	&	         18:14:13.5 	&	                 $-$17:55:32 	&	     \phn$12.81$ 	&	    \phn$-0.20$ 	&	367 -- 394	&	 92 -- 95, 126 -- 147 	&	    \phn480\phd\phn 	&	          \\
\object[GAL 012.91-00.26]{W33 A} 	&	 6,10 	&	         18:14:39.0 	&	             $-$17:52:04 	&	     \phn$12.91$ 	&	    \phn$-0.26$ 	&	375 -- 394	&	131 -- 145	&	    \phn170\phd\phn 	&	     W33 C\\
\object[LDN 483]{L483} 	&	10	&	         18:17:29.8 	&	              $-$04:39:38 	&	     \phn$24.88 $	&	 \phs\phn$5.38$ 	&	 184 -- 185, 387 -- 392 	&	142 -- 153	&	 \phn\phn31\phd\phn 	&	          \\
\object{M 17} 	&	 1,11 	&	         18:20:24.6 	&	             $-$16:13:02 	&	     \phn$15.01$ 	&	    \phn$-0.69$ 	&	370 -- 506	&	 45 -- 81, 102 -- 147 	&	    \phn700\phd\phn 	&	& 13         \\
\object[MSL2003 W43-MM1]{W43-MM1} 	&	11	&	         18:47:47.0 	&	              $-$01:54:29 	&	     \phn$30.82$ 	&	    \phn$-0.06$ 	&	507 -- 508	&	 96 -- 107, 130 -- 150 	&	    \phn590\phd\phn 	&	          \\
\object[GAL 034.3+00.2]{G34.3+0.2} 	&	 1,4,6 	&	         18:53:18.6 	&	          \phs01:14:59 	&	     \phn$34.26$ 	&	 \phs\phn$0.15$ 	&	370 -- 499	&	28 -- 154	&	       1100\phd\phn 	&	        1 \\
\object{W49 A} 	&	4	&	         19:10:13.6 	&	      \phs09:06:17 	&	     \phn$43.17$ 	&	 \phs\phn$0.01$ 	&	377 -- 418	&	 32 -- 56, 112 -- 172 	&	    \phn730\phd\phn 	&	          \\
\object[GAL 049.5-00.4]{W51 A (G49.5-0.4)} 	&	 2,4,6,8 	&	         19:23:44.1 	&	              \phs14:30:32 	&	     \phn$49.49$ 	&	    \phn$-0.39$ 	&	367 -- 402	&	14 -- 165	&	       1200\phd\phn 	&	        4 \\
\object[GAL 078.12+03.64]{IRAS 20126+4104} 	&	1	&	         20:14:29.4 	&	              \phs41:13:34 	&	     \phn$78.13$ 	&	 \phs\phn$3.62$ 	&	499 -- 500	&	96 -- 121	&	    \phn110\phd\phn 	&	          \\
W75 N 	&	2	&	         20:38:36.4 	&	              \phs42:37:34 	&	     \phn$81.87$ 	&	 \phs\phn$0.78$ 	&	265 -- 378	&	11 -- 30	&	    \phn670\phd\phn 	&	        1 \\
\object[DR 21]{DR21} 	&	 1,2,10 	&	     20:39:01.1 	&	              \phs42:19:31 	&	     \phn$81.68$ 	&	 \phs\phn$0.54$ 	&	369 -- 383	&	22 -- 145	&	    \phn820\phd\phn 	&	& 19          \\
\object[NAME EGG NEBULA]{CRL2688}	&	 6,11 	&	21:02:18.7	&	\phs36:41:38	&	\phn$80.17$	&	\phn$-6.50$	&	 127, 179 	&	 151 -- 160, 9 -- 98 	&	\phn\phn46\phd\phn	&	1 \\
\enddata
\tablenotetext{a}{Observing dates: (1)~1997~Apr~18--27;
(2)~1997~Sep~18--26; (3)~1998~Feb~14--19; (4)~1998~May~2--3;
(5)~1999~Jan~27--30; (6)~1999~Apr~4--8; (7)~2000~Jan~4--8;
(8)~2000~Jul~24; (9)~2001~Feb~21--26; (10)~2001~Apr~10--16;
(11)~2001~Jul~19--20; (12)~2002~Feb~14--18}
\tablenotetext{b}{Flux into a $20$\arcsec\ Hertz beam.}
\tablenotetext{c}{Reference or calibration source for flux
  estimate. Unreferenced objects are calibrated with respect to
  multiple sources observed with Hertz. These sources include Mars,
  Uranus, W3OH, and/or other objects in the table with known fluxes.
}
\tablenotetext{d}{See Table references below.}
\tablerefs{(1) \citealt{sandell94}; (2) \citealt{andre93}; (3) \citealt{goldsmith90}; (4) \citealt{ladd93}; (5) \citealt{das96}; (6) \citealt{das98}; (7) \citealt{houde04}; (8) \citealt{rhh02}; (9) \citealt{vaillancourt02}; (10) \citealt{cdd97}; (11) \citealt{gn97}; (12) \citealt{cdd98}; (13) \citealt{houde02}; (14) \citealt{gn98}; (15) \citealt{gn00}; (16) \citealt{dtc03}; (17) \citealt{das00}; (18) \citealt{rhh09}; (19) \citealt{kirby09}}

  \end{deluxetable*}
\clearpage
\end{landscape}

\hsize 7.1in
\vsize 11in
\tabletypesize{\normalsize}
\begin{deluxetable*}{lccccccccccc}
\tablecaption{Data}
\tablenum{2}
\tablehead{\colhead{Name} & \colhead{$\Delta$RA} & \colhead{$\Delta$Dec} & \colhead{$\Delta x$} & \colhead{$\Delta y$} & \colhead{$P$} & \colhead{$\sigma(P)$} & \colhead{$\phi$} & \colhead{$\sigma(\phi)$} & \colhead{Flux} & \colhead{$\sigma$(Flux)} & \colhead{Number of} \\ 
\colhead{} & \colhead{(arcsec)} & \colhead{(arcsec)} & \colhead{(pix)} & \colhead{(pix)} & \colhead{(\%)} & \colhead{(\%)} & \colhead{(deg)} & \colhead{(deg)} & \colhead{(Jy)} & \colhead{(Jy)} & \colhead{Observations} } 

\startdata
NGC 253  &  $-71$  &  $-18$        &  $-4.0$  &  $-1.0$     &  14.28     & 21.54    &  177.8   & \phn43.5 & \phn9.0  &  2.0  & \phn7 \\
NGC 253  &  $-71$  &  \phs \phn 0  &  $-4.0$  &  \phs 0.0   &  13.52     & 22.40    &  147.0   & \phn46.9 & \phn6.2  &  2.0  & \phn7 \\
NGC 253  &  $-53$  &  $-36$        &  $-3.0$  &  $-2.0$     &  18.78     & 17.91    & \phn37.5 & \phn27.2 & \phn4.8  &  0.5  &  11 \\
\multicolumn{1}{c}{\vdots}   & \vdots  & \vdots        & \vdots   & \vdots      & \vdots     & \vdots   & \vdots   & \vdots   & \vdots  &\vdots & \vdots \\
W3       & $-125$  & $-18$         & $-7.0$   & $-1.0$      &   0.41     & 1.20     & 112.2    & 84.0     & 101.2    &  14.3 &   \phn3 \\
W3       & $-125$  & \phs\phn0     & $-7.0$   & \phs0.0     &   0.66     & 0.44     & \phn49.7 & 19.2     & \phn94.9 & \phn2.0 &   12  \\
W3       & $-125$  & \phs18        & $-7.0$   & \phs1.0     &   1.60     & 0.45     &  30.7    & \phn8.0  & \phn89.4 & \phn1.8 &   17
\enddata
\tablecomments{Table~2 is published in its entirety in the electronic
  edition of the {\it Astrophysical Journal}.  A portion is shown here
  for guidance regarding its form and content. Pixel offsets in the
  $x$ and $y$ directions are given in units of the pixel
  center-to-center spacing, $17\farcs8$. Right Ascension and
  Declination offsets are calculated using this spacing and are
  rounded to the nearest arcsecond; offsets are given with respect to
  the object centers given in Table~1.}
\end{deluxetable*}


\begin{thebibliography}{}

\bibitem[{{Andre} {et~al.}(1993){Andre}, {Ward-Thompson}, \&
  {Barsony}}]{andre93}
{Andre}, P., {Ward-Thompson}, D., \& {Barsony}, M. 1993, \apj, 406,
122

\bibitem[Chuss et al.(2003)]{dtc03} Chuss, D. T., Davidson, J. A., Dotson, J. L., Dowell, C. D., Hildebrand, R. H., Novak, G., \& Vaillancourt, J. E. 2003, \apj, 599, 1116 

\bibitem[Dotson et al.(2000)]{jld00} Dotson, J. L., Davidson, J. A., Dowell,
C. D., Schleuning, D. A., \& Hildebrand, R. H. 2000, \apjs, 128, 335

\bibitem[Dowell(1997)]{cdd97} Dowell, C. D. 1997, \apj, 487, 237

\bibitem[Dowell et al.(1998)]{cdd98} Dowell, C. D., Hildebrand, R. H.,
Schleuning, D. A., Vaillancourt, J. E., Dotson, J. L., Novak, G., Renbarger,
T., \& Houde, M. 1998, \apj, 504, 588

\bibitem[{{Goldsmith} {et~al.}(1990){Goldsmith}, {Lis}, {Hills}, \&
  {Lasenby}}]{goldsmith90}
{Goldsmith}, P.~F., {Lis}, D.~C., {Hills}, R., \& {Lasenby}, J.
1990, \apj,  350, 186

\bibitem[Hildebrand(1988)]{rhh88} Hildebrand, R. H. 1988, \qjras, 29, 327

\bibitem[Hildebrand et al.(1990)]{rhh90} Hildebrand, R.~H., 
Gonatas, D.~P., Platt, S.~R., Wu, X.~D., Davidson, J.~A., Werner, M.~W., 
Novak, G., \& Morris, M. 1990, \apj, 362, 114 

\bibitem[Hildebrand et al.(1999)]{rhh99} Hildebrand, R. H., Dotson, J. L.,
Dowell, C. D., Schleuning, D. A., \& Vaillancourt, J. E. 1999, \apj, 516, 834

\bibitem[Hildebrand \& Dragovan(1995)]{rhh95} Hildebrand, 
R.~H. \& Dragovan, M. 1995, \apj, 450, 663

\bibitem[Hildebrand(2002)]{rhh02} Hildebrand, R. H. 2002, in Astrophysical Spectropolarimetry, ed. J. Trujillo-Bueno, F. Moreno-Insertis, and F. Sanchez, (Cambridge University Press), 265

\bibitem[Hildebrand et al.(2009)]{rhh09} Hildebrand, R. H., Kirby, L., Dotson, J. L., Houde, M., \& Vaillancourt, J. E. 2009, \apj, 696, 567

\bibitem[Houde et al.(2002)]{houde02} Houde, M., et al.\ 2002, 
\apj, 569, 803

\bibitem[Houde et al.(2004)]{houde04} Houde, M., Dowell, C. D., Hildebrand, R. H., Dotson, J. L., Vaillancourt, J. E., Phillips, T. G., Peng, R., \& Bastien, P. 2004, \apj, 604, 717

\bibitem[Kirby et al.(2005)]{lk05} Kirby, L., Davidson, 
J.~A., Dotson, J.~L., Dowell, C.~D., \& Hildebrand, R.~H. 2005, \pasp, 
117, 991 
 
\bibitem[Kirby(2009)]{kirby09} Kirby, L. 2009, \apj, 694, 1056

\bibitem[{Ladd {et~al.}(1993)Ladd, Deane, Goldader, Sanders, \&
  Wynn-Williams}]{ladd93}
Ladd, E.~F., Deane, J.~R., Goldader, J.~D., Sanders, D.~B., \&
Wynn-Williams,
  C.~G. 1993, in AIP Conf.\ Proc.\ 278, Back to the Galaxy, ed. S.~S. Holt \&
  F.~Verter (New York: AIP), 246
 
\bibitem[Li et al.(2008)]{li07} 
Li, H., Dowell, C. D., Kirby, L., Novak, G., \& Vaillancourt, J. E. 2008,  \ao, 47, 422

\bibitem[Novak et al.(1997)]{gn97}  Novak, G., Dotson, J.~L., Dowell, C.~D., 
Goldsmith, P.~F., Hildebrand, R.~H.,
  Platt, S.~R., \& Schleuning, D.~A. 1997, \apj, 487, 320

\bibitem[Novak et al.(1998)]{gn98} Novak, G., Dotson, J. L.,
  Renbarger, T., Dowell, C. D., Hildebrand, R. H., \& Schleuning,
  D. A. 1998, in Proc.\ IAU Symp.\ 184, The Central Regions of the
  Galaxy and galaxies, ed. Y. Sofue (Dordrecht: Kluwer), 349

\bibitem[Novak et al.(2000)]{gn00} 
Novak, G., Dotson, J. L., Dowell, C. D., Hildebrand, R. H., Renbarger, T., \&
2000, \apj, 529, 241

\bibitem[Novak et al.(2004)]{gn04}
Novak, G., et al.
2004, in Proc.\ SPIE 5498, Millimeter adn Submillimeter Detectors for
Astronomy II, ed. J.~Zmuidzinas, W.~S. Holland, \& S.~Withington, 278

\bibitem[Platt et al.(1991)]{srp91} Platt, S. R., Hildebrand, R. H., Pernic,
R. J., Davidson, J. A., \& Novak, G. 1991, \pasp, 103, 1193

\bibitem[Sandell(1994)]{sandell94} Sandell, G. 1994, \mnras, 271, 75

\bibitem[Schleuning et al.(1996)]{das96} Schleuning, D. A., Dowell,
  C. D., \& Platt, S. R. 1996, in ASP Conf.\ Ser.\ 97, Polarimertry of
  the Interstellar Medium, eds. W. G. Roberge \& D. C. B. Whittet (San
  Francisco: ASP), 285

\bibitem[Schleuning et al.(1997)]{das97} Schleuning, D. A., Dowell, C. D.,
Hildebrand, R. H., Platt, S. R., \& Novak, G. 1997, \pasp, 109, 307

\bibitem[Schleuning(1998)]{das98} Schleuning, D. A. 1998, \apj, 493, 811

\bibitem[Schleuning et al.(2000)]{das00} Schleuning, D.~A., 
Vaillancourt, J.~E., Hildebrand, R.~H., Dowell, C.~D., Novak, G., Dotson, 
J.~L., \& Davidson, J.~A. 2000, \apj, 535, 913

\bibitem[{Simmons \& Stewart(1985)}]{simmons85}
Simmons, J. F.~L. \& Stewart, B.~G. 1985, \aap, 142, 100

\bibitem[Vaillancourt(2002)]{vaillancourt02}Vaillancourt, J.E. 2002, \apj, 142, 53

\bibitem[Vaillancourt(2006)]{vaillancourt06}Vaillancourt, J.E. 2006, \pasp, 118, 1340

\end{thebibliography}
\end{document}